\documentclass[rmp,aps,twocolumn]{revtex4}

\usepackage{graphics}
\usepackage{epsfig}

\def\inbar{\,\vrule height1.5ex width.4pt depth0pt}
\def\IR{\relax{\rm I\kern-.18em R}}
\def\IC{\relax\hbox{$\inbar\kern-.3em{\rm C}$}}

\newcommand{\gaz}{g_A^{\mbox{$\scriptscriptstyle (Z)$}}}
\newcommand{\run}[1]{\widetilde{\alpha}_{#1}}
\newcommand{\hsp}[1]{\hspace*{#1 mm}}
\newcommand{\smallfrac}[2]{\mbox{\small ${\displaystyle \frac{#1}{#2}}$}}
\newcommand{\footfrac}[2]%

\newcommand{\cxx}{1 - {{4x^{2} P^{2}}\over{Q^{2}}} }
\def\cx{1 - {{2x    P^{2}}\over{Q^{2}}} }

\def\cut{\sqrt{1 - {{4(m^{2}+\lambda^2)}\over {s}} } }

\begin{document}
\title{The Spin Structure of the Proton}

\author{Steven D. Bass}
\email{Steven.Bass@cern.ch}
\affiliation{Institute for Theoretical Physics,
University of Innsbruck, A-6020 Innsbruck, Austria}
\affiliation {Particle Physics Theory Group, 
              Paul Scherrer Institute, 
              CH-5232 Villigen PSI, Switzerland}

\begin{abstract}  
This article reviews our present understanding of the QCD spin structure 
of the proton.
We first outline the proton spin puzzle and its possible resolution in
QCD.
We then review the present and next generation of experiments to resolve
the proton's spin-flavour structure, explaining the theoretical issues
involved, the present status of experimental investigation and the open
questions and 
challenges for future investigation.
\end{abstract}                                                                 

\maketitle

\section{INTRODUCTION}

Understanding the spin structure of the proton is one of the most challenging 
problems facing subatomic physics:
How is the spin of the proton built up out from the intrinsic spin and 
orbital angular momentum of its quark and gluonic constituents ?
What happens to spin in the transition between current and constituent
quarks in low-energy quantum chromodynamics (QCD) ?
Key issues include the role of polarized glue and gluon topology in building 
up the spin of the proton.

The story of the proton's spin dates from the discovery by
\textcite{Dennison:1927}
that the 
proton is a fermion of spin ${1 \over 2}$.
Six years later \textcite{Stern:1933} 
measured the proton's anomalous magnetic moment,
$\kappa_p = 1.79$ Bohr magnetons, 
revealing that the proton is not pointlike and has internal structure.
The challenge to understand the structure of the proton had begun!

We now understand the proton as a bound state of three confined 
valence quarks
(spin 1/2 fermions)
interacting 
through spin-one gluons, with the gauge group being colour SU(3)
\cite{Thomas:2001}.
The proton is special because of confinement, 
dynamical chiral symmetry breaking and the very strong colour gauge fields at
large distances.

Our present knowledge about the spin structure of the proton at the quark 
level comes from polarized deep inelastic scattering experiments (pDIS)
which use high-energy polarized electrons or muons to probe the structure 
of a polarized proton 
and 
new experiments in 
semi-inclusive polarized deep inelastic scattering,
polarized proton-proton collisions and polarized photoproduction experiments.

The present excitement and global programme in high energy spin 
physics was inspired by an intriguing discovery in polarized deep inelastic
scattering.
Following pioneering experiments at SLAC 
\cite{Alguard:1976,Alguard:1978,Baum:1983}, 
recent experiments
in fully inclusive polarized deep inelastic scattering 
\cite{Windmolders:1999}
have extended measurements of the nucleon's $g_1$ spin dependent 
structure function 
(the inclusive form-factor measured in these experiments)
over a broad kinematic region where one is sensitive to scattering 
on the 
{\it valence} quarks
{\it plus} the quark-antiquark {\it sea} fluctuations.
These experiments have been interpreted to imply that quarks 
and anti-quarks carry just a small fraction of 
the proton's spin 
(between about 15\% and 35\%) 
-- less than half the prediction of relativistic constituent 
quark models ($\sim 60\%$).
This result has inspired vast 
experimental and theoretical activity to understand the spin 
structure of the proton.
Before embarking on a detailed study of the spin structure of 
the proton it is essential to understand why the small value of 
this ``quark spin content''
measured in polarized deep inelastic scattering
caused such excitement and why it 
has so much challenged our understanding of the structure of the 
proton.
We give a brief survey in Section I.A.
An outline of the review is given in Section I.B.

Many elements of subatomic physics and quantum field theory are 
important in our understanding of the proton spin problem.
These include: 
\begin{itemize}
\item
the dispersion relations for polarized photon-nucleon scattering, 
\item
Regge theory and the high-energy behaviour of scattering amplitudes, 
\item
the renormalization of the operators which enter the light-cone
operator product expansion description of high energy polarized 
deep inelastic scattering, 
\item
perturbative QCD: 
the physics of large transverse momentum plus parton model factorization, 
\item
the non-perturbative and non-local topological properties of gluon 
gauge fields in QCD,
\item
the role of gluon dynamics in dynamical chiral symmetry breaking
(the large mass of the $\eta$ and $\eta'$ mesons and 
 the absence of a flavour-singlet pseudoscalar Goldstone boson in
 spontaneous chiral symmetry breaking).
\end{itemize}

The purpose of this article is to review our present understanding 
of the proton spin problem
and the physics of the new and ongoing programme 
aimed at resolving the spin-flavour structure of the nucleon.

\subsection{Spin and the proton spin problem}

Spin plays an essential role in particle interactions and the fundamental 
structure of matter, ranging from the subatomic world through 
to large-scale macroscopic effects in condensed matter physics 
(e.g. Bose-Einstein condensates, 
      superfluidity,
      and exotic phases of low temperature $^3$He) 
and the structure of dense stars. 
Spin is essential for the stability of the known Universe.
In applications,
polarized neutron beams are used to probe the structure of 
condensed matter and materials systems.
Manipulating the spin of the electron may prove 
to be a key ingredient
in designing and constructing a quantum computer 
-- the new field of ``spintronics'' \cite{Zutic:2004}.

Spin is the characteristic property of a particle besides its mass and
gauge charges.
The two invariants of the Poincare group are
\begin{eqnarray}
{\cal P}_{\mu} {\cal P}^{\mu} &=& M^2 
\nonumber \\
{\cal W}_{\mu} {\cal W}^{\mu} &=& - M^2 s (s+1) 
.
\label{eqa3}
\end{eqnarray}
Here ${\cal P}$ and ${\cal W}$ denote the momentum and Pauli-Lubanski
spin vectors respectively, $M$ is the particle mass and $s$ denotes its
spin.
The spin of a particle, whether elementary or composite, 
determines its 
equation of motion and its statistics properties.
The discovery of spin and its properties are reviewed in
\textcite{Tomonaga:1997} and \textcite{Martin:2002}.
Spin ${1 \over 2}$ particles are governed by the Dirac equation 
and Fermi-Dirac statistics
whereas 
spin 0 and spin 1 particles are governed by the Klein-Gordon equation and 
Bose-Einstein statistics.

The proton's spin vector $s_{\mu}$ is measured through the forward matrix
element of the axial-vector current
\begin{equation}
2M s_{\mu} 
=
\langle p,s |
\
{\overline \psi} \gamma_{\mu} \gamma_5 \psi 
\
| p,s \rangle 
\label{eqa4}
\end{equation}
where $\psi$ denotes the proton field operator and $M$ is the proton mass.
The quark axial charges
\begin{equation}
2M s_{\mu} \Delta q =
\langle p,s |
\ {\overline q} \gamma_{\mu} \gamma_5 q \
| p,s \rangle 
\label{eqa5}
\end{equation}
then measure information about the quark ``spin content'' of the proton.
(Here $q$ denotes the quark field operator.)
The flavour dependent axial charges 
$\Delta u$, $\Delta d$ and $\Delta s$
can be written as linear combinations 
of the isovector, SU(3) octet and flavour-singlet axial charges
\begin{eqnarray}
g_A^{(3)} &=& \Delta u - \Delta d 
\nonumber \\
g_A^{(8)} &=& \Delta u + \Delta d - 2 \Delta s 
\nonumber \\
g_A^{(0)} &=& \Delta u + \Delta d + \Delta s 
.
\label{eqa6}
\end{eqnarray}
In semi-classical quark models $\Delta q$ is interpreted as the amount of 
spin carried by quarks and antiquarks of flavour $q$.

In polarized deep inelastic scattering experiments one measures 
the nucleon's $g_1$ spin structure function as a function of 
the Bjorken variable $x$, 
the fraction of the proton's momentum which carried 
by quark, antiquark and gluon partons in incoherent 
photon-parton scattering
with the proton boosted to an infinite momentum frame.
From the first moment of $g_1$, 
these
experiments 
have been interpreted to imply a small value 
for the 
flavour-singlet axial-charge:
\begin{equation}
g_A^{(0)}\bigr|_{\rm pDIS} = 0.15 - 0.35 .
\label{eqa1}
\end{equation}
When combined with the octet axial charge measured 
in hyperon beta-decays ($g_A^{(8)} = 0.58 \pm 0.03$) 
it 
corresponds to a negative strange-quark polarization
\begin{equation}
\Delta s = -0.10 \pm 0.04 
\label{eqa2}
\end{equation}
-- that is, 
polarized in the opposite direction to the spin of the proton.

The Goldberger-Treiman relation relates the isovector axial charge 
$g_A^{(3)}$ to the product of the pion decay constant $f_{\pi}$ and 
the pion-nucleon coupling constant $g_{\pi NN}$, viz.
\begin{equation}
2 M g_A^{(3)} = f_{\pi} g_{\pi NN}
\label{eqa7}
\end{equation}
through 
spontaneously broken chiral symmetry \cite{Adler:1968}. 
The Goldberger-Treiman relation leads immediately to the result 
that the spin structure of the proton is related to the dynamics of 
chiral symmetry breaking.

{\it
What happens to gluonic degrees of freedom ?
}
The axial anomaly, a fundamental property of 
quantum field theory, tells us that the axial-vector current which
measures the quark 
``spin content'' of the proton cannot be treated independently of 
the gluon fields that the quarks live in and that the quark
``spin content''
is linked to the physics of dynamical axial U(1) symmetry breaking in 
the flavour-singlet channel.
For each flavour $q$
the gauge invariantly renormalized axial-vector current
satisfies the anomalous divergence equation \cite{Adler:1969,Bell:1969}
\begin{equation}
\partial^{\mu}
\bigl(
{\overline q} \gamma_{\mu} \gamma_5 q 
\bigr)
= 
2 m {\overline q} i \gamma_5 q 
+ 
{\alpha_s \over 4 \pi} G_{\mu \nu} {\tilde G}^{\mu \nu}
.
\label{eqa8}
\end{equation}
Here $m$ denotes the quark mass and
$
{\alpha_s \over 4 \pi} G_{\mu \nu} {\tilde G}^{\mu \nu}
$
is the topological charge density.
The anomaly is important in the flavour-singlet channel
and intrinsic to $g_A^{(0)}$.
It cancels in
the non-singlet axial-vector currents
which define $g_A^{(3)}$ and $g_A^{(8)}$.
In the QCD parton model the anomaly corresponds to physics at 
the maximum transverse momentum squared \cite{Carlitz:1988}. 
The anomaly contribution also involves non-local structure 
associated with gluon field topology 
-- see \textcite{Jaffe:1990a} and \textcite{Bass:1998a,Bass:2003b}.
In dynamical axial U(1) symmetry breaking the anomaly and gluon
topology are
associated with the large masses of the $\eta$ and $\eta'$ mesons.

{\it
What values should we expect for the $\Delta q$ ?
}
First, consider the static quark model.
The simple SU(6) proton wavefunction
\begin{eqnarray}
|p \uparrow \rangle &=&
{1 \over \sqrt{2}} | u \uparrow (ud)_{S=0} \rangle
+
{1 \over \sqrt{18}} | u \uparrow (ud)_{S=1} \rangle
\nonumber \\
& & -
{1 \over 3} | u \downarrow (ud)_{S=1} \rangle 
-
{1 \over 3} | d \uparrow (uu)_{S=1} \rangle
\nonumber \\
& & +
{\sqrt{2} \over 3} | d \downarrow (uu)_{S=1} \rangle
\label{eqa9}
\end{eqnarray}
yields the values
$g_A^{(3)} = {5 \over 3}$ and $g_A^{(8)}=g_A^{(0)}=1$.

In relativistic quark models one has to take into account
the four-component Dirac spinor
$\psi = 
{N \over \sqrt{4 \pi}}
\biggl({ f \atop i \sigma .{\hat{r}} g }\biggr)$
where $N$ is a normalization factor.
The lower component of the Dirac spinor is p-wave with intrinsic spin
primarily pointing in the opposite direction to spin of the nucleon.
Relativistic effects renormalize the axial charges obtained from SU(6)
by the factor $ N^2 \int dr r^2 (f^2 - {1 \over 3} g^2)$
with a net transfer of angular momentum
from intrinsic spin to orbital angular momentum 
-- see e.g. \textcite{Jaffe:1990a}.

Relativistic constituent quark models 
(which do not include gluonic effects associated with 
 the axial anomaly)
generally predict values of
$g_A^{(3)} \simeq 1.25$ and $g_A^{(8)} \sim g_A^{(0)} \simeq 0.6$. 
For example, consider the MIT Bag Model. 
There,
$N^2 \int_0^R dr r^2 (f^2 - {1 \over 3} g^2) = 0.65$
where $R$ is the Bag radius.
This relativistic factor reduces
$g_A^{(3)}$ from ${5 \over 3}$ to 1.09 and $g_A^{(0)}$ to 0.65.
Centre of mass motion then increases the axial charges by about 20\%
bringing $g_A^{(3)}$ close to its physical value 1.26.
Pion cloud effects are also important.
In the SU(2) Cloudy Bag model one finds renormalization factors 
equal to 0.94 for the isovector axial charge and 0.8 for the 
isosinglet axial charges \cite{Schreiber:1988} corresponding 
to a shift of total angular momentum from intrinsic spin into orbital angular
momentum.
The resultant predictions are
$g_A^{(3)} \simeq 1.25$ 
(in agreement with experiment) and
$g_A^{(0)} = g_A^{(8)} \simeq 0.6$.
(Note that, at this level, relativistic quark-pion coupling models 
 contain no explicit
 strange quark or gluon degrees of freedom with the gluonic degrees
 of freedom understood to be integrated out into the scalar confinement
 potential.)
The model prediction $g_A^{(8)} \simeq 0.6$ 
agrees with 
the value extracted from hyperon beta-decays 
[$g_A^{(8)} = 0.58 \pm 0.03$ \cite{Close:1993}]
whereas the Bag model prediction for $g_A^{(0)}$ 
exceeds the measured value of $g_A^{(0)}|_{\rm pDIS}$ by a factor of 2-4.

The overall picture of the spin structure of the proton that has emerged 
from a combination of experiment and theoretical QCD studies 
can be summarized in the following key observations:
\begin{enumerate}
\item
Constituent quark model predictions work remarkably well for the isovector
part of the nucleon's $g_1$ spin structure function $(g_1^p-g_1^n)$:
both for the first moment 
$\int_0^1 dx (g_1^p-g_1^n) \sim {1 \over 6} g_A^{(3)}$
which is predicted by the Bjorken sum rule \cite{Bjorken:1966,Bjorken:1970}
and 
also over 
the whole presently measured range of Bjorken $x$
\cite{Bass:1999}.
This includes the SLAC ``small $x$'' region ($0.02 < x < 0.1$)
-- see Section IX.B below --
where one would, a priori, not necessarily expect quark model 
results to apply.
Constituent quark model physics seems to be important in 
the spin structure of the proton probed at deep inelastic $Q^2$!
Furthermore, one finds the puzzling result that 
in the presently measured kinematics where accurate data exist
the isovector part of $g_1$ 
considerably exceeds the isoscalar part of $g_1$ at small Bjorken $x$
-- the opposite to what is observed in unpolarized deep inelastic scattering.
\item
In the singlet channel the first moment of the $g_1$ spin structure 
function for polarized
photon-gluon fusion $(\gamma^* g \rightarrow q {\bar q})$
receives a negative contribution $-{\alpha_s \over 2 \pi}$
from $k_t^2 \sim Q^2$,
where
$k_t$ is the quark transverse momentum
relative to the photon gluon direction and
$Q^2$ is the virtuality of the hard photon \cite{Carlitz:1988}.
It also receives a
positive contribution
(proportional to the mass squared of the struck quark or antiquark)
from low values of
$k_t$,
$k_t^2 \sim P^2, m^2$ where $P^2$ is the virtuality of the parent
gluon and
$m$ is the mass of the struck quark.
The contact interaction ($k_t \sim Q$) between the polarized photon and
the gluon is flavour-independent.
It is associated with the QCD axial anomaly and
measures the spin of the target gluon.
The mass dependent contribution is absorbed into the quark wavefunction
of the nucleon.
\item
Gluon topology is associated with gluonic boundary conditions in the QCD
vacuum and has the potential to induce a topological contribution to 
$g_A^{(0)}$ associated with Bjorken $x$ equal to zero: topological $x=0$ 
polarization or, essentially, a spin ``polarized condensate'' inside a 
nucleon \cite{Bass:1998a}.
This topology term is associated with a potential $J=1$ fixed 
pole in the real part of the spin dependent part of the forward 
Compton 
amplitude and, if finite, is manifest 
as a ``subtraction at infinity'' in the dispersion relation for $g_1$
\cite{Bass:2003b}.
It is associated with dynamical axial U(1) symmetry breaking 
in the transition from constituent quarks to current quarks in QCD.
\end{enumerate}
Summarising these observations,
QCD theoretical analysis leads to the formula
\begin{equation}
g_A^{(0)}
=
\biggl(
\sum_q \Delta q - 3 {\alpha_s \over 2 \pi} \Delta g \biggr)_{\rm partons}
+ {\cal C}_{\infty}
.
\label{eqa10}
\end{equation}
Here $\Delta g_{\rm partons}$ is the amount of spin carried
by polarized
gluon partons in the polarized proton and
$\Delta q_{\rm partons}$ measures the spin carried by quarks
and
antiquarks
carrying ``soft'' transverse momentum $k_t^2 \sim P^2, m^2$
where 
$P$ is a typical gluon virtuality
and
$m$ is the light quark mass 
\cite{Efremov:1988,Altarelli:1988,Carlitz:1988};
${\cal C}_{\infty}$ denotes the potential non-perturbative gluon 
topological
contribution which has support only at Bjorken $x$ equal to zero
\cite{Bass:1998a}
so that it cannot be directly measured 
 in polarized deep inelastic scattering.

Since $\Delta g \sim 1/\alpha_s$ under QCD evolution, the
polarized gluon term $[-{\alpha_s \over 2 \pi} \Delta g]$
in Eq.(\ref{eqa10})
scales as $Q^2 \rightarrow \infty$ \cite{Efremov:1988,Altarelli:1988}.
The polarized gluon contribution corresponds to two-quark-jet
events carrying large transverse momentum $k_t \sim Q$ in
the final state from photon-gluon fusion \cite{Carlitz:1988}.

The topological term ${\cal C}_{\infty}$ may be identified with
a leading twist
``subtraction at infinity'' in the dispersion relation for $g_1$,
whence
$g_A^{(0)}|_{\rm pDIS}$ is identified with $g_A^{(0)}-C_{\infty}$
\cite{Bass:2003b}.
It probes the role of gluon topology in dynamical axial U(1)
symmetry breaking in the transition from current to constituent
quarks in low energy QCD.
The deep inelastic measurement of $g_A^{(0)}$, Eq.(\ref{eqa1}),
is not necessarily inconsistent with the constituent quark model
prediction 0.6
{\it if} a substantial fraction of the spin of the constituent quark
is associated with gluon topology in the transition from constituent
to current quarks  (measured in polarized deep inelastic scattering).

A direct measurement of the strange-quark axial-charge, 
independent of the analysis of polarized deep inelastic 
scattering data 
and any possible ``subtraction at infinity'' correction,
could be made using neutrino proton elastic scattering
through the axial coupling of the $Z^0$ gauge boson.
Comparing the values of
$\Delta s$
extracted from high-energy polarized deep inelastic scattering and 
low-energy $\nu p$ elastic scattering will provide vital information
about the QCD structure of the proton.

The vital role of quark transverse momentum in the formula (\ref{eqa10})
means that it
is essential to ensure that the theory and experimental
acceptance are correctly matched when extracting 
information from
semi-inclusive measurements
aimed at disentangling
the individual
valence, sea and gluonic contributions.
For example, recent semi-inclusive measurements 
\cite{Airapetian:2004a,Airapetian:2004b}
using a forward detector and limited acceptance at
large transverse momentum ($k_t \sim Q$)
exhibit no evidence for the large negative polarized
strangeness polarization extracted from inclusive data
and may, perhaps, be more comparable with $\Delta q_{\rm partons}$
than the inclusive measurement
(\ref{eqa2}) 
which has the polarized gluon contribution included.
Further semi-inclusive measurements with increased
luminosity and a 4$\pi$ detector would be valuable.
On the theoretical side, when assessing models which attempt 
to explain 
the proton's spin structure it is important to look at
the transverse momentum dependence of the proposed dynamics
plus the model predictions for the shape of the spin structure 
functions as a function of Bjorken $x$ in addition to the first 
moment and 
the nucleon's axial charges $g_A^{(3)}$, $g_A^{(8)}$ and $g_A^{(0)}$.

New dedicated experiments are planned or underway to map out the 
spin-flavour structure of the proton and, especially, to measure
the amount of spin carried by the valence and sea quarks and 
by polarized gluons in the polarized proton.
These include semi-inclusive polarized deep inelastic scattering
(COMPASS at CERN and HERMES at DESY), 
polarized proton-proton collisions at 
the world's first polarized proton-proton collider RHIC
\cite{Bunce:2000}, 
and future polarized electron-proton collider studies 
\cite{Bass:2002c}.
Experiments at Jefferson Laboratory are mapping out 
the spin distribution of quarks carrying a 
large fraction of the proton's momentum (Bjorken $x$)
and promise to yield exciting new information on confinement related dynamics.

Further interesting information about the structure of 
the proton will come from the study of ``transversity'' 
spin distributions \cite{Barone:2002}.
Working in an infinite momentum frame, these observables 
measure the distribution of spin polarized transverse to 
the momentum of the proton in a transversely polarized proton.
Since rotations and Euclidean boosts commute and a series of
boosts can convert a longitudinally polarized nucleon into a
transversely polarized nucleon at infinite momentum, 
it follows that the difference 
between the transversity and helicity distributions reflects
the relativistic character of quark motion in the nucleon.
Furthermore, the transversity spin distribution of the nucleon
is charge-parity odd ($C=-1$) and therefore valence like  
(gluons decouple from its QCD evolution equation in contrast 
 to the evolution equation for flavour-singlet quark 
 distribution appearing in $g_1$) making a 
comparison of the different spin dependent distributions most interesting.
Studies of transversity sensitive observables in lepton nucleon and
polarized proton proton scattering are being performed 
by the HERMES \cite{Airapetian:2004c}, COMPASS 
and RHIC 
\cite{Adams:2004} experiments.

One would also like to measure the parton orbital angular momentum 
contributions to the proton's spin.
Exclusive measurements of deeply virtual Compton scattering and 
single meson production at large $Q^2$ offer a possible route to 
the quark and gluon
angular momentum contributions through the physics and formalism
of generalized parton distributions \cite{Ji:1998,Goeke:1998,Diehl:2003}.
A vigorous programme to study these reactions is being designed and
investigated at several major world laboratories.

\subsection{Outline}

This Review is organized as follows.
In the first part (Sections II - VIII) we review the present status of 
the proton spin problem focussing on the present experimental situation
for tests of polarized deep inelastic spin sum-rules and the theoretical
understanding of $g_A^{(0)}$.
In the second part (Sections IX-XII) 
we give an overview of the present global
programme aimed at disentangling the spin-flavour structure of 
the proton 
and the exciting prospects for the new generation of 
experiments aimed at resolving the proton's internal spin structure.
In Sections II and III we give an overview of the derivation of 
the spin sum-rules for polarized photon-nucleon scattering, 
detailing the assumptions that are made at each step.
Here we explain how these sum rules could be affected by potential
subtraction constants (subtractions at infinity) in the dispersion
relations for the spin dependent part of the forward Compton amplitude.
We next give a brief review of the partonic (Section IV) and possible
fixed pole (Section V) contributions to deep inelastic scattering.
Fixed poles are well known to play a vital role in the Adler sum-rule
for W-boson nucleon scattering \cite{Adler:1966} and the Schwinger term 
sum-rule
for the longitudinal structure function measured in unpolarized deep
inelastic $ep$ scattering \cite{Broadhurst:1973}.
We explain how fixed poles could, in principle, affect the sum-rules
for the first moments of the $g_1$ and $g_2$ spin structure functions.
For example, a subtraction constant correction to the Ellis-Jaffe
sum rule for the first moment of the nucleon's $g_1$ spin dependent
structure function would follow if there is a constant real term in
the spin dependent part of the deeply virtual forward Compton scattering
amplitude.
Section VI discusses the QCD axial anomaly and its possible role in 
understanding
the first moment of $g_1$.
The relationship between the spin structure of the proton and chiral
symmetry is outlined in Section VII.
This first part of the paper concludes with an overview in Section VIII
of the different possible explanations of the small value of $g_A^{(0)}$
that have been proposed in the literature, how they relate to QCD, and
possible future experimental tests which could help clarrify the key issues.
We next
focus on the new programme to
disentangle the proton's spin-flavour structure and 
the Bjorken $x$ dependence of the separate valence, sea and gluonic 
contributions (Section IX), the theory and experimental investigation of
transversity observables (Section X), quark orbital angular momentum and
exclusive reactions (Section XI) and the $g_1$ spin structure function of 
the
polarized photon (Section XII).
A summary of key issues and challenging questions for the next generation 
of experiments is given in Section XIII.

Complementary review articles on the spin structure of the proton, 
each with a different emphasis,
are given in
\textcite{Anselmino:1995}, \textcite{Cheng:1996}, 
\textcite{Shore:1998}, 
\textcite{Lampe:2000}, 
\textcite{Fillipone:2001}, \textcite{Jaffe:2001},
\textcite{Barone:2002} and \textcite{Stoesslein:2002}.

\section{SCATTERING AMPLITUDES}

In photon-nucleon scattering 
the spin dependent structure functions $g_1$ and $g_2$
are defined through
the imaginary part of the forward Compton scattering amplitude.
Consider the amplitude for forward scattering of a photon
carrying momentum $q_{\mu}$ ($q^2 = -Q^2 \leq 0$)
from a polarized nucleon
with momentum $p_{\mu}$, mass $M$ and spin $s_{\mu}$.
Let $J_{\mu}(z)$ denote the electromagnetic current in QCD.
The forward Compton amplitude
\begin{equation}
T_{\mu \nu}(q,p) =
i \int d^{4}z \ e^{iq.z}
               \langle p, s | \ T(J_{\mu}(z) J_{\nu}(0) ) \ | p, s \rangle
\\
\label{eqb11}
\end{equation}
is given by the sum of spin independent
(symmetric in $\mu$ and $\nu$)
and
spin dependent (antisymmetric in $\mu$ and $\nu$)
contributions, {\it viz.}
\begin{eqnarray}
T_{\mu \nu}^{S}
&=&
{1 \over 2} (T_{\mu \nu} + T_{\nu \mu})
\nonumber \\
&=&
- T_1 (g_{\mu \nu} + {q_{\mu} q_{\nu} \over Q^{2} })
\nonumber \\
& &              
+ {1 \over M^2} T_2
             (p_{\mu} + {p.q \over Q^{2}} q_{\mu})
             (p_{\nu} + {p.q \over Q^{2}} q_{\nu})
\nonumber \\
\label{eqb12}
\end{eqnarray}
and
\begin{eqnarray}
T_{\mu \nu}^{A}
&=&
{1 \over 2} (T_{\mu \nu} - T_{\nu \mu})
\nonumber \\
&=&
{i \over M^2}
\epsilon_{\mu \nu \lambda \sigma}
q^{\lambda}
\biggl[
s^{\sigma}
( A_1 + {\nu \over M} A_2 ) - {1 \over M^2} s.q p^{\sigma} A_2 \biggr] 
.
\nonumber \\
\label{eqb13}
\end{eqnarray}
Here $\nu = p.q / M$ and $\epsilon_{0123} =+1$;
the proton spin vector is normalized to $s^2 = -1$.
The form-factors $T_1$, $T_2$, $A_1$ and $A_2$ are functions of $\nu$
and $Q^2$.

The hadron tensor for inclusive photon nucleon scattering which contains
the spin dependent structure functions
is obtained from the imaginary part of $T_{\mu \nu}$
\begin{equation}
W_{\mu \nu} = {1 \over \pi} {\rm Im} T_{\mu \nu}
= {1 \over 2 \pi}
     \int d^{4}z \ e^{iq.z}
\langle p, s | \ [J_{\mu}(z), J_{\nu}(0)] \ |p, s \rangle
.
\label{eqb14}
\end{equation}
Here the connected matrix element is understood
(indicating that the photon interacts with the target and not the vacuum).
The spin independent and spin dependent components of $W_{\mu \nu}$
are
\begin{equation}
W_{\mu \nu}^{S} = - W_1 (g_{\mu \nu} + {q_{\mu} q_{\nu} \over Q^{2} })
             + {1 \over M^2} W_2
             (p_{\mu} + {p.q \over Q^{2}} q_{\mu})
             (p_{\nu} + {p.q \over Q^{2}} q_{\nu})
\label{eqb15}
\end{equation}
and
\begin{equation}
W_{\mu \nu}^{A} =
{i \over M^2}
\epsilon_{\mu \nu \lambda \sigma}
q^{\lambda}
\biggl[
s^{\sigma}
( G_1 + {\nu \over M} G_2 ) - {1 \over M^2} s.q p^{\sigma} G_2 \biggr]
\label{eqb16}
\end{equation}
respectively.
The structure functions contain all of the target dependent information 
in the deep inelastic process.

The cross sections for the absorption of a transversely polarized
photon with spin polarized parallel
$\sigma_{3 \over 2}$ and anti-parallel
$\sigma_{1 \over 2}$ to the spin of the
(longitudinally polarized) target nucleon are
\begin{eqnarray}
\sigma_{3 \over 2} &=& {4 \pi^2 \alpha \over \sqrt{\nu^2 + Q^2}}
\biggl[ W_1 - {\nu \over M^2} G_1 + {Q^2 \over M^3} G_2 \biggr]
\nonumber \\
\sigma_{1 \over 2} &=& {4 \pi^2 \alpha \over \sqrt{\nu^2 + Q^2}}
\biggl[ W_1 + {\nu \over M^2} G_1 - {Q^2 \over M^3} G_2 \biggr] ,
\nonumber \\
\label{eqb17}
\end{eqnarray}
where we use usual conventions for the virtual photon flux factor
\cite{Roberts:1990}.
The spin dependent and spin independent parts of the inclusive 
photon nucleon cross section
are
\begin{equation}
\sigma_{1 \over 2} - \sigma_{3 \over 2}
= {8 \pi^2 \alpha \over \sqrt{\nu^2 + Q^2}}
\biggl[ {\nu \over M^2} G_1 - {Q^2 \over M^3} G_2 \biggr]
\label{eqb18}
\end{equation}
and
\begin{equation}
\sigma_{1 \over 2} + \sigma_{3 \over 2}
= {8 \pi^2 \alpha \over \sqrt{\nu^2 + Q^2}} 
W_1
.
\label{eqb19}
\end{equation}
The $G_2$ spin structure function decouples from polarized
photoproduction.
For real photons ($Q^2=0$) one finds the equation 
$
\sigma_{1 \over 2} - \sigma_{3 \over 2}
= {8 \pi^2 \alpha \over M^2} G_1
$.
The cross section for the absorption of a longitudinally polarized
photon is
\begin{equation}
\sigma_{0} 
= {4 \pi^2 \alpha \over \sqrt{\nu^2 + Q^2}} W_L 
= {4 \pi^2 \alpha \over \sqrt{\nu^2 + Q^2}}
\biggl[ 
\biggl\{ 1 + {\nu^2 \over Q^2} \biggr\} W_2 - W_1 
\biggr]
.
\label{eqb20}
\end{equation}
The $W_2$ structure function is measured in unpolarized lepton nucleon
scattering through the absorption of transversely and longitudinally 
polarized photons.

Our present knowledge about the high-energy spin structure of 
the nucleon comes from 
polarized deep inelastic scattering experiments.
These experiments involve scattering a high-energy charged 
lepton beam from a nucleon target at large momentum transfer squared.
One measures the inclusive cross-section.
The lepton beam
(electrons at DESY, JLab and SLAC and muons at CERN) 
is longitudinally polarized.
The nucleon target may be either longitudinally or transversely polarized.

The relation between deep inelastic lepton nucleon cross-sections 
and 
the 
virtual-photon nucleon cross-sections discussed above is discussed
and derived in various textbooks
-- e.g. \textcite{Roberts:1990}.
Polarized deep inelastic scattering experiments have so far all 
been performed using a fixed target.
Consider polarized $ep$ scattering.
We specialize to the target rest frame and let $E$ 
denote the energy of the incident electron which is scattered 
through an angle $\theta$ to emerge in the final state with energy $E'$.
Let $\uparrow \downarrow$ denote the longitudinal polarization of
the electron beam. 
For a longitudinally polarized proton target
(with spin denoted $\Uparrow \Downarrow$)
the unpolarized and polarized differential cross-sections are
\begin{eqnarray}
\Biggl(
{d^2 \sigma \uparrow \Downarrow \over d\Omega dE^{'} } 
&+&
{d^2 \sigma \uparrow \Uparrow \over d\Omega dE^{'} }
\Biggr)
\nonumber \\
&=&
{\alpha^2 \over 4 E^2 \sin^4 {\theta \over 2} }
\biggl[ 2  \sin^{2} {\theta \over 2} \ W_1  
 + 
 \cos^{2} {\theta \over 2} \ W_2 \biggr]
\nonumber \\
\label{eqb21}
\end{eqnarray}
and
\begin{eqnarray}
\Biggl(
{d^2 \sigma \uparrow \Downarrow \over d\Omega dE^{'} } 
&-&
{d^2 \sigma \uparrow \Uparrow \over d\Omega dE^{'} }
\Biggr)
\nonumber \\
&=&
{4 \alpha^2 \over M^3 Q^2} \ {E^{'} \over E} \
\biggl[ M (E+E^{'} \cos \theta ) \ G_1 - Q^2 \ G_2
\biggr] 
.
\nonumber \\
\label{eqb22}
\end{eqnarray}
For a target polarized transverse to the electron beam the spin dependent
part of the differential cross-section is
\begin{equation}
\Biggl(
{d^2 \sigma \uparrow \Rightarrow \over d\Omega dE^{'} }  -
{d^2 \sigma \uparrow \Leftarrow \over d\Omega dE^{'} }
\Biggr)
=
{4 \alpha^2 \over M^3 Q^2} {E^{' 2} \over E }
\sin \theta
\biggl[ M \ G_1 + 2 E \ G_2 \biggr] 
.
\label{eqb23}
\end{equation}

\subsection{Scaling and polarized deep inelastic scattering}

In high $Q^2$ deep inelastic scattering the structure functions 
exhibit approximate scaling.
One finds
\begin{eqnarray}
  M \ W_1 (\nu, Q^2) &\rightarrow& F_1 (x, Q^2)  \nonumber \\
\nu \ W_2 (\nu, Q^2) &\rightarrow& F_2 (x, Q^2) \nonumber \\
{\nu \over M} \ G_1 (\nu, Q^2) &\rightarrow& g_1 (x, Q^2) \nonumber \\
{\nu^2 \over M^2} \ G_2(\nu, Q^2) &\rightarrow& g_2 (x, Q^2)
.
\label{eqb24}
\end{eqnarray}
The structure functions 
$F_1$, $F_2$, $g_1$ and $g_2$ 
are to a very good approximation independent of $Q^2$ and depend 
{\it only} on $x$.
(The small $Q^2$ dependence which is present in these structure
 functions 
 is
 logarithmic and determined by perturbative QCD evolution.)
Substituting (\ref{eqb24}) in the cross-section formula (\ref{eqb22})
for the longitudinally polarized target 
one finds that the $g_2$ 
contribution 
to the differential cross section 
and the longitudinal spin asymmetry
is suppressed relative to the $g_1$ 
contribution 
by the kinematic factor ${M \over E} \sim 0$, 
{\it viz.}
\begin{eqnarray}
{\cal A}_1 = 
{\sigma_{1 \over 2} - \sigma_{3 \over 2} 
\over 
 \sigma_{1 \over 2} + \sigma_{3 \over 2}}
 = 
{M \nu G_1 - Q^2 G_2 \over M^3 W_1}
 = 
{g_1 - {Q^2 \over \nu^2} \ g_2 \over F_1} 
\rightarrow 
{g_1 \over F_1}
.
\nonumber \\
\label{eqb25}
\end{eqnarray}
For a transverse polarized target
this kinematic suppression factor 
for $g_2$ is missing 
meaning that
transverse polarization is vital to measure $g_2$.
We refer to \textcite{Roberts:1990} and \textcite{Windmolders:2002} 
for the procedure how the spin dependent structure functions 
are extracted from the spin asymmetries measured in polarized 
deep inelastic scattering.

In the (pre-QCD) parton model the deep inelastic structure functions
$F_1$ and $F_2$ 
are
written as
\begin{equation}
F_1 (x) = {1 \over 2x} F_2 (x) = {1 \over 2} \sum_q e_q^2 \{q + {\bar q} \}(x)
\label{eqb26}
\end{equation}
and the polarized structure function $g_1$ is
\begin{equation}
g_1 (x) = {1 \over 2} \sum_q e_q^2 \Delta q(x)
.
\label{eqb27}
\end{equation}
Here
$e_q$ denotes the electric charge of the struck quark
and
\begin{eqnarray}
\{q + {\bar q} \}(x) 
&=& (q^{\uparrow} + {\overline q}^{\uparrow})(x) +
                        (q^{\downarrow} + {\overline q}^{\downarrow})(x)
\nonumber \\
\Delta q(x) 
&=& (q^{\uparrow} + {\overline q}^{\uparrow})(x) -
                        (q^{\downarrow} + {\overline q}^{\downarrow})(x)
\nonumber \\
\label{eqb28}
\end{eqnarray}
denote the spin-independent (unpolarized) and spin-dependent 
quark parton distributions
which measure the distribution of quark momentum and spin in
the proton.
For example, 
${\overline q}^{\uparrow}(x)$
is interpreted as the probability to find an anti-quark of 
flavour $q$
with plus component of momentum $x p_+$ 
($p_+$ is the plus component of the target proton's momentum)
and spin polarized in the same direction as the spin of the
target proton.
When we integrate out the momentum fraction $x$
the quantity
$
\Delta q = \int_0^1 dx \ \Delta q(x)
$
is interpreted as the fraction of the proton's spin which 
is carried by quarks (and anti-quarks) of flavour $q$
-- hence the parton model interpretation of $g_A^{(0)}$ as 
the
total fraction of the proton's spin carried by up, down and
strange quarks.
In QCD 
the flavour-singlet combination of these quark parton 
distributions 
mixes with 
the spin independent and spin dependent gluon distributions 
respectively under $Q^2$ evolution.
The gluon parton distributions measure the momentum and spin
dependence of glue in the proton.
The second spin structure function $g_2$ 
has a non-trivial parton interpretation \cite{Jaffe:1990b}
and
vanishes without the effect of quark transverse momentum 
-- see e.g. \textcite{Roberts:1990}.

\begin{figure}[t!]
\begin{center}
\includegraphics{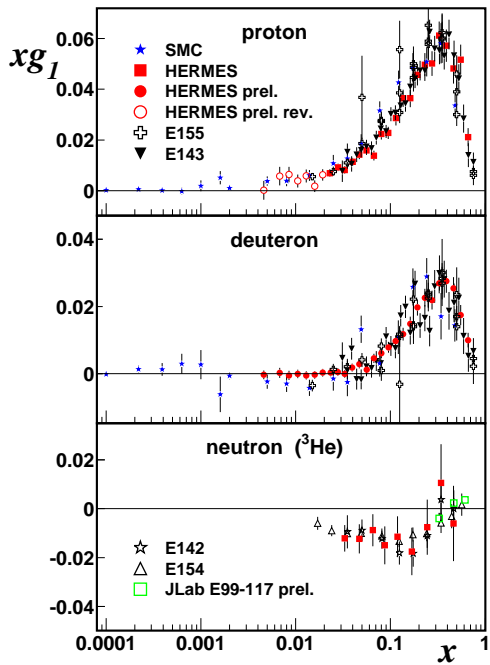}
\vspace{9.0cm}
\parbox{8.0cm}
{\caption[Delta]
{The world data on $xg_1$ with data points shown at the $Q^2$
 they were measured at. Figure courtesy of U. Stoesslein.}
\label{fig:fig1}}
\end{center}
\end{figure}

\begin{figure}[h]
\begin{center}
\includegraphics{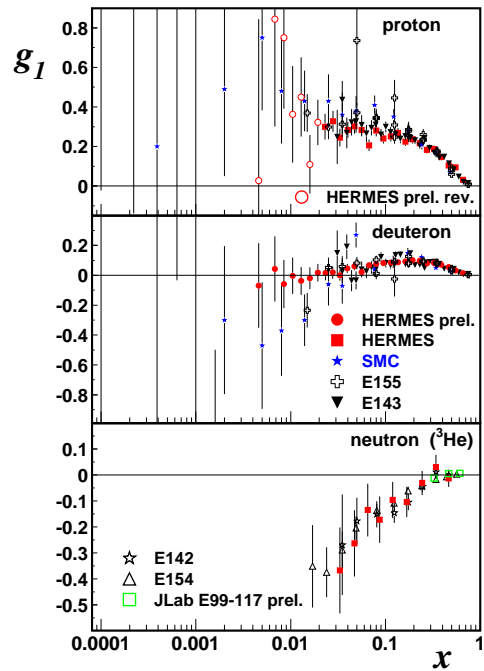}
\vspace{9.0cm}
\parbox{8.0cm}
{\caption[Delta]
{The world data on $g_1$ with data points shown at the $Q^2$
 they were measured at. Figure courtesy of U. Stoesslein.}
\label{fig:fig2}}
\end{center}
\end{figure}

An overview of the world data on the nucleon's $g_1$ spin structure
function is shown in Figure~\ref{fig:fig1} (which shows $x g_1$) 
and 
Figure~\ref{fig:fig2} (which shows $g_1$).
There is a general consistency between all data sets.
The largest range is provided by the SMC experiment
\cite{Adeva:1998a,Adeva:1999}, 
namely
$0.00006 < x < 0.8$ and $0.02 < Q^2< 100$ GeV$^2$.
This experiment
used proton and deuteron targets with 100-200 GeV muon beams. 
The 
final results are given in ~\cite{Adeva:1998a}.
The low $x$ data from SMC
\cite{Adeva:1999}
are at a $Q^2$ well below 1 GeV$^2$, and the asymmetries are found
to be compatible with zero.
The most precise data comes from the electron scattering experiments  
at SLAC 
(E154 on the neutron \cite{Abe:1997} 
 and E155 on the proton \cite{Anthony:1999,Anthony:2000}), 
JLab \cite{Zheng:2004a,Zheng:2004b}
(on the neutron)
and HERMES at DESY \cite{Airapetian:1998,Ackerstaff:1997}
(on the proton and neutron), 
with JLab focussed on the large $x$ region.
The recipes for extracting the neutron's spin structure function
from experiments using a deuteron or $^3$He target are discussed 
in 
\textcite{Piller:2000} and \textcite{Thomas:2002}.

Note the large isovector component in the data at small $x$
(between 0.01 and 0.1)
which considerably exceeds the isoscalar component 
in the measured kinematics.
This result is in stark contrast to the situation in the unpolarized
structure function $F_2$ where the small $x$ region is dominated 
by isoscalar pomeron exchange.
Given the large experimental 
errors on the data little can 
presently be concluded about $g_1$ at the smallest $x$ values 
($x$ less than about 0.006).

\begin{figure}[t]
\begin{center}
\includegraphics{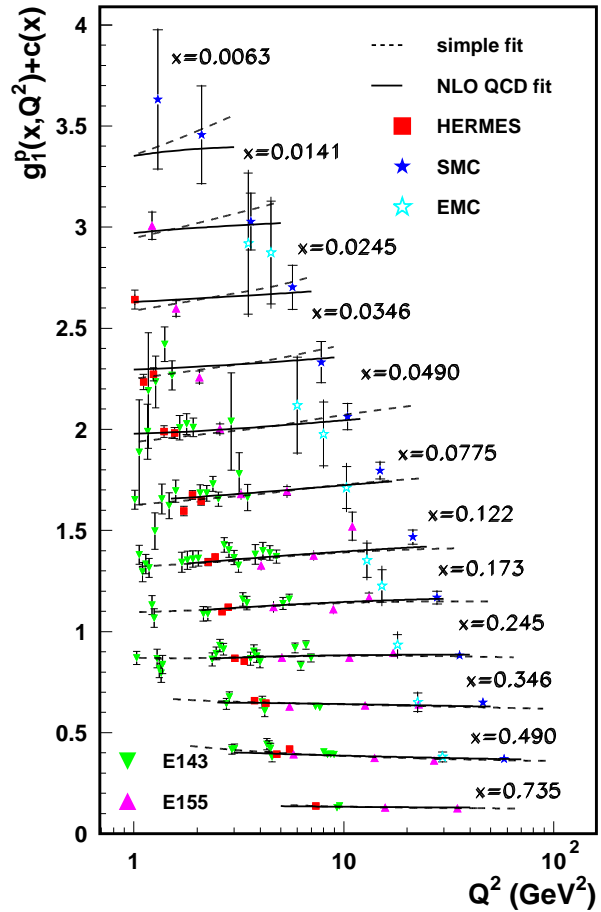}
\vspace{13.0cm}
\parbox{8.0cm}
{\caption[Delta]
{$Q^2$ dependence of $g_1^p$ for $Q^2 > 1$GeV$^2$ 
 together
 with a simple fit according to \textcite{Anthony:2000} and a 
 NLO perturbative QCD fit from \textcite{Stoesslein:2002}. 
}
\label{fig:fig3}}
\end{center}
\end{figure}

The structure function data at different values of $x$ 
(Figs.\ref{fig:fig1} and \ref{fig:fig2})
are measured at different $Q^2$ values in the experiments,
viz. $x_{\rm expt.} = x(Q^2)$.
For the ratios $g_1 / F_1$ there is no experimental evidence of 
$Q^2$ dependence in any given $x$ bin.
The E155 Collaboration at SLAC 
found the following good
phenomenological fit to their final data set with 
$Q^2 > 1$GeV$^2$ 
and energy of the hadronic final state $W > 2$GeV
\cite{Anthony:2000}:
\begin{eqnarray}
{g_1^p \over F_1^p} &=& x^{0.700} (0.817 + 1.014 x - 1.489 x^2)
\times
(1 + {c^p \over Q^2})
\nonumber \\
{g_1^n \over F_1^n} &=& x^{-0.335} (-0.013 - 0.330 x + 0.761 x^2)
\times
(1 + {c^n \over Q^2})
.
\nonumber \\
\label{eqb29}
\end{eqnarray}
The coefficients $c^p = -0.04 \pm 0.06$ and $c^n = 0.13 \pm 0.45$
describing the $Q^2$ dependence are found to be small and consistent
with zero.
The $Q^2$ dependence of the $g_1$ spin structure function is shown in
Fig.\ref{fig:fig3}.
It is useful to compare data at the same $Q^2$, 
e.g. for the comparison of experimental data with the predictions
of deep inelastic sum-rules.
To this end, the measured $x$ points are shifted to the same $Q^2$
using either the (approximate) $Q^2$ independence of the asymmetry
or 
performing
next-to-leading-order (NLO) QCD motivated fits 
\cite{Adeva:1998b,Anthony:2000,Gehrmann:1996,Altarelli:1997,
      Gluck:2001,Blumlein:2002,Goto:2000,Hirai:2004,Leader:2002}
to the measured data and evolving the measured data points all to 
the same value of $Q^2$.

\subsection{Regge theory and the small $x$ behaviour of spin 
structure
functions}

The small $x$ or high energy behaviour of spin structure functions is an 
important issue
both for the extrapolation of data needed 
to test spin sum rules for the first moment of $g_1$ and also in its own 
right.

Regge theory makes predictions for the high-energy asymptotic behaviour of
the structure functions:
\begin{eqnarray}
&W_1& \sim \nu^{\alpha}   \nonumber \\
&W_2& \sim \nu^{\alpha -2} \nonumber \\
&G_1& \sim \nu^{\alpha -1}  \nonumber \\
&G_2& \sim \nu^{\alpha -1}
.
\label{eqb30}
\end{eqnarray}
Here $\alpha$ denotes the (effective) intercept for the 
leading Regge exchange contributions.
The Regge predictions for the leading exchanges include
$\alpha = 1.08$
for the pomeron contributions to $W_1$ and $W_2$,
and $\alpha \simeq 0.5$ for the $\rho$ and $\omega$
exchange contributions to the spin independent structure functions.

For $G_1$ the leading gluonic exchange behaves as $\{ \ln \nu \} / \nu$
\cite{Close:1994,Bass:1994}.
In the isovector and isoscalar channels there are also isovector $a_1$ 
and isoscalar $f_1$ Regge exchanges plus contributions from the 
pomeron-$a_1$ and pomeron-$f_1$ cuts \cite{Heimann:1973}.
If one makes the usual assumption that the $a_1$ and $f_1$ Regge
trajectories are straight lines parallel to the $(\rho, \omega)$
trajectories then one finds
$\alpha_{a_1} \simeq \alpha_{f_1} \simeq -0.4$,
within the phenomenological range $-0.5 \leq \alpha_{a_1} \leq 0$
discussed in \textcite{Ellis:1988}.
Taking the masses of the $a_1(1260)$ and $a_3(2070)$ states
plus the $a_1(1640)$ and $a_3(2310)$ states from the \textcite{PDG:2004}
yields two parallel $a_1$ trajectories with slope $\sim 0.75$GeV$^{-2}$
and a leading trajectory with slightly lower intercept: 
$\alpha_{a_1} \simeq -0.18$.

For this value of the $a_1$ intercept the effective 
intercepts corresponding to the soft-pomeron $a_1$ cut 
and the hard-pomeron $a_1$ cut 
are $\simeq -0.1$ and $\simeq +0.25$ 
respectively 
{\it if} one takes the
soft and hard pomerons as two distinct exchanges 
\cite{Cudell:1999}
\footnote{I thank P.V. Landshoff for valuable discussions on this issue.}.
In the framework of the Donnachie-Landshoff-Nachtmann model of
soft pomeron physics \cite{Landshoff:1987,Donnachie:1988}
the logarithm in the ${\ln \nu \over \nu}$
contribution comes from the region of internal momentum where
two non-perturbative gluons are radiated collinear with the proton 
\cite{Bass:1994}.

For $G_2$ one expects contributions from possible multi-pomeron
(three or more) cuts ($\sim (\ln \nu)^{-5}$)
and Regge-pomeron cuts ($\sim \nu^{\alpha_i(0)-1} / \ln \nu $)
with
$\alpha_i(0) <1$
(since the pomeron does not couple to $A_1$ or $A_2$ 
 as a single gluonic exchange) -- see \textcite{Ioffe:1984}.

In terms of the scaling structure functions of deep inelastic scattering
the relations (\ref{eqb30}) become
\begin{eqnarray}
&F_1& \sim {1 \over x}^{\alpha}   \nonumber \\
&F_2& \sim {1 \over x}^{\alpha -1} \nonumber \\
&g_1& \sim {1 \over x}^{\alpha}  \nonumber \\
&g_2& \sim {1 \over x}^{\alpha +1}
.
\label{eqb31}
\end{eqnarray}
For deep inelastic values of $Q^2$ there is some debate about the 
application of Regge arguments.
In the conventional approach
the effective intercepts for small $x$, or high $\nu$, physics tend 
to increase with increasing $Q^2$ through perturbative QCD 
evolution which acts to shift the weight of the structure functions
to smaller $x$.
The polarized isovector combination $g_1^p-g_1^n$ 
is observed to rise in the small $x$ data from SLAC and SMC
like $\sim x^{-0.5}$ 
although it should be noted that, in the measured $x$ range,
this exponent could be softened through multiplication by a
$(1-x)^n$ factor -- 
for example associated with perturbative QCD counting 
rules at large $x$ ($x$ close to one).
For example, the exponent $x^{-0.5}$ could be modified 
to about 
$x^{-0.25}$ through multiplication by a factor $(1-x)^6$.
In an alternative approach 
\textcite{Cudell:1999}
have argued
that the Regge intercepts should be independent of $Q^2$ and that the
``hard pomeron'' revealed in unpolarized deep inelastic scattering at 
HERA is a distinct exchange independent of the soft 
pomeron which should also be present in low $Q^2$ photoproduction data.

Detailed investigation of spin dependent Regge theory and the low $x$
behaviour of spin structure functions 
could be performed at SLAC or
using a 
future polarized $ep$ collider (e-RHIC)
where measurements could be obtained through a broad range of $Q^2$
from photoproduction through the ``transition region'' to polarized
deep inelastic scattering.
These measurements would provide a baseline for investigations of 
perturbative QCD motivated small $x$ behaviour in $g_1$.
Open questions include: 
Does the rise in $g_1^p-g_1^n$ at small Bjorken $x$ persist to small
values of $Q^2$ ?
How does this rise develop as a function of $Q^2$ ?
Further possible exchange contributions in the flavour-singlet
sector associated with polarized glue could also be looked for.
For example, colour coherence predicts that the ratio of 
polarized to unpolarized gluon distributions
$\Delta g(x) / g(x) \propto x$ as $x \rightarrow 0$ 
\cite{Brodsky:1995a}
suggesting
that, perhaps, there is a spin analogue of the hard pomeron with
intercept about 0.45 corresponding to the polarized gluon distribution.

The $s$ and $t$ dependence of spin dependent Regge theory is being
investigated by the pp2pp experiment 
\cite{Bultmann:2003}
at RHIC 
which is studying polarized proton proton elastic 
scattering at 
centre of mass energies $50 < \sqrt{s} < 500$GeV 
and four momentum transfer $0.0004 < |t| < 1.3$GeV$^2$.

\section{DISPERSION RELATIONS AND SPIN SUM RULES}

Sum rules for the (spin) structure functions measured in deep
inelastic scattering
are derived using
dispersion relations and the operator product expansion.
For fixed $Q^2$ the forward Compton scattering amplitude
$T_{\mu \nu}(\nu,Q^2)$
is analytic in the photon energy $\nu$ except for branch cuts
along the positive real axis for $|\nu| \geq Q^2/2M$.
Crossing symmetry implies that
\begin{eqnarray}
A_1^* (Q^2, -\nu) &=&   A_1 (Q^2, \nu) \nonumber \\
A_2^* (Q^2, -\nu) &=& - A_2 (Q^2, \nu)
.
\label{eqc32}
\end{eqnarray}
The spin structure functions in the imaginary parts of $A_1$ and $A_2$
satisfy the crossing relations
\begin{eqnarray}
G_1 (Q^2, -\nu) &=&  - G_1 (Q^2, \nu) \nonumber \\
G_2 (Q^2, -\nu) &=&  + G_2 (Q^2, \nu)
.
\label{eqc33}
\end{eqnarray}
For $g_1$ and $g_2$ these relations become
\begin{eqnarray}
g_1 (x, Q^2) &=&  + g_1 (-x, Q^2) \nonumber \\
g_2 (x, Q^2) &=&  + g_2 (-x, Q^2)
.
\label{eqc34}
\end{eqnarray}
We use Cauchy's integral theorem and the crossing relations to derive
dispersion relations for $A_1$ and $A_2$.
Assuming that the asymptotic behaviour of the spin structure functions
$G_1$ and $G_2$ yield convergent integrals 
we are tempted
to write the two unsubtracted dispersion relations:
\begin{eqnarray}
A_1 (Q^2, \nu)
&=&
{2 \over \pi} \int_{Q^2/2M}^{\infty} \ {\nu' d \nu' \over \nu'^2 - \nu^2}
\ {\rm Im} A_1 (Q^2, \nu')
\nonumber \\
A_2 (Q^2, \nu)
&=&
{2 \over \pi} \nu \int_{Q^2/2M}^{\infty} \ {d \nu' \over \nu'^2 - \nu^2}
\ {\rm Im} A_2 (Q^2, \nu')
.
\nonumber \\
\label{eqc35}
\end{eqnarray}
These expressions can be rewritten as dispersion relations involving
$g_1$ and $g_2$.
We define:
\begin{eqnarray}
\alpha_1 (\omega, Q^2) &=& {\nu \over M} \ A_1 \nonumber \\
\alpha_2 (\omega, Q^2) &=& {\nu^2 \over M^2} \ A_2
.
\label{eqc36}
\end{eqnarray}
Then, the formulae in (\ref{eqc35}) become
\begin{eqnarray}
\alpha_1 (\omega, Q^2)
&=&
2 \omega
\int_1^{\infty} \ {d \omega' \over \omega'^2 - \omega^2}
\ g_1 (\omega', Q^2)
\nonumber \\
\alpha_2 (\omega, Q^2)
&=&
2 \omega^3
\int_1^{\infty} \ {d \omega' \over \omega'^2 (\omega'^2 - \omega^2)}
\ g_2 (\omega', Q^2)
\nonumber \\
\label{eqc37}
\end{eqnarray}
where $\omega = {1 \over x} = {2 M \nu \over Q^2}$.

In general there are two alternatives to an unsubtracted dispersion relation.
\begin{enumerate}
\item
First, if the high energy behaviour of $G_1$ and/or $G_2$ (at some fixed
$Q^2$)
produced a divergent integral, then the dispersion relation would require
a subtraction.
Regge predictions for the high energy behaviour of $G_1$ and $G_2$
-- see Eq.(\ref{eqb30}) --
each lead to convergent integrals so this scenario is not expected 
to occur, even after including the possible effects of QCD evolution.

\item
Second,
even if the integral in the unsubtracted relation converges, there is
still
the potential for a ``subtraction at infinity''.
This scenario would occur if the real part of $A_1$ and/or $A_2$
does not vanish sufficiently fast enough when
$\nu \rightarrow \infty$ so that we pick up a finite contribution
from the contour
(or
``circle at infinity'').
In the context of Regge theory such subtractions can arise from fixed
poles
(with
$J = \alpha(t) = 0$ in $A_2$
 or
$J = \alpha(t) = 1$ in $A_1$
for all $t$)
in the real part of
the
forward Compton amplitude.
We shall discuss these fixed poles and potential subtractions in Section V.
\end{enumerate}

In the presence of a potential ``subtraction at infinity'' the dispersion 
relations (\ref{eqc35}) are modified to:
\begin{eqnarray}
A_1 (Q^2, \nu)
&=&
{\cal P}_1 (\nu, Q^2) 
\nonumber \\
& & 
\ \ \ \ \ 
+
{2 \over \pi} \int_{Q^2/2M}^{\infty} \ {\nu' d \nu' \over \nu'^2 - \nu^2}
\ {\rm Im} A_1 (q^2, \nu')
\nonumber \\
A_2 (Q^2, \nu)
&=&
{\cal P}_2 (\nu, Q^2) 
\nonumber \\
& & \ \ \ \ \ 
+
{2 \over \pi} \nu \int_{Q^2/2M}^{\infty} \ {d \nu' \over \nu'^2 - \nu^2}
\ {\rm Im} A_2 (q^2, \nu')
.
\nonumber \\
\label{eqc38}
\end{eqnarray}
Here ${\cal P}_1 (\nu, Q^2)$ and 
${\cal P}_2 (\nu, Q^2)$ 
denote the subtraction constants.
Factoring out the $\nu$ dependence of these subtraction
constants,
we define two $\nu$ independent quantities $\beta_1 (Q^2)$
and $\beta_2 (Q^2)$:
\begin{eqnarray}
{\cal P}_1 (\nu, Q^2) &=& \beta_1 (Q^2)
\nonumber \\
{\cal P}_2 (\nu, Q^2) &=& \beta_2 (Q^2) {M \over \nu} .
\label{eqc39}
\end{eqnarray}
The crossing relations (\ref{eqc32})
for $A_1$ and $A_2$
are observed by the functions ${\cal P}_i$.
Scaling requires that
$\beta_1 (Q^2)$ and $\beta_2 (Q^2)$
(if finite) must be nonpolynomial in $Q^2$ -- see Section V.
The equations (\ref{eqc38}) can be rewritten:
\begin{eqnarray}
\alpha_1 (\omega, Q^2)
&=&
{Q^2 \over 2 M^2 } \ \beta_1(Q^2) \ \omega  +
\nonumber \\
& & 
2 \omega
\int_1^{\infty} \ {d \omega' \over \omega'^2 - \omega^2}
\ g_1 (\omega', Q^2)
\nonumber \\
\alpha_2 (\omega, Q^2)
&=&
{Q^2 \over 2 M^2 } \ \beta_2 (Q^2) \ \omega +
\nonumber \\
& & 
2 \omega^3
\int_1^{\infty} \ {d \omega' \over \omega'^2 (\omega'^2 - \omega^2)}
\ g_2 (\omega', Q^2)
.
\nonumber \\
\label{eqc40}
\end{eqnarray}
Next, the fact that both $\alpha_1$ and $\alpha_2$ are analytic
for $| \omega | \leq 1$ allows us to make the Taylor series expansions
(about $\omega = 0$)
\begin{eqnarray}
\alpha_1 (x, Q^2)
&=&
{Q^2 \over 2 M^2 } \ \beta_1 (Q^2) \ {1 \over x}  
\nonumber \\
& & 
+
{2 \over x}
\sum_{n=0,2,4,..} \biggl( {1 \over x^n} \biggr) \int_0^1 dy \ y^n g_1 (y, Q^2)
\nonumber \\
\alpha_2 (x, Q^2)
&=&
{Q^2 \over 2 M^2 } \ \beta_2 (Q^2) \ {1 \over x} 
\nonumber \\
& &
+
{2 \over x^3}
\sum_{n=0,2,4,..} \biggl( {1 \over x^n} \biggr) \int_0^1 dy \ y^{n+2}
g_2 (y, Q^2)
\nonumber \\
\label{eqc41}
\end{eqnarray}
with $x={1 \over \omega}$.

These equations form the basis for the spin sum rules for polarized photon
nucleon scattering.
We next outline the derivation of the Bjorken \cite{Bjorken:1966,Bjorken:1970} 
and
Ellis-Jaffe \cite{Ellis:1974}
sum rules for the isovector and flavour-singlet parts of $g_1$
in polarized deep inelastic scattering,
the Burkhardt-Cottingham sum rule for $G_2$ \cite{Burkhardt:1970},
and
the Gerasimov-Drell-Hearn sum rule 
for polarized photoproduction \cite{Gerasimov:1965,Drell:1966}.
Each of these spin sum rules assumes no subtraction at infinity.

\subsection{Deep inelastic spin sum rules}

Sum rules for polarized deep inelastic scattering are derived by combining
the dispersion relation expressions (\ref{eqc41})
with the light cone operator production expansion.
When $Q^2 \rightarrow \infty$ the leading contribution to the spin dependent
part of the forward
Compton amplitude comes from the nucleon matrix elements of a tower of gauge
invariant local operators multiplied by Wilson coefficients, 
{\it viz.}
\begin{eqnarray}
T_{\mu \nu}^A 
&=&
i \epsilon_{\mu \nu \lambda \sigma} q^{\lambda} 
\sum_{n=0,2,4,..}
\biggl( - {2 \over q^2} \biggr)^{n+1}
q^{\mu_1} q^{\mu_2} ... q^{\mu_n}
\nonumber \\
& & 
\ \ \ \ \ \ \ \ \ \ \ \ \ \ \
\sum_{i=q,g}
\Theta^{(i)}_{\sigma \{ \mu_1 ... \mu_n \} }
E^i_n ({Q^2 \over \mu^2}, \alpha_s)
\nonumber \\
\label{eqc42}
\end{eqnarray}
where
\begin{equation}
\Theta^{(q)}_{\sigma \{\mu_1 ... \mu_n \} } \equiv i^n
\bar{\psi} \gamma_{\sigma} \gamma_5 D_{ \{ \mu_1} ... D_{\mu_n \} } \psi
- {\rm traces}
\label{eqc43}
\end{equation}
and
\begin{equation}
\Theta^{(g)}_{\sigma \{\mu_1 ... \mu_n \} } \equiv
i^{n-1}
\epsilon_{\alpha \beta \gamma \sigma}
G^{\beta \gamma}
 D_{ \{ \mu_1} ... D_{\mu_{n-1}} G_{\ \mu_{n} \}}^{\alpha}
- {\rm traces}
\label{eqc44}
\end{equation}
are local operators.
Here $D_{\mu} = \partial_{\mu} + ig A_{\mu}$ is the gauge covariant 
derivative and the sum over even values of $n$ in Eq.(\ref{eqc42})
reflects the crossing symmetry properties of $T_{\mu \nu}$.
The functions
$E^q_n ({Q^2 \over \mu^2}, \alpha_s)$ and
$E^g_n ({Q^2 \over \mu^2}, \alpha_s)$ are the respective Wilson coefficients.
(Note that,
 for simplicity, in this discussion we consider the case of a single quark
 flavour with unit charge and zero quark mass.
 The results quoted in Section III.B below
 include the extra steps of using the full electromagnetic current in QCD.)

The operators in Eq.(\ref{eqc42})
may each be written as the sum of a
totally symmetric operator and an operator with mixed symmetry
\begin{equation}
\Theta_{\sigma \{\mu_1 ... \mu_n \} }
=
\Theta_{ \{ \sigma \mu_1 ... \mu_n \} }
+
\Theta_{ [ \sigma , \{ \mu_1 ] ... \mu_n \} }
.
\label{eqc45}
\end{equation}
These operators have the matrix elements:
\begin{eqnarray}
& & 
\langle p, s | \Theta_{\{\sigma \mu_1 ... \mu_n \}} | p,s \rangle
\nonumber \\
& & \ \ \ \ \ \ \ =
\{ s_{\sigma} p_{\mu_1} ... p_{\mu_n}
 + s_{\mu_1} p_{\sigma} p_{\mu_2} ... p_{\mu_n}
 + ... \}
{a_n \over n+1}
\nonumber \\
& & \langle p,s | \Theta_{[ \{ \sigma \mu_1 ]... \mu_n \}} | p,s \rangle
\nonumber \\
& & \ \ \ \ \ \ \ = 
\{ ( s_{\sigma} p_{\mu_1} - s_{\mu_1} p_{\sigma} ) p_{\mu_2} ... p_{\mu_n}
\nonumber \\
& &
\ \ \ \ \ \ \ \ \ \ 
 + (s_\sigma p_{\mu_2} - s_{\mu_2} p_{\sigma}) p_{\mu_1} ... p_{\mu_n}
 + ... \}
{d_n \over n+1}
.
\nonumber \\
\label{eqc46}
\end{eqnarray}
Now define
${\tilde a}_n = a_n^{(q)} E_{1 n}^q +  a_n^{(g)} E_{1 n}^g$
and
${\tilde d}_n = d_n^{(q)} E_{2 n}^q +  d_n^{(g)} E_{2 n}^g$
where $E_{1 n}^i$ and $E_{2 n}^i$ are
the Wilson coefficients
for $a_n^i$ and $d_n^i$ respectively.
Combining equations (\ref{eqc42}) and (\ref{eqc46}) one obtains 
the following equations
for $\alpha_1$ and $\alpha_2$:
\begin{eqnarray}
\alpha_1(x,Q^2) + \alpha_2(x,Q^2)
&=&
\sum_{n=0,2,4,...} {{\tilde a}_n + n {\tilde d}_n \over n+1} {1 \over x^{n+1}}
\nonumber \\
\alpha_2(x,Q^2)
&=&
\sum_{n=2,4,...} {n ({\tilde d}_n - {\tilde a}_n) \over n+1} {1 \over x^{n+1}}
.
\nonumber \\
\label{eqc47}
\end{eqnarray}
These equations are compared with the Taylor series expansions
(\ref{eqc41}),
whence we obtain
the moment sum rules for $g_1$ and $g_2$:
\begin{equation}
\int_0^1 dx x^n g_1 = {1 \over 2} {\tilde a}_n
\
  - 
\
\delta_{n0} \ {1 \over 2} {Q^2 \over 2 M^2} \beta_1 (Q^2)
\label{eqc48}
\end{equation}
for $= 0,2,4,...$
and
\begin{equation}
\int_0^1 dx x^n g_2 = {1 \over 2} {n \over n+1} ({\tilde d}_n - {\tilde a}_n)
\label{eqc49}
\end{equation}
for $n = 2,4,6,...$
\\

Note:
\begin{enumerate}
\item
The first moment of $g_1$ is given by the nucleon matrix element of
the axial vector current ${\bar \psi} \gamma_{\sigma} \gamma_5 \psi$.
There is no twist-two, spin-one, gauge-invariant, local gluon operator
to contribute to the first moment of $g_1$ \cite{Jaffe:1990a}.
\item
The potential subtraction term ${Q^2 \over 2 M^2} \beta_1 (Q^2)$
in the dispersion relation in (\ref{eqc41})
multiplies a ${1 \over x}$ term in the series expansion
on the left hand side,
and thus provides a
potential correction factor to sum rules for the first moment of $g_1$.
It follows that the first moment of $g_1$ measured in polarized
deep inelastic scattering measures the nucleon matrix element of
the axial vector current
up to this potential ``subtraction at infinity'' term,
which corresponds to the residue of any $J=1$
fixed pole with nonpolynomial residue contribution to the real part of $A_1$.
\item
There is no ${1 \over x}$ term in the operator product expansion formula
(\ref{eqc47}) for $\alpha_2 (x, Q^2)$.
This is matched by the lack of any ${1 \over x}$ term in the unsubtracted
version of the dispersion relation (\ref{eqc41}).
The operator product expansion provides no information about the first
moment of $g_2$ without
additional assumptions concerning analytic continuation and the $x \sim 0$
behaviour of $g_2$ \cite{Jaffe:1990b}.
We shall return to this discussion in the context of the 
Burkhardt-Cottingham sum rule for $g_2$ in Section III.D
below.
\end{enumerate}
If there are finite subtraction constant corrections
to one (or more) spin sum rules,
one can include the correction by re-interpreting
the relevant structure function as a distribution
with the subtraction constant included as twice the coefficient of a
$\delta (x)$ term \cite{Broadhurst:1973}.

\subsection{$g_1$ spin sum rules in polarized deep inelastic 
scattering}

The value of $g_A^{(0)}$ extracted from polarized deep inelastic
scattering is obtained as follows.
One includes the sum over quark charges squared in $W_{\mu \nu}$
and assumes no twist-two subtraction constant
($\beta_1 (Q^2) = O(1/Q^4)$).
The first moment of  the structure function $g_1$
is then related
to the scale-invariant axial charges of the target nucleon by:
\begin{eqnarray}
& &
\int_0^1 dx \ g_1^p (x,Q^2) =
\nonumber \\
& & \ \ \ \ \ \ \ \ \ \ \ \ 
\Biggl( {1 \over 12} g_A^{(3)} + {1 \over 36} g_A^{(8)} \Biggr)
\Bigl\{1 + \sum_{\ell\geq 1} c_{{\rm NS} \ell\,}
\alpha_s^{\ell}(Q)\Bigr\} 
\nonumber \\
& & \ \ \ \ \ \ \ \ \ \ \ \ 
+ {1 \over 9} g_A^{(0)}|_{\rm inv}
\Bigl\{1 + \sum_{\ell\geq 1} c_{{\rm S} \ell\,}
\alpha_s^{\ell}(Q)\Bigr\}  +  {\cal O}({1 \over Q^2})
\nonumber \\
& & \ \ \ \ \ \ \ \ \ \ \ \ 
 - \ \beta_1 (Q^2) {Q^2 \over 4 M^2}
.
\nonumber \\
\label{eqc50}
\end{eqnarray}
Here $g_A^{(3)}$, $g_A^{(8)}$ and $g_A^{(0)}|_{\rm inv}$ are
the isotriplet, SU(3)
octet and scale-invariant  flavour-singlet axial charges respectively.
The flavour non-singlet $c_{{\rm NS} \ell}$ and singlet
$c_{{\rm S} \ell}$
Wilson coefficients are calculable in $\ell$-loop perturbative QCD
\cite{Larin:1997}.
One then assumes no twist-two subtraction constant
($\beta_1 (Q^2) = O(1/Q^4)$)
so that the axial charge contributions saturate the first moment
at leading twist.

The first moment of $g_1$ is constrained by low energy weak
interactions.
For proton states $|p,s\rangle$ with momentum $p_\mu$ and spin $s_\mu$
\begin{eqnarray}
2 M s_{\mu} \ g_A^{(3)} &=&
\langle p,s |
\left(\bar{u}\gamma_\mu\gamma_5u - \bar{d}\gamma_\mu\gamma_5d \right)
| p,s \rangle   \nonumber \\
2 M s_{\mu} \ g_A^{(8)} &=&
\langle p,s |
\left(\bar{u}\gamma_\mu\gamma_5u + \bar{d}\gamma_\mu\gamma_5d
                   - 2 \bar{s}\gamma_\mu\gamma_5s\right)
| p,s \rangle
.
\nonumber \\
\label{eqc51}
\end{eqnarray}
Here $g_A^{\scriptscriptstyle (3)} = 1.2695 \pm 0.0029$
is the isotriplet axial charge measured in neutron beta-decay;
$g_A^{\scriptscriptstyle (8)} = 0.58 \pm 0.03$
is the octet charge measured independently in hyperon beta decays 
(and SU(3)) \cite{Close:1993}.
The assumption of good SU(3) here
 is supported
 by the recent KTeV measurement \cite{KTeV:2001} of
the $\Xi^0$ beta decay $\Xi^0 \rightarrow \Sigma^+ e {\bar \nu}$.
The non-singlet axial charges are scale invariant.

The scale-invariant flavour-singlet axial charge $g_A^{(0)}|_{\rm inv}$
is defined by
\begin{equation}
2M s_\mu g_A^{(0)}|_{\rm inv} =
\langle p, s|
\ E(\alpha_s) J^{GI}_{\mu5} \ |p, s \rangle
\label{eqc52}
\end{equation}
where
\begin{equation}
J^{GI}_{\mu5} = \left(\bar{u}\gamma_\mu\gamma_5u
                  + \bar{d}\gamma_\mu\gamma_5d
                  + \bar{s}\gamma_\mu\gamma_5s\right)_{GI}
\label{eqc53}
\end{equation}
is the
gauge-invariantly renormalized singlet axial-vector operator
and
\begin{equation}
E(\alpha_s) = \exp \int^{\alpha_s}_0 \! d{\tilde \alpha_s}\,
\gamma({\tilde \alpha_s})/\beta({\tilde \alpha_s})
\label{eqc54}
\end{equation}
is a renormalization group factor
which corrects
for the (two loop) non-zero anomalous dimension
$\gamma(\alpha_s)$ 
\cite{Kodaira:1980}
of $J_{\mu 5}^{GI}$
and which goes to one in the limit
$Q^2 \rightarrow \infty$;
$\beta (\alpha_s)$ is the QCD beta function.
We are free to choose
the QCD coupling $\alpha_s(\mu)$ at either a hard or a soft scale 
$\mu$.
The singlet axial charge $g_A^{(0)}|_{\rm inv}$
is independent of the renormalization scale $\mu$
and corresponds
to the three flavour
$g_A^{(0)}(Q^2)$ evaluated in the limit $Q^2 \rightarrow \infty$.
If we take $\alpha_s (\mu_0^2) \sim 0.6$ as typical of the infrared
region of QCD, then the renormalization group factor
$E(\alpha_s) \simeq 1 - 0.13 - 0.03 = 0.84$
where -0.13 and -0.03
are the ${\cal O}(\alpha_s)$ and ${\cal O}(\alpha_s^2)$ corrections
respectively.

In terms of the flavour dependent axial-charges
\begin{equation}
2M s_{\mu} \Delta q =
\langle p,s |
{\overline q} \gamma_{\mu} \gamma_5 q 
| p,s \rangle 
\label{eqc55}
\end{equation}
the isovector, octet and singlet axial charges are:
\begin{eqnarray}
g_A^{(3)} &=& \Delta u - \Delta d 
\nonumber \\
g_A^{(8)} &=& \Delta u + \Delta d - 2 \Delta s 
\nonumber \\
g_A^{(0)} \equiv
g_A^{(0)}|_{\rm inv}/E(\alpha_s) &=& \Delta u + \Delta d + \Delta s 
.
\label{eqc56}
\end{eqnarray}
The perturbative QCD coefficients in Eq.(\ref{eqc50})
have been calculated 
to $O(\alpha_s^3)$ precision \cite{Larin:1997}.
For three flavours they evaluate as:
\begin{eqnarray}
\Bigl\{1 &+& \sum_{\ell\geq 1} c_{{\rm NS} \ell\,}
\alpha_s^{\ell}(Q)\Bigr\} 
\nonumber \\
&=&
\biggl[
1 - ({\alpha_s \over \pi})
- 3.58333 ({\alpha_s \over \pi})^2 -20.21527 ({\alpha_s \over \pi})^3 
+ ... \biggr]
\nonumber \\
\Bigl\{1  &+& \sum_{\ell\geq 1} c_{{\rm S} \ell\,}
\alpha_s^{\ell}(Q)\Bigr\}
\nonumber \\
&=&
\biggl[
1 - 0.33333 ({\alpha_s \over \pi})
- 0.54959 ({\alpha_s \over \pi})^2 - 4.44725 ({\alpha_s \over \pi})^3 
\nonumber \\
& & \ \ \ \ \ \ \ \ \ \ \ \ \ \ \ \ \ \ \ \ \ \ + ... \biggr]
.
\label{eqc57}
\end{eqnarray}

In the isovector channel the Bjorken sum rule \cite{Bjorken:1966,Bjorken:1970}
\begin{eqnarray}
I_{Bj} 
&=&
\int_0^1 dx \Biggl( g_1^p - g_1^n \Biggr)
\nonumber \\
&=&
\frac{g_A^{(3)}}{6}
\left[1 - \frac{\alpha_s}{\pi} - 3.583 \left(\frac{\alpha_s}{\pi} \right)^2
        - 20.215 \left(\frac{\alpha_s}{\pi} \right)^3 \right]
\nonumber \\
\label{eqc58}
\end{eqnarray}
has been confirmed in polarized deep inelastic scattering 
experiments at the level of 10\%
(where the perturbative QCD coefficient expansion is truncated at 
 $O(\alpha_s^3)$).
The E155 Collaboration at SLAC found
$\int_0^1 dx (g_1^p - g_1^n) = 0.176 \pm 0.003 \pm 0.007$
using a next-to-leading order QCD motivated fit to evolve
$g_1$ data from the E154 and E155 experiments to $Q^2=5$GeV$^2$
-- in good agreement with the theoretical prediction 
$0.182 \pm 0.005$ from the Bjorken sum-rule \cite{Anthony:2000}.
Using a similar procedure
the SMC experiment obtained 
$\int_0^1 dx (g_1^p - g_1^n) = 0.174 ^{+0.024}_{-0.012}$, 
also at 5GeV$^2$ \cite{Adeva:1998b} and also in agreement with 
the theoretical prediction.

\begin{figure}
\centerline{\psfig{figure=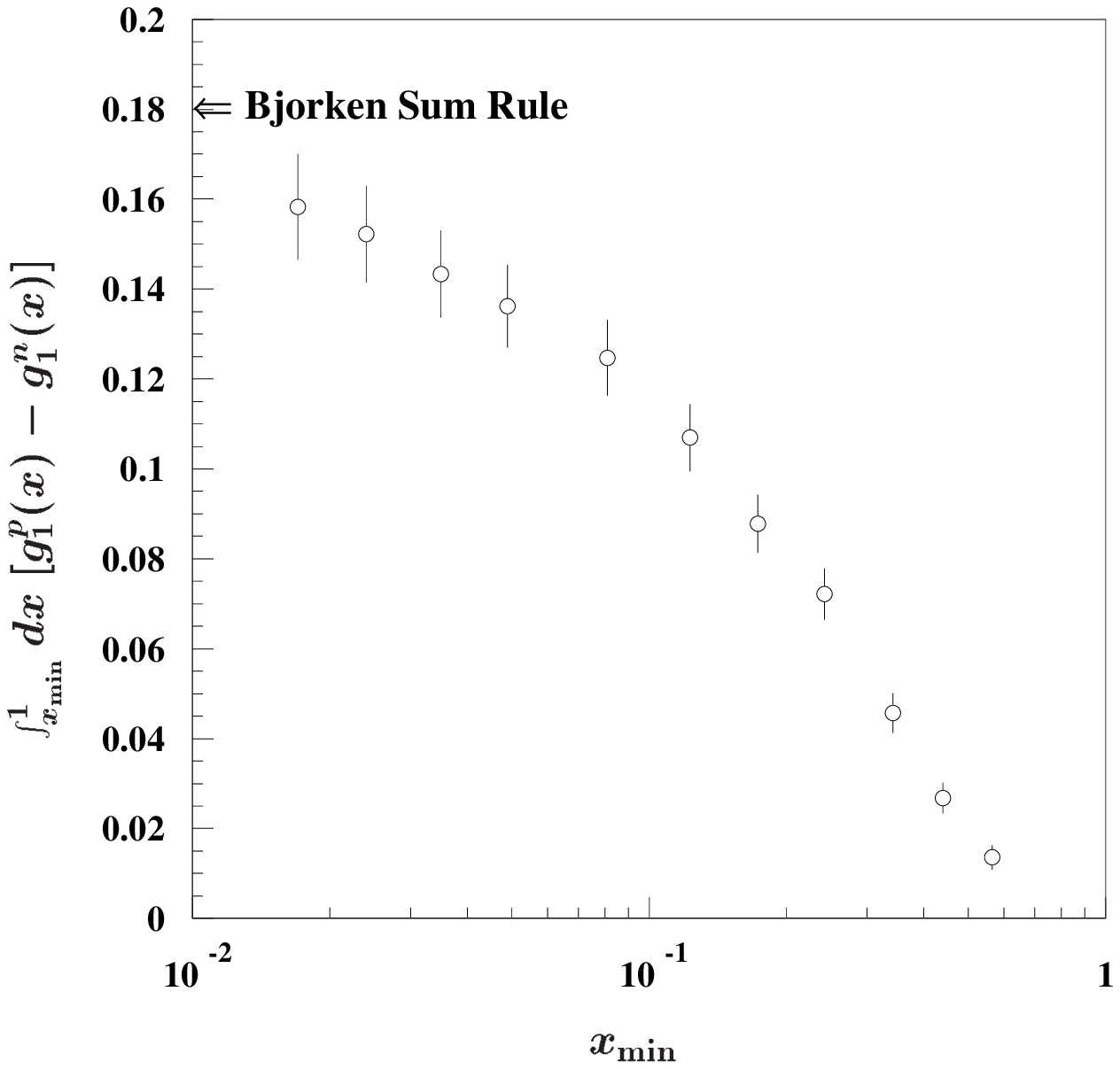,width=4.5in}}
\caption{Difference between the measured proton (SLAC E-143) and 
neutron (SLAC E-154) integrals calculated from
a minimum $x$ value, $x_{\rm min}$ up to $x$ of 1.  The value is compared to
the theoretical prediction from the Bjorken sum rule which makes a prediction
over the full $x$ range.  For the prediction, the Bjorken sum rule is evaluated
up to third order in $\alpha_s$\protect\cite{Larin:1997} and 
at $Q^2$ = 5 (GeV/c)$^2$.
Error bars on the data are dominated by systematic uncertainties
and are highly correlated point-to-point. 
Figure from \textcite{Abe:1997}. }
\label{fig:fig4}
\end{figure}

The evolution of the Bjorken integral \cite{Abe:1997}
$\int_{x_{\rm min}}^1 dx (g_1^p - g_1^n)$ as a function of $x_{min}$ 
is shown for the SLAC data (E143 and E154)
in Fig.\ref{fig:fig4}.
Note that about 50\% of the sum-rule comes from $x$ values below
about 0.12 and that about 10-20\% comes from values of $x$ less than 
about 0.01.

Substituting the values of $g_A^{(3)}$ and $g_A^{(8)}$
from beta-decays
(and assuming no subtraction constant correction)
in the first moment equation
(\ref{eqc50})
polarized deep inelastic data implies
\begin{equation}
g^{(0)}_A |_{\rm pDIS} = 0.15 - 0.35
\label{eqc59}
\end{equation}
for the flavour singlet (Ellis Jaffe) moment
corresponding to the
polarized strangeness
$\Delta s = -0.10 \pm 0.04$ quoted in Section I.
The measured value of 
$
g^{(0)}_A |_{\rm pDIS}
$
compares with the value 0.6 predicted by relativistic quark 
models and is less than 50\% the value one would expect if 
strangeness were not important ({\it viz.} $g_A^0 = g_A^8$)
and the value predicted by relativistic quark models without
additional gluonic input.

The small $x$ extrapolation of $g_1$ data is the largest source
of experimental error on measurements of the nucleon's axial charges from
deep inelastic scattering.
The first polarized deep inelastic experiments \cite{Ashman:1988,Ashman:1989}
used a simple Regge motivated 
extrapolation 
$\{ g_1 \sim {\tt constant} \}$
to evaluate the first moment sum-rules. 
More recent measurements quoted 
in the literature frequently use the technique of 
performing next-to-leading-order QCD motivated 
fits to the $g_1$ data, evolving the data points all to the same value of
$Q^2$ and then extrapolating these fits to $x=0$.
Values extracted from 
these fits using the ``$\overline{\rm MS}$ scheme''
include
$g_A^{(0)} = 0.23 \pm 0.04 \pm 0.06$ 
(SLAC experiment E155 at $Q^2=5$GeV$^2$ \cite{Anthony:2000}),
$g_A^{(0)} = 0.19 \pm 0.05 \pm 0.04$ (SMC at $Q^2=1$GeV$^2$ 
\cite{Adeva:1998b}), 
and
$g_A^{(0)} = 0.29 \pm 0.10$ 
(the mean value at $Q^2=4$GeV$^2$ obtained in \textcite{Blumlein:2002}).

Note that polarized deep inelastic scattering experiments 
measure $g_1$ between some small but finite value $x_{\rm min}$
and an upper value
$x_{\rm max}$ which is close to one.
As we decrease $x_{\rm min} \rightarrow 0$ we measure the first moment
\begin{equation}
\Gamma \equiv \lim_{x_{\rm min} \rightarrow 0} \
\int^1_{x_{\rm min}} dx \ g_1 (x,Q^2).
\label{eqc60}
\end{equation}
Polarized deep inelastic experiments cannot, even in principle, measure at
$x=0$ with finite $Q^2$.
They miss any possible $\delta (x)$ terms which might exist in $g_1$ at
large $Q^2$.
That is, they miss any potential (leading twist)
fixed pole correction to the deep inelastic spin sum rules.

Measurements of $g_1$ could be extended to smaller $x$ with a future 
polarized $ep$ collider.
The low $x$ behaviour of $g_1$ is itself an interesting topic.
Small $x$ measurements,
besides reducing the error on the first moment 
(and gluon polarization, $\Delta g$, in the proton -- see Section IX.E 
 below), 
would provide valuable information about Regge and QCD 
dynamics at low $x$ 
where the shape of $g_1$ is particularly sensitive to 
the different theoretical inputs discussed in the literature:
e.g. 
$(\alpha_s \ln^2 {1 \over x})^k$ resummation and DGLAP evolution
\cite{Kwiecinski:1999}, 
possible $Q^2$ independent Regge intercepts \cite{Cudell:1999},
and the non-perturbative ``confinement physics'' to hard 
(perturbative QCD) 
scale transition. 
Does the colour glass condensate of small $x$ physics
\cite{Iancu:2002} 
carry net spin polarization ?
We refer to Ziaja \cite{Ziaja:2003}
for a recent discussion of perturbative QCD predictions 
for the small $x$ behaviour of $g_1$ in deep inelastic scattering.
In the conventional picture based on QCD evolution and no separate
hard pomeron trajectory much larger changes in the effective intercepts 
which 
describe the shape of the structure functions at small Bjorken $x$ are 
expected in $g_1$ than in the unpolarized structure function $F_2$ so far
studied at HERA as one increases $Q^2$ through the transition region from 
photoproduction to deep inelastic values of $Q^2$ \cite{Bass:2001}.
It will be fascinating to study this physics in 
future experiments, perhaps using a future polarized $ep$ collider.

\subsection{$\nu p$ elastic scattering}

Neutrino proton elastic scattering measures the proton's weak axial
charge $\gaz$ through elastic Z$^0$ exchange.
Because of anomaly cancellation in the Standard Model
the weak neutral current couples to the combination $u-d+c-s+t-b$,
{\it viz.}
\begin{equation}
J_{\mu5}^Z\
=\ \smallfrac{1}{2} \biggl\{\,\sum_{q=u,c,t} - \sum_{q=d,s,b}\,\biggr\}\:
        \bar{q}\gamma_\mu\gamma_5q
.
\label{eqc61}
\end{equation}
It measures the combination
\begin{equation}
2\gaz = \bigl( \Delta u - \Delta d - \Delta s \bigr)
       + \bigl( \Delta c - \Delta b + \Delta t \bigr)
.
\label{eqc62}
\end{equation}
Heavy quark renormalization group arguments can be used
to calculate the heavy $t$, $b$ and $c$ quark contributions 
to $\gaz$
both at leading-order \cite{Collins:1978,Kaplan:1988,Chetyrkin:1993}
and at next-to-leading-order (NLO) \cite{Bass:2002d}.
Working to NLO it is
necessary to introduce 
``matching functions'' \cite{Bass:2003c}
to maintain renormalization group invariance through-out.
The result is:
\begin{eqnarray}
2\gaz = \bigl(\Delta u - \Delta d - \Delta s\bigr)_{\rm inv}
           &+&\hsp{0.2} {\cal H}\hsp{0.1}\bigl(
               \Delta u + \Delta d + \Delta s\bigr)_{\rm inv}
\nonumber \\
    &+& \ \ \ \ \ O(m_{t,b,c}^{-1})
\nonumber \\
\label{eqc63}
\end{eqnarray}
where ${\cal H}$ is a polynomial in the running couplings
$\run{h}$,
\begin{eqnarray}
{\cal H}
 &=&  \smallfrac{6}{23\pi}\bigl(\run{b}-\run{t}\bigr)
             \Bigl\{1 + \smallfrac{125663}{82800\pi}\run{b}
                      + \smallfrac{6167}{3312\pi}\run{t}
                      - \smallfrac{22}{75\pi}\run{c}  \Bigr\}
\nonumber \\
& &
\ \ \ \ \ \ \ 
- \smallfrac{6}{27\pi} \run{c}
                      - \smallfrac{181}{648 \pi^2}\run{c}^2
                      + O\bigl(\run{t,b,c}^3\bigr)
.
\nonumber \\
\label{eqc64}
\end{eqnarray}
Here $(\Delta q)_{\rm inv}$ denotes the scale-invariant version of
$\Delta q$ 
which are obtained from linear combinations of
$g_A^{(3)}$, $g_A^{(8)}$ and $g_A^{(0)}|_{\rm pDIS}$
and
$\run{h}$ denotes Witten's renormalization-group-invariant 
running couplings for heavy quark physics \cite{Witten:1976}.
Taking $\widetilde{\alpha}_t = 0.1$, $\widetilde{\alpha}_b = 0.2$
and $\widetilde{\alpha}_c = 0.35$ in (\ref{eqc64}), one finds a small
heavy-quark correction factor ${\cal H}= -0.02$, with leading-order terms 
dominant.
The factor $\bigl(\run{b}-\run{t}\bigr)$ ensures that all contributions
from $b$ and $t$ quarks cancel for $m_t=m_b$ (as they should).

Modulo the small heavy-quark corrections quoted above, a precision
measurement of $g_A^{(Z)}$,
together with
$g_A^{(3)}$ and $g_A^{(8)}$,
would
provide a weak interaction determination of $(\Delta s)_{\rm inv}$,
complementary to the deep inelastic measurement of ``$\Delta s$''
in Eq.(\ref{eqa2}).
The singlet axial charge in principle measurable in $\nu p$ elastic 
scattering is independent of any assumptions about the presence or
absence of a subtraction at infinity correction to the Ellis-Jaffe
deep inelastic first moment of $g_1$, 
the $x \sim 0$ behaviour of $g_1$
or SU(3) flavour breaking.
Modulo any ``subtraction at infinity'' correction to the first 
moment of $g_1$, 
one obtains a rigorous sum-rule relating 
deep inelastic scattering in the Bjorken
region of high-energy and high-momentum-transfer to three independent,
low-energy measurements in weak interaction physics:
the neutron and hyperon beta decays plus $\nu p$ elastic scattering.

A precision measurement of the $Z^0$ axial coupling to the proton is
therefore of very high priority.
Ideas are being discussed for a dedicated experiment 
\cite{Tayloe:2002}.
Key issues are the ability to measure close to the elastic point and
a very low duty factor ($\sim 10^{-5}$) neutrino beam to control 
backgrounds, e.g. from cosmic rays.

The experiment E734 at BNL made the first attempt to measure $\Delta s$ 
in $\nu p$ and ${\overline \nu} p$ elastic scattering 
\cite{Ahrens:1987}.
This experiment extracted differential cross-sctions 
$d\sigma /d Q^2$ in the range $0.4 < Q^2 < 1.1$GeV$^2$. 
Extrapolating the axial form factor 
$(1 - 2 \Delta s|_{\rm inv}/g_A^{(3)}) 
 / 
 (1 + Q^2/M_A^2)^2
$
to the elastic limit
one obtains the value for $\Delta s$ \cite{Kaplan:1988}:
$
\Delta s = -0.15 \pm 0.09 
$
taking the mass parameter in the dipole form factor 
to be
$
M_A = 1.032 \pm 0.036 {\rm GeV}$.
However, 
the data is also consistent with 
$
\Delta s = 0 
$
if one takes the mass parameter to be
$
M_A = 1.06 \pm 0.05 {\rm GeV}
$
which is consistent with the world average and therefore
equally valid as a solution.
That is, there is a strong correlation between the value
of $\Delta s$ and the dipole mass parameter $M_A$ used in
the analysis which prevents an unambiguous extraction of $\Delta s$ 
from the E734 data \cite{Garvey:1993}.
A new dedicated precision experiment is required.

The neutral-current axial-charge $g_A^{(Z)}$ could also be measured
through
parity violation in light atoms 
\cite{Fortson:1984,Khriplovich:1991,Missimer:1985,Bruss:1998,Bruss:1999,
Alberico:2002,Campbell:1989}.

\subsection{The Burkhardt-Cottingham sum rule}

The Burkhardt-Cottingham sum rule \cite{Burkhardt:1970} reads:
\begin{equation}
\int_{Q^2/2M}^{\infty} d \nu G_2 (Q^2, \nu)
= {2 M^3 \over Q^2} \int_0^1 dx g_2
= 0, 
 \ \ \ \ \ \forall Q^2
.
\label{eqc65}
\end{equation}
For deep inelastic scattering, this sum rule is derived
by assuming that the moment formula (\ref{eqc49})
can be analytically continued to $n=0$.
In general, the Burkhardt-Cottingham sum rule is derived
by assuming no $\alpha \geq 0$
singularity in $G_2$
(or, equivalently, no ${1 \over x}$ or more singular small
 behaviour in $g_2$)
and no ``subtraction at infinity''
(from an
 $\alpha = J = 0$ fixed pole in the real part of $G_2$)
\cite{Jaffe:1990b}.
The most precise measurements of $g_2$
to date
in polarized deep inelastic scattering
come from
the SLAC E-155 and E-143 experiments, which report
$\int_{0.02}^{0.8} dx  \ g_2^p = -0.042 \pm 0.008$
for the proton and
$\int_{0.02}^{0.8} dx  \ g_2^d = -0.006 \pm 0.011$
for the deuteron at $Q^2=5$GeV$^2$ \cite{Anthony:2003}.
New, even more accurate,
measurements
of $g_2$
(for the neutron using a $^3$He target)
from
Jefferson Laboratory \cite{Amarian:2004} 
for $Q^2$ between 0.1 and 0.9 GeV$^2$
are consistent with the sum rule.
Further measurements to test the Burkhardt-Cottingham sum rule would
be most valuable, particularly given the SLAC proton result quoted above.

\begin{figure}[t]
\includegraphics{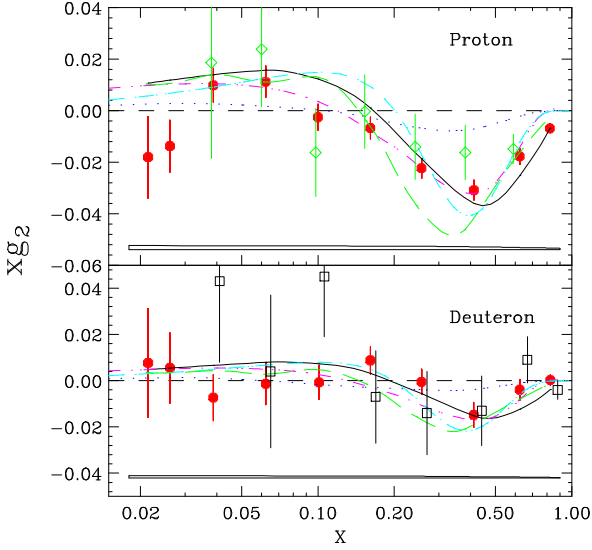}
\vspace{8.0cm}
\parbox{8.5cm}
{\caption[Delta]
{
The $Q^2$ averaged measured of $x g_2$ (SLAC data)
compared with the twist-two Wandzura-Wilczek contribution $g_2^{WW}$
term (solid line) and several quark model calculations.
Figure from \textcite{Anthony:2003}.
}
\label{fig:fig5}}
\end{figure}

The formula (\ref{eqc49}) indicates that $g_2$ can be written as the sum 
\begin{equation}
g_2 = g_2^{\rm WW} (x) + {\overline g}_2 (x)
\label{eqc66}
\end{equation}
of a twist-two term \cite{Wandzura:1977}, 
denoted $g_2^{\rm WW}$
\begin{equation}
g_2^{\rm WW} = - g_1 (x) + \int_x^1 {dy \over y} g_1 (y)
\label{eqc67}
\end{equation}
and a second contribution ${\bar g}_2$ 
which is the sum of a 
higher-twist (twist 3) contribution $\xi(x,Q^2)$
and a ``transversity'' term $h_1(x,Q^2)$
which is suppressed by the ratio of the quark to target nucleon masses
and therefore negligible for light $u$ and $d$ quarks
\begin{equation}
{\overline g}_2 (x,Q^2) 
= 
- \int_x^1 
{dy \over y}
{\partial \over \partial y} 
\biggl( {m_q \over M} h_1 (y,Q^2) + \xi (y,Q^2) \biggr) 
\label{eqc68}
\end{equation}
-- see \textcite{Cortes:1992}.
The first moment of the twist 2 contribution 
$g_2^{\rm WW}$ vanishes 
through integrating the convolution formula (\ref{eqc67}).
If one drops the transversity contribution 
from the formalism 
(being proportional to the light quark mass), 
one obtains the equation
\begin{equation}
{\tilde d}_2 (Q^2)
=
3 \int_0^1 dx x^2 \biggl[ g_2 (x,Q^2) - g_2^{\rm WW} (x,Q^2) \biggr]
\label{eqc69}
\end{equation}
for the leading twist 3 matrix element in Eq.(\ref{eqc49}).
The values extracted from dedicated SLAC measurements 
are
$d_2^p = 0.0032 \pm 0.0017$ for the proton 
and 
$d_2^n = 0.0079 \pm 0.0048$ for the neutron
-- that is, consistent with zero (no twist-3) at two standard 
deviations \cite{Anthony:2003}.
These twist 3 matrix elements are related in part to the response of 
the collective colour electric and magnetic fields to the spin of the nucleon. 
Recent analyses attempt to extract the twist-four 
corrections to $g_1$.
The results and the gluon field polarizabilities are small and consistent 
with zero
\cite{Deur:2004}.

\subsection{The Gerasimov-Drell-Hearn sum rule}

The Gerasimov-Drell-Hearn (GDH)
sum-rule \cite{Gerasimov:1965,Drell:1966} 
for spin dependent
photoproduction
relates the difference of the two cross-sections for the absorption
of a real photon with spin polarized anti-parallel, 
$\sigma_{1 \over 2}$, and 
parallel,
$\sigma_{3 \over 2}$ ,
to the target spin to the square of the anomalous magnetic
moment of the target.
The GDH sum rule reads:
\begin{eqnarray}
\int_{\rm threshold}^{\infty} {d \nu \over \nu}
( \sigma_{1 \over 2} - \sigma_{3 \over 2} )
&=& {8 \pi^2 \alpha \over M^2}
\int_{\rm threshold}^{\infty} {d \nu \over \nu}
G_1
\nonumber \\
&=& - {2 \pi^2 \alpha \over M^2} \kappa^2
\label{eqc71}
\end{eqnarray}
where $\kappa$ is the anomalous magnetic moment.
The sum rule
follows from the very general principles of causality, unitarity,
Lorentz and electromagnetic gauge invariance and one assumption:
that the $g_1$ spin structure function satisfies an unsubtracted
dispersion relation.
Modulo the no-subtraction hypothesis,
the Gerasimov-Drell-Hearn sum-rule is valid for a target of
arbitrary spin $S$, whether elementary or composite 
\cite{Brodsky:1969} -- for reviews see \textcite{Bass:1997} and
\textcite{Drechsel:2004}.

\begin{figure}[h]
\includegraphics{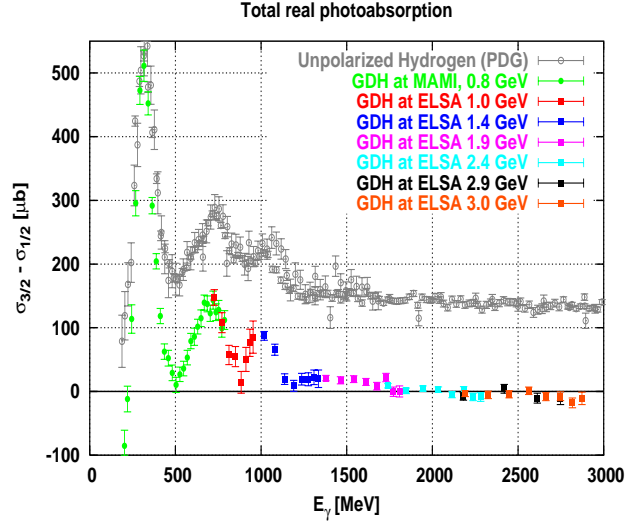}
\vspace{7.0cm}
\parbox{8.0cm}
{\caption[Delta] 
{The spin dependent photoproduction cross-section for the proton
 target (ELSA and MAMI data). Figure courtesy of K. Helbing.}
\label{fig:fig6}}
\end{figure}

\begin{figure}[h]
\includegraphics{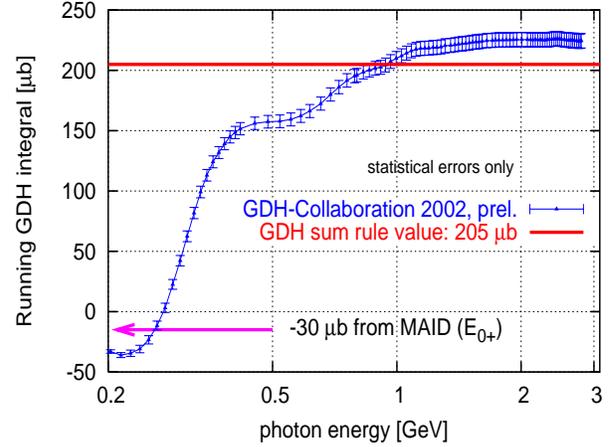}
\vspace{6.0cm}
\parbox{8.0cm}
{\caption[Delta] 
{Running GDH integral for the proton (ELSA and MAMI). Figure 
 courtesy of K. Helbing.}
\label{fig:fig7}}
\end{figure}

The GDH sum-rule is derived by setting $\nu=0$ in the dispersion 
relation for $A_1$, Eq.(\ref{eqc38}).
For small photon energy $\nu \rightarrow 0$
\begin{equation}
A_1(0,\nu) 
= 
- {1 \over 2} \kappa^2 + {\tilde \gamma} \nu^2 + O(\nu^4) 
.
\label{eqc70}
\end{equation}
Here
$\gamma_N = {\alpha \over M^2} {\tilde \gamma}$
is the spin polarizability
which measures the stiffness of the nucleon spin against 
electromagnetic induced deformations relative to the axis
defined by the nucleon's spin.
This low-energy theorem follows from Lorentz invariance and
electromagnetic gauge invariance
(plus the existence of a finite mass gap between the
 ground state and continuum
 contributions to forward Compton scattering) 
\cite{Brodsky:1969,Low:1954,Gell-Mann:1954}.

The integral in Eq.(\ref{eqc71}) converges for each of the leading
Regge contributions (discussed in Section II.B).
If the sum rule were observed
to fail
(with a finite integral)
the interpretation would be a ``subtraction at infinity''
induced by a $J=1$ fixed pole in the real part of the spin amplitude
$A_1$
\cite{Abarbanel:1968}.

Present experiments at ELSA and MAMI are aimed at measuring 
the GDH integrand through the range of incident photon energies 
$E_{\gamma} =$ 
0.14 - 0.8 GeV (MAMI) \cite{Ahrens:2000,Ahrens:2001,Ahrens:2002} 
and 0.7 - 3.1 GeV (ELSA) \cite{Dutz:2003}.
The inclusive cross-section for the proton target
${\sigma_{3 \over 2} - \sigma_{1 \over 2}}$
is shown in Fig.~\ref{fig:fig6}.
The presently analysed GDH integral on 
the proton
is shown in Fig.~\ref{fig:fig7} 
and
is dominated by the $\Delta$ resonance contribution.
(The contribution to the sum-rule from the unmeasured 
 region close to threshold is estimated from the MAID
 model \cite{Drechsel:2002}.)
The combined data from the ELSA-MAMI experiments
suggest that the contribution to the GDH integral
for a proton target
from energies $\nu < 3$GeV
exceeds
the total sum rule prediction (-204.5$\mu$b) by about 5-10\%
\cite{Helbing:2002}.
Phenomenological estimates
suggest that about $+25 \pm 10 \mu$b of the sum rule 
may reside at higher energies \cite{Bass:1999a,Bianchi:1999} and 
that this high energy contribution is predominantly in the isovector
channel.
(It should be noted, however, that any 10\% fixed pole correction would
 be competitive with this high energy contribution within the errors.)
Further measurements, including at higher energy, would be valuable.
Preliminary data on the neutron has just been released from MAMI and
ELSA \cite{Helbing:2004}.
This data, if confirmed, suggests that the neutron GDH integal, if it
indeed obeys the GDH sum-rule, 
will require a 
large (mainly isovector) contribution (perhaps 45$\mu$b) 
from photon energies $E_{\gamma}$ greater than about 1800 MeV.
With the caution that these data are still preliminary, it is interesting
to note that, just like the measured $g_1$ at deep inelastic $Q^2$, 
the high-energy part of the spin dependent cross-section 
$\sigma_{1 \over 2} - \sigma_{3 \over 2}$ at $Q^2=0$ 
seems to be 
largely isovector prompting the question whether 
there is some physics conspiracy to suppress the singlet term.
It should be noted however that perturbative QCD motivated
fits to $g_1$ data with a positive polarized gluon distribution 
(and no node in it)
predict that $g_1$ should develop a strong 
negative contribution at $x < 0.0001$ at deep inelastic $Q^2$ 
-- see e.g. \textcite{DeRoeck:1999} and references therein.

In addition to the GDH sum-rule, one also finds a second sum-rule
for the nucleon's spin polarizability.
This spin polarizability sum-rule is derived by taking the second
derivative of $A_1(Q^2,\nu)$ in the dispersion relation
(\ref{eqc38}) and evaluating the resulting expression at $\nu=0$, 
viz. 
${\partial^2 \over \partial \nu^2} A_1 (Q^2, \nu)|_{\nu=0}$.
One finds
\begin{equation}
\int_0^{\infty} {d \nu' \over \nu'^3}
\Biggl (\sigma_{1 \over 2} - \sigma_{3 \over 2} \Biggr)(\nu')
=
{4 \pi^2} \ \gamma_N
.
\label{eqc72}
\end{equation}
In comparison with the GDH sum-rule the relevant information is 
now concentrated more on the low energy side because of the 
$1/\nu'^3$
weighting factor under the integral.
Main contributions come from the $\Delta$(1232) resonance and
the low energy pion photoproduction continuum described by the
electric dipole amplitude $E_{0+}$.
The value extracted from MAMI data \cite{Drechsel:2002}
\begin{equation}
\gamma_p = (-1.01 \pm 0.13) .  10^{-4} \ {\rm fm}^4
\label{eqc73}
\end{equation}
is within the range of predictions of chiral perturbation theory.

Further experiments to test the GDH sum-rule and to measure the
$\sigma_{1 \over 2} - \sigma_{3 \over 2}$
at and close to $Q^2=0$ are being carried out at
Jefferson Laboratory, GRAAL at Grenoble, LEGS at BNL, and SPRING-8
in Japan.

We note two interesting properties of the GDH sum rule.

First, 
we write the anomalous magnetic moment $\kappa$ as the 
sum of its isovector $\kappa_V$ and isoscalar $\kappa_S$ 
contributions,
viz.
$
\kappa_N = \kappa_S + \tau_3 \kappa_V.
$
One then obtains the isospin dependent expressions:
\begin{eqnarray}
  ({\rm GDH})_{I=0} = ({\rm GDH})_{VV} + ({\rm GDH})_{SS} 
                    &=& - {2 \pi^2 \alpha \over m^2} 
                         (\kappa_V^2 + \kappa_S^2)     
\nonumber \\
  ({\rm GDH})_{I=1} = ({\rm GDH})_{VS}  
                    &=& - {2 \pi^2 \alpha \over m^2} 
                         2 \kappa_V \kappa_S
.
\nonumber \\
\label{eqc74}
\end{eqnarray}
The physical values of the proton and nucleon anomalous magnetic 
moments $\kappa_p = 1.79$ and $\kappa_n = -1.91$ 
correspond to
$\kappa_S = -0.06$ and $\kappa_V = +1.85$.
Since 
$\kappa_S /  \kappa_V \simeq - {1 \over 30}$, 
it follows that $({\rm GDH})_{SS}$ 
is negligible compared to $({\rm GDH})_{VV}$.
That is, to good approximation,
the isoscalar sum-rule $({\rm GDH})_{I=0}$ 
measures the isovector anomalous magnetic moment $\kappa_V$.
Given this isoscalar measurement,
the isovector sum-rule $({\rm GDH})_{I=1}$ 
then measures the 
isoscalar anomalous magnetic moment $\kappa_S$.

Second,
the anomalous magnetic moment is measured in the matrix element of 
the vector current.
Furry's theorem tells us that the real-photon GDH integral 
for a gluon or a photon target vanishes. 
Indeed, this is the reason that the first moment of the $g_1$ spin 
structure function for a real polarized photon target vanishes to all 
orders and at every twist: 
$\int_0^1 dx \ g_1^{\gamma}(x,Q^2)$ 
independent of the virtuality $Q^2$ of the second photon that it is 
probed with
\cite{Bass:1998b}.
Assuming correction to the 
GDH sum rule,
this result implies that the two non-perturbative gluon exchange 
contribution to $\sigma_{1 \over 2} - \sigma_{3 \over 2}$
which behaves as $\ln \nu / \nu$ 
in the high energy Regge limit has a node at some value $\nu = \nu_0$
so that it does not contribute to the GDH integral.
There is no axial anomaly contribution to the anomalous magnetic
moment and hence no axial anomaly contribution to the GDH sum-rule.

\subsection{The transition region}

\begin{figure}[h]
\includegraphics{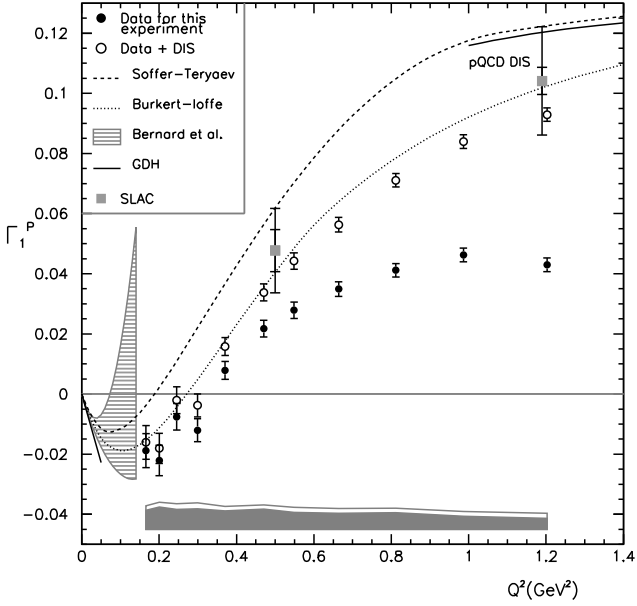}
\vspace{8.5cm}
\parbox{8.5cm}
{\caption[Delta] 
{
Data from JLab (CLAS) and SLAC on the low $Q^2$ behaviour of 
$\int_0^1 dx g_1^p$
compared to various theoretical models interpolating 
the scaling and photoproduction limits \cite{Fatemi:2003}.}
\label{fig:fig8}}
\end{figure}

Several experiments have explored the transition region between 
polarized photoproduction (the physics of the GDH sum-rule) and
polarized deep inelastic scattering (the physics of the Bjorken
sum-rule and $g_A^{(0)}$ through the Ellis-Jaffe moment).

The $Q^2$ dependent quantity \cite{Anselmino:1989}
\begin{eqnarray}
\Gamma (Q^2) \equiv I(Q^2) 
&=& \int_{ {Q^2 \over 2M} }^{\infty} {d \nu \over \nu} 
G_1 (\nu, Q^2)
\nonumber \\
&=&
{2 M^2 \over Q^2}
\int_0^1 dx g_1 (x, Q^2)
\label{eqc75}
\end{eqnarray}
interpolates between the two limits with
$I(0) = - {1 \over 4} \kappa^2_N$
implied by the GDH sum rule.
Measurements of $\int_0^1 dx g_1^p = {Q^2 \over 2 M^2} I(Q^2)$
are shown in Fig.~\ref{fig:fig8}.
Note the negative slope predicted at $Q^2=0$ by the GDH
sum rule and the sign change around $Q^2 \sim 0.3$GeV$^2$.
The shape of the curve is driven predominantly by the role
of the $\Delta$ resonance and the $1/Q^2$ pole in Eq.(\ref{eqc75}).
Fig.~\ref{fig:fig8} shows also the predictions of various models 
\cite{Soffer:1993,Burkert:1994}
which try to describe the intermediate $Q^2$ range through a combination
of resonance physics and vector-meson dominance at low $Q^2$ and scaling
parton physics at DIS $Q^2$.
Chiral perturbation theory \cite{Bernard:2003} 
may describe the behaviour of this ``generalized GDH integral'' 
close to threshold -- see the shaded band in Fig. \ref{fig:fig8}.

In the model of Ioffe and collaborators \cite{Anselmino:1989,Burkert:1994} 
the integral at
low to intermediate $Q^2$ for the inelastic part of 
$(\sigma_A - \sigma_P)$
is given as the sum of a contribution from resonance 
production,
denoted $I^{\rm res}(Q^2)$,
which has a strong $Q^2$ dependence for small $Q^2$ and 
then drops
rapidly with $Q^2$, and a non-resonant vector-meson dominance
contribution
which they took as the sum of a monopole and a dipole term, viz.
\begin{equation}
I(Q^2) = 
I^{\rm res} (Q^2) + 2 M^2 \Gamma^{\rm as}
\Biggl( {1 \over Q^2 + \mu^2} 
- {C \mu^2 \over (Q^2 + \mu^2)^2} \Biggr).
\label{eqc76}
\end{equation}
Here $\Gamma^{\rm as}$ is taken as
\begin{equation}
\Gamma^{\rm as} = \int_0^1 dx g_1 (x, \infty)
\label{eqc77}
\end{equation}
and 
\begin{equation}
C = 1 + {1 \over 2} 
\ {\mu^2 \over M^2} \ {1 \over \Gamma^{\rm as}} \
\Bigl( {1 \over 4} \kappa^2 + I^{\rm res}(0) \Bigr)
.
\label{eqc78}
\end{equation}
The mass parameter 
$\mu$ is identified with rho meson mass, $\mu^2 \simeq m_{\rho}^2$.

\section{PARTONS AND SPIN STRUCTURE FUNCTIONS}

\subsection{The QCD parton model}

We now return to $g_1$ in the scaling regime of polarized deep inelastic
scattering.
As noted in Section II.A above, in the (pre-QCD) parton model $g_1$ is 
written as
\begin{equation}
g_1 (x) = {1 \over 2} \sum_q e_q^2 \Delta q(x)
\label{eqd79}
\end{equation}
where
$e_q$ denotes the quark charge
and
$
\Delta q(x)$
is the polarized quark distribution.

In QCD we have to consider the effects of gluon radiation and
(renormalization group)
mixing of the flavour-singlet 
quark distribution 
with the polarized gluon distribution of the proton.
The parton model description of polarized deep inelastic scattering
involves writing the deep inelastic structure functions as the sum
over the convolution of ``soft'' quark and gluon parton distributions
with ``hard'' photon-parton scattering coefficients:
\begin{eqnarray}
g_1 (x) 
&=&
\Biggl\{
         {1 \over 12} (\Delta u - \Delta d) 
       + {1 \over 36} (\Delta u + \Delta d - 2 \Delta s) 
\Biggr\}
\otimes C^q_{ns}
\nonumber \\
& &        + {1 \over 9}  \Biggl\{
(\Delta u + \Delta d + \Delta s) \otimes C^q_s
+ 
f \Delta g \otimes C^g
\Biggr\}       
.
\nonumber \\
\label{eqd80}
\end{eqnarray}
Here
$\Delta q(x)$ and $\Delta g(x)$ denote the polarized quark and gluon 
parton distributions, 
$C^q(z)$ and $C^g(z)$ denote the corresponding hard
scattering coefficients, 
and $f$ is the number of quark flavours liberated into the final state
($f=3$ below the charm production threshold).
The parton distributions contain all the target dependent information and 
describe a flux of quark and gluon partons into the (target independent) 
interaction between the hard photon and the parton which is described by 
the coefficients and which is calculable using perturbative QCD.
The perturbative coefficients are independent of infra-red mass
singularities in the photon-parton collision which are absorbed 
into the soft parton distributions (and softened by confinement
related physics).

The separation of $g_1$ into ``hard'' and ``soft'' 
is not unique and depends on the choice of ``factorization scheme''. 
For example, one might use a kinematic cut-off on the partons' 
transverse momentum squared ($k_t^2 > \lambda^2$) to define 
the factorization scheme and thus separate the hard and soft 
parts of the phase space for the photon-parton collision.
The cut-off $\lambda^2$ is called the factorization scale.
The coefficients have the perturbative expansion
$C^q = \delta(1-x) +
         {\alpha_s \over 2\pi} f^q(x, {Q^2 / \lambda^2})$
and
  $C^g = {\alpha_s \over 2\pi} f^g(x, {Q^2 / \lambda^2})$
where
the strongest singularities in the functions $f^q$ and $f^g$ 
as $x \rightarrow 1$ are
$\ln (1-x)/(1-x)_+$ and $\ln (1-x)$ 
respectively
-- see e.g. \textcite{Lampe:2000}.
The deep inelastic structure functions are dependent on $Q^2$ and 
independent of
the factorization scale $\lambda^2$ and the ``scheme'' used to
separate
the $\gamma^{*}$-parton cross-section into ``hard'' and ``soft''
contributions.
Examples of different ``schemes'' one might use include using 
modified minimal subtraction (${\rm \overline{MS}}$) 
\cite{tHooft:1972,Bodwin:1990}
to regulate the mass singularities which arise in scattering 
from massless partons, 
and cut-offs on other kinematic variables such as the invariant 
mass squared or the virtuality of the struck quark.
Other schemes which have been widely used in the literature and
analysis of polarized deep inelastic scattering data are the ``AB''
\cite{Ball:1996}
and ``CI'' (chiral invariant) \cite{Cheng:1996} or ``JET'' \cite{Leader:1998}
schemes.
We illustrate factorization scheme dependence and the use of 
these schemes in the analysis of $g_1$ data in Sections VI.D and IX.C below.

If the same ``scheme'' is applied consistently to all hard processes 
then the factorization theorem asserts that the parton distributions
that one extracts from experiments should be process independent
\cite{Collins:1993a}.
In other words, the same polarized quark and gluon distributions
should be obtained from future experiments involving polarized
hard QCD processes in polarized proton proton collisions (e.g. 
at RHIC) and polarized deep inelastic scattering experiments.
The factorization theorem for unpolarized hard processes has
been successfully tested in a large number of experiments involving
different reactions at various laboratories.
Tests of the polarized version await future independent 
measurements of the polarized gluon and sea-quark distributions 
from a variety of different hard scattering processes with polarized
beams.

\subsection{Light-cone correlation functions}

The (spin-dependent) parton distributions may also be defined via
the operator product expansion.
For $g_1$ 
this means that the odd moments of the polarized quark
and gluon distributions
project out the target matrix elements of 
the 
renormalized, spin-odd, composite operators which appear in 
the operator product expansion, 
{\it viz.}
\begin{eqnarray}
& &
2M s_+ (p_+)^{2n} \int^1_0 dx \ x^{2n} \Delta q (x, \mu^2) 
\nonumber \\
& &
\ \ \ \ \ \ \ \ =
\langle p,s | 
\biggl[ 
{\overline q}(0) \gamma_+ \gamma_5 (i D_+)^{2n} q(0) \biggr]_{\mu^2}
|p,s \rangle
\nonumber \\
\\
& &
2M s_+ (p_+)^{2n} \int^1_0 dx \ x^{2n} \Delta g (x, \mu^2) 
\nonumber \\
& &
\ \ \ \ \ \ \ \ =
\langle p,s | 
\biggl[ {\bf \rm Tr} \ G_{+ \alpha}(0) (iD_+)^{2n-1}
{\tilde G}^{\alpha}_{\ +}(0) \biggr]_{\mu^2} |p,s \rangle 
\nonumber \\
& &
\ \ \ \ \ \ \ \ \ \ \ \ \ \ \ \ \ \ \ \ \ \ \ \ \ \ \ \ 
\ \ \ \ \ \ \ \ \ \ \ \ \ \
(n \geq 1).
\nonumber \\
\label{eqd82}
\end{eqnarray}
The association of $\Delta q (x, \mu^2)$ with quarks and 
$\Delta g(x, \mu^2)$
with gluons follows
when we evaluate the target matrix elements in Eqs.(81) and 
(\ref{eqd82})
in the light-cone gauge, 
where $D_+ \rightarrow \partial_+$ and 
the explicit dependence of $D_+$ on the gluon field drops out.
The operator product expansion involves writing the product of
electromagnetic currents
$J_{\mu} (z) J_{\nu} (0)$
in Eq.(\ref{eqb11})
as the expansion over gauge invariantly renormalized, local, 
composite quark and gluonic operators at lightlike separation 
$z^2 \rightarrow 0$
--
the realm of deep inelastic scattering \cite{Muta:1998}.
The subscript $\mu^2$ on the operators in Eq.(\ref{eqd82})
 emphasises 
 the dependence on the renormalization scale.
\footnote{
Note that the parton distributions defined through the operator product 
expansion include the effect of renormalization effects such as 
the axial anomaly 
(and the trace anomaly for the spin-independent 
distributions which appear in $F_1$ and $F_2$)
in addition to absorbing the mass singularities in photon-parton
scattering.}

Mathematically, the relation between the parton distributions and the 
operator product expansion is given in terms of light-cone correlation 
functions of point split operator matrix elements along the light-cone.
Define 
\begin{equation}
\psi^{\pm} = P^{\pm} \psi
\label{eqd83}
\end{equation}
where
\begin{equation}
P^{\pm} = {1 \over 2} (1 \pm \alpha_3) = {1 \over 2} \gamma^{\pm} \gamma^{\mp}
.
\label{eqd84}
\end{equation}
The polarized quark and antiquark distributions are given by
\begin{eqnarray}
\Delta \psi (x) 
&=& 
{1 \over 2 \sqrt{2} \pi}
\int d \xi^-
e^{- ix M \xi^- / \sqrt{2}}
\nonumber \\
& & 
\langle p,s |
(\psi^{+R})^{\dagger} (\xi^-) \psi^{+R}(0) 
\nonumber \\
& & 
\ \ \ \ \ \ \ \ \ \ \ \ \ \ 
-
(\psi^{+L})^{\dagger} (\xi^-) \psi^{+L}(0) 
| p,s \rangle 
,
\nonumber \\
\Delta {\bar \psi} (x) 
&=& 
{1 \over 2 \sqrt{2} \pi}
\int d \xi^-
e^{- ix M \xi^- / \sqrt{2}}
\nonumber \\
& & 
\langle
p, s |
\psi^{+L} (\xi^-) (\psi^{+L})^{\dagger}(0) 
\nonumber \\
& & 
\ \ \ \ \ \ \ \ \ \ \ \ \ \ 
-
\psi^{+R} (\xi^-) (\psi^{+R})^{\dagger}(0) 
| p,s \rangle
.
\nonumber \\
\label{eqd85}
\end{eqnarray}
In this notation $\Delta q = \Delta \psi + \Delta {\bar \psi}$.
The non-local operator in the correlation function is rendered
gauge invariant through a path ordered exponential which 
simplifies to unity in the light-cone gauge $A_+=0$.
Taking the moments of these
distributions reproduces the results of the operator product expansion
in Eq.~(\ref{eqc48}).
\footnote{
Some care has to be taken regarding renormalization of the light-cone
correlation functions. 
The bare correlation function from which we project out moments as 
local operators 
is ultra-violet divergent.
\textcite{Llewellyn:1988} proposed a solution of this problem 
by defining 
the renormalized light cone correlation function as a series expansion 
in the proton matrix elements of gauge invariant local operators.
For the polarized quark distribution this becomes:
\begin{equation}
\langle 
\overline{\psi}(z_{-}) \gamma_{+} \gamma_{5} \psi (0) 
\rangle
=
\sum_{n} {(-z_{-})^{n} \over n!} 
\langle 
\bigl[
\overline{\psi} \gamma_{+} \gamma_{5} (D_{+})^{n} \psi 
\bigr] 
(0)
\rangle
.
\label{eqd86}
\end{equation}
}
The light-cone correlation function for the polarized gluon distribution
is
\begin{eqnarray}
x \Delta g(x) 
&=& {i \over 2 \sqrt{2} M \pi} 
\int d \xi^- \ 
e^{-i x \xi^- M/\sqrt{2}}
\nonumber \\
& & 
\langle 
p,s |
G_{+ \nu}(\xi^-) {\tilde G}_{\ +}^{\nu}(0) 
-
G_{+ \nu}(0) {\tilde G}_{\ +}^{\nu}(\xi^-) 
| p,s 
\rangle
.
\nonumber \\
\label{eqd87}
\end{eqnarray}
In the light-cone gauge ($A_+=0$) one finds
$
G^{+ \nu}_a = \partial^{+} A^{\nu}_a - \partial^{\nu} A^{+}_a
            = \partial^{+} A^{\nu}_a
$
so that
\begin{equation}
G^{+ \nu} {\tilde G}_{\nu}^{\ +}
= 
G^+_R G_{-L} - G^+_L G_{-R}
=
G^+_R G^{+R} - G^+_L G^{+L}
.
\label{eqd88}
\end{equation}
Thus $\Delta g(x)$ measures the distribution of gluon polarization
in the nucleon.
One can evaluate the first moment of $\Delta g(x)$ 
from its light-cone correlation function. 
One first assumes that
\begin{equation}
\lim_{x \rightarrow 0^+} x \Delta g (x) = 0
.
\label{eqd89}
\end{equation}
In $A_+=0$ gauge the first moment becomes
\begin{eqnarray}
& & 
\int_0^1 dx \Delta g(x) =
\nonumber \\
& & 
{1 \over \sqrt{2} M} 
\biggl[ 
\langle 
A^{\nu} (\xi^-) {\tilde G}^{\ +}_{\nu} (0) 
\rangle |_{\xi^- \rightarrow \infty}
-
\langle 
A^{\nu} (0) {\tilde G}^{\ +}_{\nu} (0)
\rangle 
\biggr]
\nonumber \\
\label{eqd90}
\end{eqnarray}
-- that is,
the sum of the forward matrix element of the gluonic Chern Simons
current $K_+$ plus a surface term \cite{Manohar:1990}
which may or may not vanish in QCD.

\section{FIXED POLES}

Fixed poles are exchanges in Regge phenomenology with no $t$ dependence:
the trajectories are described by
$J = \alpha(t) = 0$ or 1 for all $t$
\cite{Abarbanel:1967,Brodsky:1972,Landshoff:1972}.
For example, for fixed $Q^2$
a $t-$independent real constant term in the spin amplitude $A_1$ would
correspond to a $J=1$ fixed pole.
Fixed poles are excluded in hadron-hadron scattering
by unitarity but are not
excluded from Compton amplitudes (or parton distribution functions)
because these are calculated only to lowest order in the current-hadron
coupling.
Indeed, there are two famous examples where fixed poles are required:
(by current algebra) in the Adler sum rule for W-boson nucleon scattering,
and to reproduce the Schwinger term sum rule for the longitudinal structure
function
measured in unpolarized deep inelastic $ep$ scattering.
We review the derivation of
these fixed pole contributions,
and then discuss potential fixed pole
corrections
to the Burkhardt-Cottingham, $g_1$ and Gerasimov-Drell-Hearn sum-rules.
\footnote
{We refer to \textcite{Efremov:2003} for a
 recent discussion of an ``$x=0$'' fixed pole contribution
 to the twist 3, chiral-odd structure function $e(x)$.}
Fixed poles in the real part of the forward Compton amplitude have the
potential to induce
``subtraction at infinity''
corrections to sum rules for photon nucleon (or lepton nucleon) scattering.
For example, a $\nu$ independent term in the real part of $A_1$ would
induce a subtraction constant correction to the spin sum rule for the first
moment of $g_1$.
Bjorken scaling at large $Q^2$ constrains the $Q^2$ dependence of
the residue
of any fixed pole in the real of the forward Compton amplitude
(e.g. $\beta_1(Q^2)$ and $\beta_2(Q^2)$ in the dispersion relations
 (\ref{eqc41}) ).
To be consistent with scaling
these residues must decay as or faster
than $1/Q^2$ as $Q^2\to\infty$.
That is, they must be nonpolynomial in $Q^2$.

\subsection{Adler sum rule}

The first example we consider is the Adler sum rule for W-boson nucleon
scattering \cite{Adler:1966}:
\begin{eqnarray}
& &\int_{Q^2/2M}^{+ \infty} d \nu
\biggl[
W_2^{\bar{\nu} p} (\nu, Q^2) - W_2^{\nu p} (\nu, Q^2)
\biggr]
\nonumber \\
& &
\ \ \ \ \ 
= \int_0^1 {dx \over x}
\biggl[
F_2^{\bar{\nu} p} (x, Q^2) - F_2^{\nu p} (x, Q^2)
\biggr]
\nonumber \\
& &
\ \ \ \ \ 
=
\begin{array}{c}
4 - 2 \cos^2 \theta_c \ \ \ \ \ ({\rm BCT}) \\
2 \ \ \ \ \ \ \ \ \ \ \ \ \ \ \ \ \ \ \ ({\rm ACT})
.
\end{array}
\label{eqe91}
\end{eqnarray}
Here
$\theta_c$ is the Cabibbo angle, and
BCT and ACT refer to below and above the charm production threshold.

The Adler sum rule is derived from current algebra.
The right hand side of the sum rule is the coefficient of a $J=1$
fixed pole term
\begin{equation}
{i \over \pi} f_{abc} \ F_c \
\biggl[ (p_{\mu} q_{\nu} + q_{\mu} p_{\nu}) - M \nu g_{\mu \nu}
\biggr] / Q^2
\label{eqe92}
\end{equation}
in the imaginary part of the forward Compton amplitude for W-boson nucleon
scattering \cite{Heimann:1972}.
This fixed pole term is required by the
commutation relations between the charge raising and lowering weak currents
\begin{eqnarray}
q_{\mu} T^{\mu \nu}_{ab}
&=&
- {1 \over \pi}
\int d^4 x \  e^{i q.x} \
\langle p,s | \biggl[ J_a^{0} (x),
                     J^{\nu}_b (0) \biggr] | p,s \rangle
\delta (x^0)
\nonumber \\
&=& - {i \over \pi} f_{abc} \langle ps |  J_c^{\nu} (0)| ps \rangle
.
\label{eqe93}
\end{eqnarray}
Here
$F_c$ is a generalized form factor at zero momentum transfer:
\begin{equation}
\langle p,s | J_c^{\nu} (0) | p,s \rangle \equiv p^{\nu} F_c
.
\label{eqe94}
\end{equation}
The fixed pole term appears in lowest order perturbation theory,
and is not
renormalized because it is a consequence of the charge algebra.
The Adler sum rule is protected against radiative QCD corrections

\subsection{Schwinger term sum rule}

Our second example is the Schwinger term sum rule \cite{Broadhurst:1973}
which relates the
logarithmic integral in $\omega$
(or Bjorken $x$)
of the longitudinal structure function $F_L (\omega, Q^2)$
($F_L = {1 \over 2} \omega F_2 - F_1$)
measured in unpolarized deep inelastic scattering to the
target matrix element of the operator Schwinger term ${\cal S}$
defined through the equal-time
commutator of the electromagnetic charge and current densities
\begin{equation}
\langle p,s |
\biggl[ J_0({\vec y}, 0), J_i(0) \biggr]
| p,s \rangle
= i \ \partial_i \ \delta^3 ({\vec y}) \ {\cal S} .
\label{eqe95}
\end{equation}
The Schwinger term sum rule reads
\begin{equation}
{\cal S}
=
\lim_{Q^2 \rightarrow \infty}
\biggl[
4 \int_{1}^{\infty}
  {d \omega \over \omega} {\tilde F}_L (\omega, Q^2)
-
4 \sum_{\alpha >0} \gamma (\alpha,Q^2) / \alpha
- C(q^2)
\biggr]
.
\label{eqe96}
\end{equation}
Here $C(Q^2)$ is the nonpolynomial residue of any $J=0$ fixed
pole contribution in the real part of $T_2$
and
\begin{equation}
{\tilde F}_L (\omega, Q^2)
=
F_L (\omega, Q^2)
-
\sum_{\alpha  \geq 0} \gamma (\alpha, Q^2) \omega^{\alpha} 
\label{eqe97}
\end{equation}
represents $F_L$ with the leading ($\alpha > 0$) Regge behaviour subtracted.
The integral in Eq.(\ref{eqe96})
is convergent because ${\tilde F}_L (\omega, Q^2)$
is defined with all Regge contributions with effective intercept
greater than or equal to zero removed from $F_L (Q^2, \omega)$.
The Schwinger term ${\cal S}$ vanishes in vector gauge theories
like QCD.

Since $F_L (\omega, Q^2)$ is positive definite, it follows that
QCD possesses the required non-vanishing $J=0$ fixed pole in the
real part of $T_2$.

\subsection{Burkhardt-Cottingham sum rule}

The third example, and the first in connection with spin, is 
the Burkhardt-Cottingham sum rule for the first moment of $g_2$
\cite{Burkhardt:1970}:
\begin{equation}
\int_{Q^2/2M}^{\infty} d \nu \ G_2 (Q^2, \nu)
= {2 M^3 \over Q^2} \int_0^1 dx g_2
= 0
.
\nonumber
\label{eqe98}
\end{equation}
Suppose that future experiments find that the sum rule is violated and
that the integral is finite.
The conclusion \cite{Jaffe:1990b}
would be a $J=0$ fixed pole with nonpolynomial residue in the real part
of $A_2$.
To see this
work at fixed $Q^2$ and assume that all Regge-like singularities
contributing to $A_2 (\nu, Q^2)$ have intercept less than zero so that
\begin{equation}
A_2 (\nu, Q^2) \sim \nu^{-1 - \epsilon}
\label{eqe99}
\end{equation}
as $\nu \rightarrow \infty$ for some $\epsilon > 0$.
Then the large $\nu$ behaviour of $A_2$ is obtained
by taking
$\nu \rightarrow \infty$ under the $\nu'$ integral giving
\begin{equation}
A_2(Q^2, \nu) \sim - {2 \over \pi \nu}
\int_{Q^2/2M}^{\infty} d \nu' \ {\rm Im} A_2 (Q^2, \nu')
\label{eqe100}
\end{equation}
which contradicts the assumed behaviour unless the integral vanishes;
hence the sum rule.
{\it If} there is an $\alpha(0)=0$ fixed pole in
the real part of $A_2$
the fixed pole will not
contribute to ${\rm Im} A_2$
and therefore not spoil the convergence of the integral.

One finds
\begin{equation}
\beta_2(Q^2) \sim - {2 \over \pi M}
\int_{Q^2/2M}^{\infty} d \nu' \ {\rm Im} A_2 (Q^2, \nu')
\label{eqe101}
\end{equation}
for the residue of any $J=0$ fixed pole coupling to $A_2(Q^2, \nu)$.

\subsection{$g_1$ spin sum rules}

Scaling requires that any fixed pole correction to the Ellis-Jaffe $g_1$
sum rule must have nonpolynomial residue.
Through Eq.(\ref{eqc41}), the fixed pole coefficient $\beta_1 (Q^2)$
must decay as or faster than $O(1/Q^2)$ as $Q^2 \rightarrow \infty$.
The coefficient is further constrained
by the requirement that $G_1$ contains no kinematic singularities
(for example at $Q^2=0$).
In Section VI.C 
we will identify a potential leading-twist topological
$x=0$ contribution to the first moment of $g_1$ through analysis of
the axial
anomaly contribution to $g_A^{(0)}$.
This zero-mode topological contribution (if finite) generates a
leading twist fixed pole correction to the flavour-singlet part of
$\int_0^1 dx g_1$.
{\it If} present, this fixed pole will also violate
the Gerasimov-Drell-Hearn sum rule
(since the two sum rules are derived from $A_1$)
{\it unless} 
the underlying dynamics suppress the fixed pole's residue at $Q^2=0$.
The possibility of a fixed pole correction to $g_1$ 
spin sum-rules
was raised in pre-QCD work as early as
\textcite{Abarbanel:1968} and \textcite{Heimann:1973}.

Note that any fixed pole correction to the Gerasimov-Drell-Hearn sum rule
is most probably a non-perturbative effect.
The sum rule (\ref{eqc41})
has been verified
to $O(\alpha^2)$ for all
$2 \rightarrow 2$ processes $\gamma a \rightarrow bc$
where $a$ is either a real lepton, quark, gluon
or elementary Higgs target \cite{Altarelli:1972,Brodsky:1995b},
and for electrons in QED to $O(\alpha^3)$ \cite{Dicus:2001}.

One could test for a fixed pole correction to the Ellis-Jaffe moment
through a precision measurement of the flavour singlet axial charge
from an independent process where one is not sensitive to theoretical
assumptions about the presence or absence of a $J=1$ fixed pole in $A_1$.
Here the natural choice is elastic neutrino proton scattering
where the parity violating part of the cross-section includes a 
direct weak interaction measurement of the scale invariant 
flavour-singlet 
axial charge $g_A^{(0)}|_{\rm inv}$.

A further test could come from a precision measurement of the $Q^2$
dependence of the polarized gluon distribution at next-to-next-to-leading 
order accuracy where one becomes sensitive to any possible leading-twist
subtraction constant -- see below Eq.(\ref{eqf125}).

The subtraction constant fixed pole correction hypothesis could
also, in principle, be tested through measurement of the real
part of the spin dependent part of the forward deeply virtual
Compton amplitude.
While this measurement may seem extremely difficult at
the present time one should not forget that Bjorken believed
when writing his original
Bjorken sum rule paper that the sum rule would never be tested!

\section{THE AXIAL ANOMALY, GLUON TOPOLOGY 
AND THE FLAVOUR SINGLET AXIAL CHARGE $g_A^{(0)}$}

We next discuss the role of the axial anomaly in the interpretation of
$g_A^{(0)}$.

\subsection{The axial anomaly}

In QCD one has to consider the effects of renormalization.
The flavour singlet axial vector current $J_{\mu 5}^{GI}$
in Eq.(\ref{eqc53})
satisfies the anomalous divergence equation 
\cite{Adler:1969,Bell:1969,Crewther:1978}
\begin{equation}
\partial^\mu J^{GI}_{\mu5}
= 2f\partial^\mu K_\mu + \sum_{i=1}^{f} 2im_i \bar{q}_i\gamma_5 q_i
\label{eqf102}
\end{equation}
where
\begin{equation}
K_{\mu} = {g^2 \over 32 \pi^2}
\epsilon_{\mu \nu \rho \sigma}
\biggl[ A^{\nu}_a \biggl( \partial^{\rho} A^{\sigma}_a
- {1 \over 3} g
f_{abc} A^{\rho}_b A^{\sigma}_c \biggr) \biggr]
\label{eqf103}
\end{equation}
is the gluonic Chern-Simons current and the number of light flavours 
$f$ is $3$.
Here $A^{\mu}_a$ is the gluon field and
$
\partial^{\mu} K_{\mu}
= {g^2 \over 32 \pi^2} G_{\mu \nu} {\tilde G}^{\mu \nu}
$
is the topological charge density.
Eq.(\ref{eqf102}) allows us to define a partially conserved current
\begin{equation}
J_{\mu 5}^{GI} = J_{\mu 5}^{\rm con} + 2f K_{\mu}
\label{eqf104}
\end{equation}
viz.
$
\partial^\mu J^{\rm con}_{\mu5}
= \sum_{i=1}^{f} 2im_i \bar{q}_i\gamma_5 q_i
$.

When we make a gauge transformation $U$
the gluon field transforms as
\begin{equation}
A_{\mu} \rightarrow U A_{\mu} U^{-1} + {i \over g} (\partial_{\mu} U) U^{-1}
\label{eqf105}
\end{equation}
and the operator $K_{\mu}$
transforms as
\begin{eqnarray}
K_{\mu} \rightarrow K_{\mu}
&+&
 i {g \over 8 \pi^2} \epsilon_{\mu \nu \alpha \beta}
\partial^{\nu}
\biggl( U^{\dagger} \partial^{\alpha} U A^{\beta} \biggr)
\nonumber \\
&+& {1 \over 24 \pi^2} \epsilon_{\mu \nu \alpha \beta}
\biggl[
(U^{\dagger} \partial^{\nu} U)
(U^{\dagger} \partial^{\alpha} U)
(U^{\dagger} \partial^{\beta} U)
\biggr]
.
\nonumber \\
\label{eqf106}
\end{eqnarray}
(Partially) conserved currents are not renormalized.
It follows that
$J_{\mu 5}^{\rm con}$
is renormalization scale invariant and the scale dependence of 
$J_{\mu 5}^{GI}$ associated with the factor $E(\alpha_s)$ 
is carried 
by $K_{\mu}$.
This is summarized in the equations:
\begin{eqnarray}
J_{\mu 5} &=& Z_5 \ J_{\mu 5} \bigg|_{\rm bare}
\nonumber \\
K_{\mu} &=& K_{\mu}|_{\rm bare} \ + \ 
            {1 \over 2f} (Z_5 - 1) J_{\mu 5} \bigg|_{\rm bare}
\nonumber \\
J_{\mu 5}^{\rm con} &=& J_{\mu 5}^{\rm con} \bigg|_{\rm bare}
\label{eqf107}
\end{eqnarray}
where $Z_5$ denotes the renormalization factor for $J_{\mu 5}$.
Gauge transformations shuffle a scale invariant operator 
quantity between the two operators $J_{\mu 5}^{\rm con}$ 
and $K_{\mu}$ whilst keeping $J_{\mu 5}^{GI}$ invariant.

The nucleon matrix element of $J_{\mu 5}^{GI}$ is
\begin{equation}
\langle p,s|J^{GI}_{5 \mu}|p',s'\rangle
= 2M \biggl[ {\tilde s}_\mu G_A (l^2) + l_\mu l.{\tilde s} G_P (l^2) \biggr]
\label{eqf108}
\end{equation}
where $l_{\mu} = (p'-p)_{\mu}$
and
${\tilde s}_{\mu}
= {\overline u}_{(p,s)} \gamma_{\mu} \gamma_5 u_{(p',s')} / 2M $.
Since $J^{GI}_{5 \mu}$ does not couple to a massless
Goldstone
boson it follows that $G_A(l^2)$ and $G_P(l^2)$ contain
no massless pole terms.
The forward matrix element of $J^{GI}_{5 \mu}$ is well
defined and
\begin{equation}
g_A^{(0)}|_{\rm inv} = E(\alpha_s) G_A (0).
\label{eqf109}
\end{equation}

We would like to isolate the gluonic contribution to $G_A (0)$
associated with $K_{\mu}$ and thus write $g_A^{(0)}$
as the sum of
(measurable)
``quark'' and ``gluonic'' contributions.
Here one has to be careful because of the gauge dependence of
the operator $K_{\mu}$.
To understand the gluonic contributions to $g_A^{(0)}$ it is
helpful to go back to the deep inelastic cross-section in Section II.

\subsection{The anomaly and the first moment of $g_1$}

We specialise to the target rest frame and let $E$ denote the
energy of the incident charged lepton
which is scattered through an angle $\theta$
to emerge in the final state with energy $E'$.
Let $\uparrow \downarrow$ denote the longitudinal polarization of
the beam
and $\Uparrow \Downarrow$ denote a longitudinally polarized proton
target.
The spin dependent part of the differential cross-sections is
\begin{eqnarray}
& &
\Biggl(
{d^2 \sigma \uparrow \Downarrow \over d\Omega dE^{'} } 
-
{d^2 \sigma \uparrow \Uparrow \over d\Omega dE^{'} }
\Biggr)
\nonumber \\
& & =
{4 \alpha^2 E^{'} \over Q^2 E \nu }
\biggl[ (E+E^{'} \cos \theta ) \ g_1 (x, Q^2) - {2 x M}
\ g_2 (x, Q^2) \biggr]
\nonumber \\
\label{eqf110}
\end{eqnarray}
which is obtained from the product of the lepton and hadron tensors
\begin{equation}
{d^2 \sigma \over d\Omega dE'}
= {\alpha^2 \over Q^4} {E' \over E} \ L_{\mu \nu}^A \ W^{\mu \nu}_A
.
\label{eqf111}
\end{equation}
Here the lepton tensor
\begin{equation}
L_{\mu \nu}^A = 2 i \epsilon_{\mu \nu \alpha \beta} k^{\alpha} q^{\beta}
\label{eqf112}
\end{equation}
describes the lepton-photon vertex and
the hadronic tensor
\begin{eqnarray}
{1 \over M} W^{\mu \nu}_A 
&=&
i \epsilon^{\mu \nu \rho \sigma} q_{\rho}
\biggl(
s_{\sigma} {1 \over p.q} g_1 (x,Q^2)
\nonumber \\
& & 
\ \ \ \ \ 
+ [ p.q s_{\sigma} - s.q p_{\sigma} ] {1 \over M^2 p.q} g_2 (x,Q^2) \biggr)
\nonumber \\
\label{eqf113}
\end{eqnarray}
describes the photon-nucleon interaction.

Deep inelastic scattering involves the Bjorken limit:
$Q^2 = - q^2$ and $p.q = M\nu$ both $\rightarrow \infty$
with
$x = {Q^2 \over 2 M \nu}$ held fixed.
In terms of light-cone coordinates this corresponds to taking
$q_- \rightarrow \infty$
with
$q_+ = -x p_+$ held finite.
The leading term in $W_A^{\mu \nu}$
is obtained by taking the Lorentz index of $s_{\sigma}$
as $\sigma = +$.
(Other terms are suppressed by powers of ${1 \over q_-}$.)

If we wish to understand the first moment of $g_1$ in terms
of the
matrix elements of anomalous currents
($J_{\mu 5}^{\rm con}$ and $K_{\mu}$),
then we have to understand
the forward matrix element of $K_+$ and its contribution to
$G_A(0)$.

Here we are fortunate in that the parton model is formulated in the
light-cone gauge ($A_+=0$) where the forward matrix elements of $K_+$
are invariant.
In the light-cone gauge the non-abelian three-gluon part of $K_+$
vanishes. The forward matrix elements of $K_+$ are then invariant
under all residual gauge degrees of freedom.
Furthermore,
in this gauge, $K_+$ measures the gluonic ``spin'' content of the
polarized target \cite{Jaffe:1996,Manohar:1990}
 -- strictly speaking, 
 up to the non-perturbative surface term
          we find from integrating the light-cone correlation function,
 Eq.(\ref{eqd90}).
One finds
\begin{equation}
G_A^{(\rm A_+ = 0)}(0) = \sum_q \Delta q_{\rm con}
- f {\alpha_s \over 2 \pi} \Delta g
\label{eqf114}
\end{equation}
where
$\Delta q_{\rm con}$ is measured by the partially conserved 
current
$J_{+5}^{\rm con}$
and
$- {\alpha_s \over 2 \pi} \Delta g$ is measured by $K_+$.
Positive gluon polarization tends to reduce the value of
$g_A^{(0)}$
and offers a
possible source for OZI violation in $g_A^{(0)}|_{\rm inv}$.
The connection between this more formal derivation and the QCD
parton model will 
be explored in Section VI.D below.
In perturbative QCD $\Delta q_{\rm con}$ is identified
with
$\Delta q_{\rm partons}$ and
$\Delta g$ is identified
with $\Delta g_{\rm partons}$
-- see Section VI.D below and 
\textcite{Carlitz:1988,Efremov:1988,Altarelli:1988} and
\textcite{Bass:1991}.

\subsection{Gluon topology, large gauge transformations and 
            connection to the axial U(1) problem}

If we were to work only in the light-cone gauge we might think
that we have a complete parton model description of the first
moment of $g_1$.
However, one is free to work in any gauge including a covariant
gauge where the forward matrix elements of $K_+$ 
are not necessarily invariant under the residual gauge degrees
of freedom \cite{Jaffe:1990a}.
Understanding the interplay between spin and 
gauge invariance leads to rich and interesting physics possibilities.

We illustrate this by an example in covariant gauge.

The matrix elements of $K_{\mu}$ need to be specified with
respect to a specific gauge.
In a covariant gauge we can write
\begin{equation}
\langle p,s|K_\mu |p',s'\rangle
= 2M \biggl[ {\tilde s}_\mu K_A(l^2) + l_\mu l.{\tilde s} K_P(l^2) \biggr]
\label{eqf115}
\end{equation}
where $K_P$ contains a massless Kogut-Susskind pole \cite{Kogut:1974}.
This massless pole is an essential ingredient in the solution of
the axial U(1) problem \cite{Crewther:1978}
(the absence of any near massless Goldstone boson in the singlet
 channel associated with spontaneous axial U(1) symmetry breaking)
and
cancels with a corresponding massless pole term in $(G_P - K_P)$.
The Kogut Susskind pole is associated with the (unphysical) 
massless boson that one expects to couple to $J_{\mu 5}^{\rm con}$
in the chiral limit and which is not seen in the physical spectrum.

We next define gauge-invariant form-factors $\chi^{g}(l^2)$ for the 
topological charge density 
and $\chi^q(l^2)$ for the quark chiralities in the divergence of
$J_{\mu 5}$:
\begin{eqnarray}
2M \ l.{\tilde s} \ \chi^g(l^2) &=&
\langle p,s | {g^2 \over 32 \pi^2} G_{\mu \nu} {\tilde G}^{\mu \nu}
 | p', s' \rangle
\nonumber \\
2M \ l.{\tilde s} \ \chi^q(l^2) &=&
\langle p,s | 
\sum_{i=1}^{f} 2im_i \bar{q}_i\gamma_5 q_i
 | p', s' \rangle
.
\label{eqf116}
\end{eqnarray}
Working in a covariant gauge, we find
\begin{equation}
\chi^{g}(l^2) = K_A(l^2) + l^2 K_P(l^2)
\label{eqf117}
\end{equation}
by contracting Eq.(\ref{eqf116}) with $l^{\mu}$.
(Also, note
 the general gauge invariant
 formula
 $g_A^{(0)} = \chi^q(0) + f \chi^g(0)$.)

When we make a gauge transformation any change
$\delta_{\rm gt}$
in $K_A(0)$ is compensated
by a corresponding change in the residue of the Kogut-Susskind
pole in $K_P$, viz.
\begin{equation}
\delta_{\rm gt} [ K_A(0) ]
+ \lim_{l^2 \rightarrow 0} \delta_{\rm gt} [ l^2 K_P(l^2) ] = 0
.
\label{eqf118}
\end{equation}
As emphasised above, 
the Kogut-Susskind pole corresponds to the Goldstone boson associated 
with spontaneously broken $U_A(1)$ symmetry
\cite{Crewther:1978}.
There is no Kogut-Susskind pole in perturbative QCD.
It follows that the quantity which is shuffled
between the $J_{+5}^{\rm con}$ and $K_+$
contributions to $g_A^{(0)}$ is strictly non-perturbative;
it vanishes in perturbative QCD and is not present in the QCD
parton model.

The QCD vacuum is understood to be a Bloch superposition of 
states characterised by different topological winding number 
\cite{Callan:1976,Jackiw:1976} 
\begin{equation}
| {\rm vac}, \theta \rangle = \sum_n e^{i n \theta}
 \  | n \rangle 
\label{eqf119}
\end{equation}
where the QCD $\theta$ angle is zero 
(experimentally less than $10^{-10}$)
-- see e.g. \textcite{Quinn:2004}.

One can show \cite{Jaffe:1990a} 
that the forward matrix elements of
$K_{\mu}$ are invariant under ``small'' gauge transformations
(which are topologically deformable to the identity)
but not invariant under ``large'' gauge transformations which
change the topological winding number.
Perturbative QCD involves only ``small'' gauge transformations;
``large'' gauge transformations involve strictly non-perturbative physics.
The second term on the right hand side of Eq.(\ref{eqf106}) 
is a total derivative;
its matrix elements vanish in the forward direction.
The third term on the right hand side of 
Eq.(\ref{eqf106}) is associated with the gluon topology \cite{Cronstrom:1983}.

The topological winding number is determined by the gluonic 
boundary conditions at ``infinity''
(
 a large surface with boundary which is spacelike with respect
 to the positions $z_k$ of any operators or fields in the physical
 problem)
\cite{Crewther:1978}.
It is insensitive to local deformations of the gluon
field $A_{\mu}(z)$ or of the gauge transformation $U(z)$.
When we take the Fourier transform to momentum space
the topological structure induces a light-cone zero-mode which
can contribute to $g_1$ only at $x=0$.
Hence, we are led
to consider the possibility that there may be a
term in $g_1$ which is proportional to $\delta(x)$ \cite{Bass:1998a}.

It remains an open question whether the net non-perturbative
quantity which
is shuffled between $K_A(0)$ and $(G_A - K_A)(0)$ under ``large''
gauge transformations
is finite or not.
If it is finite and, therefore, physical, then, when we choose
$A_+ =0$,
this non-perturbative quantity must be contained in
some combination of the $\Delta q_{\rm con}$ and $\Delta g$ 
in Eq.(\ref{eqf114}).

Previously, in Sections III and V, 
we found that a $J=1$ fixed pole in the real
part of $A_1$ in the forward Compton amplitude
could also induce a ``$\delta (x)$ correction''
to the sum rule for the first moment of $g_1$
through a subtraction at infinity in the dispersion relation
(\ref{eqc40}).
Both the topological $x=0$ term and the subtraction constant
${Q^ 2 \over 2M^2} \beta_1 (Q^2)$
(if finite)
give real coefficients of
${1 \over x}$ terms in Eq.(\ref{eqc41}).
It seems reasonable therefore to conjecture that the physics of
gluon topology
may induce a $J=1$ fixed pole correction to the Ellis-Jaffe sum rule.
Whether this correction is finite or not is an issue for future experiments.

Instantons provide an example how to generate topological $x=0$
polarization \cite{Bass:1998a}.
Quarks instanton interactions flip chirality, thus connecting
left and right handed quarks.
Whether instantons spontaneously or explicitly break axial U(1)
symmetry depends on the role of zero modes in the quark instanton
interaction and how one should include non local structure in the
local anomalous Ward identity.
Topological $x=0$ polarization is natural in theories of spontaneous
axial U(1) symmetry breaking by instantons \cite{Crewther:1978}
where any instanton induced suppression of $g_A^{(0)}|_{\rm pDIS}$
is compensated by a shift of flavour-singlet
axial charge from quarks carrying finite momentum to a zero mode 
($x=0$).
It is not generated by mechanisms \cite{tHooft:1986} of explicit U(1)
symmetry breaking by instantons.
Experimental evidence for or against a ``subtraction at infinity''
correction to the Ellis-Jaffe sum rule would provide valuable
information about gluon topology and vital clues to the nature of 
dynamical axial U(1) symmetry breaking in QCD.

\begin{widetext}

\subsection{Photon gluon fusion}

We next consider the role of the axial anomaly in the QCD parton model 
and its relation to semi-inclusive measurements of jets and high $k_t$
hadrons in polarized deep inelastic scattering.

Consider the polarized photon-gluon fusion process
$\gamma^* g \rightarrow q {\bar q}$.
We evaluate the $g_1$ spin structure function for this process as a
function of the transverse momentum squared of the struck quark, $k_t^2$,
with respect to the photon-gluon direction.
We use $q$ and $p$ to denote the photon and gluon momenta and
use the cut-off $k_t^2 \geq \lambda^2$ to separate the total
phase space into ``hard'' ($k_t^2 \geq \lambda^2$) and ``soft''
($k_t^2 < \lambda^2$) contributions.
One finds \cite{Bass:1998b}:
\begin{eqnarray}
g_1^{(\gamma^* g)}
|_{\rm hard}
&=&
-{\alpha_s \over 2 \pi } {\cut \over \cxx} \Biggl[ (2x-1)(\cx)
\nonumber \\
& &
\ \ \ \ \ \ \ \ 
\biggl\{
1 - {1 \over {\cut \sqrt{\cxx} }}
\ln \biggl({ {1+\sqrt{\cxx} \cut}\over {1-\sqrt{\cxx} \cut}}
\biggr) \biggr\}
\nonumber \\
& &
\ \ \ \ \ \ \ \ \ \ \ \ \ \ \ \ \ \ \ \
+ (x-1+{{x P^{2}}\over{Q^{2}}})
{{\left( 2m^{2}(\cxx)- P^{2}x(2x-1)(\cx)\right)}
\over {(m^{2} + \lambda^2) (\cxx) - P^{2}x(x-1+{{x P^{2}}\over{Q^{2}}})}}
\Biggr]
\nonumber \\
\label{eqf120}
\end{eqnarray}
for each flavour of quark liberated into the final state.
Here $m$ is the quark mass, $Q^2 =-q^2$ is the virtuality of the hard photon,
$P^2=-p^2$ is the virtuality of the gluon target, $x$ is the Bjorken variable 
($x= {Q^2 \over 2 p.q}$) and $s$ is the centre of mass energy squared,
$s= (p+q)^2 = Q^2 \bigl( {1 - x \over x} \bigr) - P^2$, for the photon-gluon 
collision.

When $Q^2 \rightarrow \infty$
the expression for $g_1^{(\gamma^* g)}|_{\rm hard}$
simplifies to the leading twist (=2) contribution:
\begin{equation}
g_1^{(\gamma^* g)}|_{\rm hard}
= {\alpha_s \over 2 \pi} \Biggl[ (2x-1) \Biggl\{
\ln {1-x \over x} - 1
+ \ln {Q^2 \over {x(1-x) P^2 + (m^2 + \lambda^2)} } \Biggr\}
+ (1 -x) { {2m^2 - P^2x(2x-1)} \over { m^2 + \lambda^2 - P^2 x(x-1)} }
\Biggr] .
\label{eqf121}
\end{equation}
Here we take $\lambda$ to be independent of $x$.
Note that for finite quark masses,
phase space limits Bjorken $x$ to
$x_{max} = Q^2 / (Q^2 + P^2 + 4 (m^2 + \lambda^2))$
and protects
$g_1^{(\gamma^* g)}|_{\rm hard}$
from reaching the $\ln (1-x)$ singularity in Eq. (\ref{eqf121}).
For this photon-gluon fusion process,
the first moment of the ``hard'' contribution is:
\begin{equation}
\int_0^1 dx g_1^{(\gamma^{*} g)}|_{\rm hard}
= - {\alpha_s \over 2 \pi}
\left[1 + \frac{2m^2}{P^2}
\frac{1}{\sqrt{ 1 + {4 (m^2 + \lambda^2) \over P^2} } }
\ln \left(
\frac{\sqrt{1 + {4 (m^2+\lambda^2) \over P^2} -1} }
{\sqrt{1 + {4 (m^2+\lambda^2) \over P^2} +1} } \right) \right]
.
\label{eqf122}
\end{equation}
The ``soft'' contribution to the first moment of $g_1$ is then
obtained by subtracting Eq. (\ref{eqf122})
from the inclusive first moment (obtained by setting $\lambda =0$).

\end{widetext}

For fixed gluon virtuality $P^2$ the photon-gluon fusion process
induces two distinct contributions to the first moment of $g_1$.
Consider the leading twist contribution, Eq. (\ref{eqf122}).
The first term, $-{\alpha_s \over 2 \pi}$, in Eq.(\ref{eqf122})
is mass-independent and comes from the region of phase space
where the struck quark carries large transverse momentum squared
$k_t^2 \sim Q^2$.
It measures a contact photon-gluon interaction and is associated
\cite{Carlitz:1988,Bass:1991}
with the
axial anomaly though the $K_+$ Chern-Simons current contribution
to $J_{\mu 5}^{GI}$.
The second mass-dependent term comes from the region of phase-space
where the struck quark carries transverse momentum $k_t^2 \sim m^2,P^2$.
This
positive mass dependent term is proportional to the mass
squared of the struck quark.
The mass-dependent in Eq.~(\ref{eqf122})
can safely be neglected for light-quark flavor (up and down) production.
It is very important for strangeness and charm production
\cite{Bass:1999b}.
For vanishing cut-off ($\lambda^2=0$) this term vanishes in the limit
$m^2 \ll P^2$ and tends to $+{\alpha_s \over 2 \pi}$ when $m^2 \gg P^2$
(so that the first moment of $g_1^{(\gamma^* g)}$
 vanishes in this limit).
The vanishing of
 $\int_0^1 dx g_1^{(\gamma^* g)}$ in the limit $m^2 \ll P^2$
 to leading order in $\alpha_s (Q^2)$
 follows from an application \cite{Bass:1998b} 
of the fundamental GDH sum-rule.

One can also analyse the photon-gluon fusion process using $x$ dependent
cut-offs.
Examples include
the virtuality of the struck quark 
\begin{equation}
m^2 - k^2 
= P^2 x + {k_t^2 + m^2 \over (1-x)} > \lambda_0^2 
= {\rm constant}(x)
\end{equation}
or the invariant 
mass squared of the
quark-antiquark pair produced in the photon-gluon collision
\begin{equation}
{\cal M}^2_{q {\overline q}} 
= {k_t^2 + m^2 \over x (1-x)} + P^2 \geq \lambda_0^2 
= {\rm constant}(x) 
.
\end{equation}
These different choices of infrared cut-offs correspond to different
jet definitions and different factorization schemes for photon-gluon 
fusion in the QCD parton model
-- see \textcite{Bass:1991,Bass:1998b,Mankiewicz:1991} and
\textcite{Manohar:1991}.
If we evaluate the first moment of $g_1^{(\gamma^{*} g)}$ 
using the cut-off on the quarks' virtuality,
then we find
``half of the anomaly'' 
in the gluon coefficient through the mixing 
of transverse and longitudinal momentum components.
The anomaly coefficient for the first moment is recovered with the
invariant mass squared cut-off through a sensitive cancellation of
large and
small $x$ contributions \cite{Bass:1991}.

We noted above that when one applies the operator product expansion 
the first term in Eq.(\ref{eqf122}) corresponds to the gluon matrix 
element of the anomalous gluonic current $K_{+}$.
This operator product expansion analysis can be generalized 
to the 
higher moments of 
$g_1^{(\gamma^{*}g)}$.
The anomalous contribution to the higher moments is controlled
by choosing the correct prescription for $\gamma_5$. 
One finds \cite{Bass:1992a,Cheng:1996}
that the axial anomaly contribution to the {\it shape} of $g_1$ 
at finite $x$ is given by the convolution of the polarized gluon 
distribution $\Delta g(x,Q^2)$ with the hard coefficient
\begin{equation}
{\tilde C}^{(g)}|_{\rm anom} = - {\alpha_s \over \pi} (1-x) .
\end{equation}
This anomaly contribution is a small $x$ effect in $g_1$; 
it is
essentially negligible for $x$ less than 0.05.
The hard coefficient ${\tilde C}^{(g)}|_{\rm anom}$ 
is normally included as a term in the gluonic Wilson coefficent $C^g$ 
-- see Section IX.C below.
It is associated with two-quark jet events carrying $k_t^2 \sim Q^2$
in the final state.

Eq. (\ref{eqf122}) leads to the well known formula quoted in Section I
\begin{equation}
g_A^{(0)} =
\Biggl( \sum_q \Delta q
  - 3 {\alpha_s \over 2 \pi} \Delta g \Biggr)_{\rm partons}
\ + \ {\cal C}_{\infty}
.
\label{eqf123}
\end{equation}
Here $\Delta g$ is the amount of spin carried by polarized gluon
partons in the polarized proton and
$\Delta q_{\rm partons}$ measures
the spin carried by quarks and antiquarks
carrying ``soft'' transverse momentum $k_t^2 \sim m^2, P^2$.
Note that the mass independent contact interaction in Eq.(\ref{eqf122})
is flavour independent.
The mass dependent term associated with low $k_t$ breaks flavour SU(3)
in the perturbative sea.
The third term 
${\cal C}_{\infty} = {1 \over 2} 
\lim_{Q^2 \rightarrow \infty} {Q^2 \over 2 M^2} \beta_1(Q^2)$
describes any fixed pole ``subtraction at infinity'' correction 
to $g_A^{(0)}$.

Equations (\ref{eqf107}) yield the renormalization group equation
\begin{equation}
\biggl\{
{\alpha_s \over 2 \pi} \Delta g \biggr\}_{Q^2}
= 
\biggl\{ 
{\alpha_s \over 2 \pi} \Delta g \biggr\}_{\infty}
+ \ {1 \over 3} \ 
\biggl\{
1/E(\alpha_s) -1 \biggr\} \ g_A^{(0)} \bigg|_{\rm inv}
.
\label{eqf124}
\end{equation}
It follows that the polarized gluon term satisfies
\begin{equation}
\alpha_s \Delta g \sim {\tt constant}, \ \ \ Q^2 \rightarrow \infty
.
\label{eqf125}
\end{equation}
This key result, first noted in the context of the QCD parton model
by Altarelli and Ross \cite{Altarelli:1988} and Efremov and Teryaev 
\cite{Efremov:1988}
means that the polarized gluon contribution makes a scaling 
contribution to the first moment of $g_1$ at next-to-leading order.
(In higher orders the $Q^2$ evolution of 
 $\Delta g$ depends on the value of $g_A^{(0)}|_{\rm inv}$
 suggesting one, in principle, method to search for any finite 
 ${\cal C}_{\infty}$.)

\begin{widetext}

\begin{figure}[h!]
\includegraphics{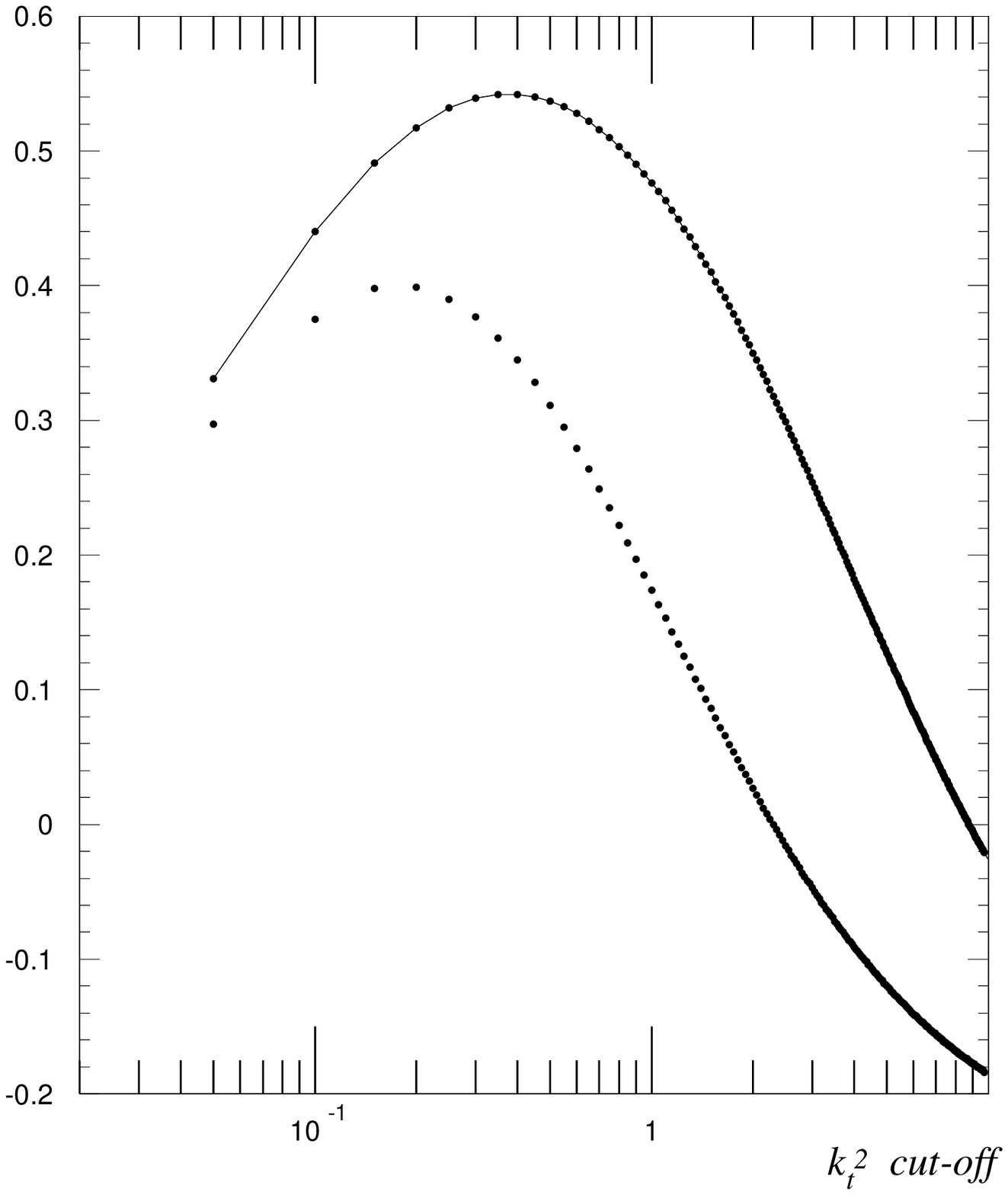} 
\includegraphics{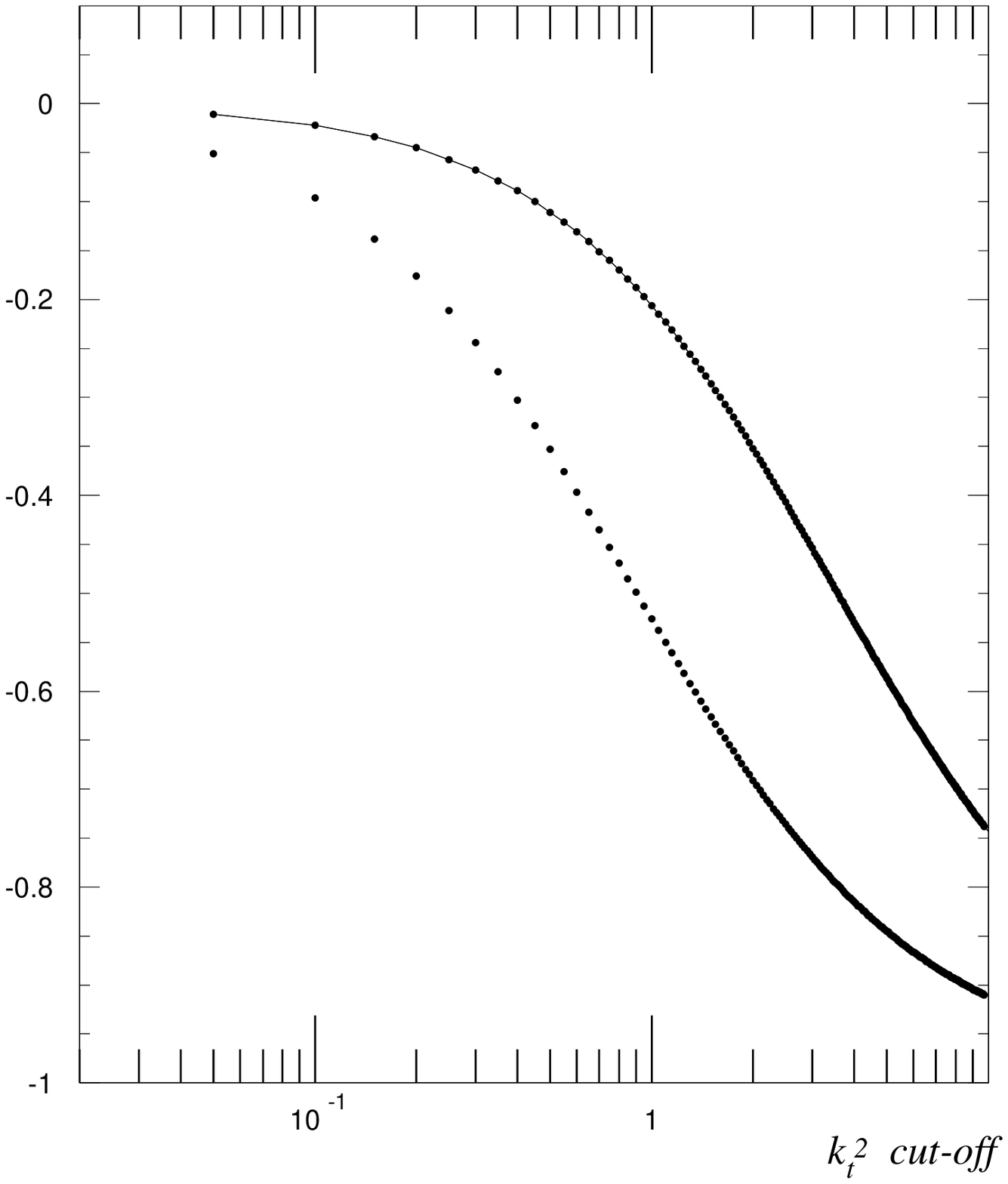} 
\begin{center}
\vspace{7.5cm}
\parbox{16.5cm}
{\caption[Delta]
{
$\int_0^1 dx \ g_1^{(\gamma^* g)}|_{\rm soft}$
for polarized strangeness production (left)
and light-flavor ($u$ or $d$) production (right)
with $k_t^2 < \lambda^2$
in units of ${\alpha_s \over 2 \pi}$ \cite{Bass:2003a}.
Here
$Q^2=2.5$GeV$^2$ (dotted line) and 10GeV$^2$ (solid line).

}
\label{fig:fig9}}
\end{center}
\end{figure}
\end{widetext}

The transverse momentum dependence of the gluonic and sea quark
partonic contributions to $g_A^{(0)}$
suggests the interpretation of measurements of quark sea polarization
will depend on the large $k_t$ acceptance of the apparatus.
Let
$g_1^{(\gamma^* g)}|_{\rm soft} (\lambda)$ denote the contribution
to
$g_1^{(\gamma^* g)}$ for photon-gluon fusion where the hard photon
scatters on the struck quark or antiquark carrying
transverse momentum $k_t^2 < \lambda^2$.
Fig. \ref{fig:fig9} shows the first moment of 
$g_1^{(\gamma^* g)}|_{\rm soft}$
for the strange and light (up and down) flavour production
respectively as a function of the transverse momentum cut-off 
$\lambda^2$.
Here we set $Q^2 =2.5$GeV$^2$
(corresponding to the HERMES experiment) and 10GeV$^2$ (SMC).
Following \cite{Carlitz:1988},
we take $P^2 \sim \Lambda_{\rm qcd}^2$ and set $P^2 = 0.1$GeV$^2$.
Observe the small value for the light-quark
sea polarization at low transverse momentum and
the positive value for the integrated
strange sea polarization at low $k_t^2$:
$k_t < 1.5$GeV at the HERMES $Q^2=2.5$GeV$^2$.

\subsection{Choice of currents and ``spin''}

The axial anomaly presents us with three candidate currents we might 
try to use to define the quark spin content:
$J_{\mu 5}$, the renormalization scale invariant 
$E(\alpha_s) J_{\mu 5}$
and 
$J_{\mu 5}^{\rm con}$.
One might also consider using the chiralities $\chi^{(q)}$ 
from Eq.(\ref{eqf116}).
We next 
explain how each current yields gauge invariant possible definitions.

First, note that if we try to define intrinsic spin operators
\begin{equation}
S_{k} = \int d^{3}x \ ({\overline q} \gamma_{k} \gamma_{5} q) 
\ \ \ \ \  k=1,2,3
\label{eqf126}
\end{equation}
using the axial vector current operators, then we find that
the operators constructed using the gauge invariantly renormalized
current $J_{\mu 5}$ cannot satisfy the (spin) commutation relations
of SU(2)
$
[ S_{i}, S_{j} ](\mu^{2}) = i \epsilon_{ijk}
S_{k} (\mu^{2})
$
at more than one scale $\mu^2$ because of the anomalous dimension
and the renormalization group factor associated with $E(\alpha_s)$
and the axial anomaly \cite{Bass:1993b}.
The most natural scale to normalize the axial vector
current operators 
to satisfy SU(2) is, perhaps, $\mu \rightarrow \infty$
-- that is, using the scale invariant current $\{ E(\alpha_s) J_{\mu 5} \}$.
Then we find
$
[ S_{i}, S_{j} ](\mu^{2}) = i \epsilon_{ijk}
E(\alpha_s) S_{k} (\mu^{2})
$
if we use the gauge invariant current renormalized at another scale.
One might argue that gluon spin is renormalization scale dependent,
Eq.(\ref{eqf124}), 
so not worry too much about this issue but there are further points 
to consider.

Next choose the $A_0=0$ gauge and define two operator charges:
\begin{eqnarray}
X(t) &=& \int d^3z \ J_{0 5} (z)
\nonumber \\
Q_5  &=& \int d^3z \ J_{0 5}^{\rm con}(z)
.
\nonumber \\
\label{eqf127}
\end{eqnarray}
Because partially conserved currents are not renormalized it follows 
that $Q_5$ 
is a time independent operator.
The charge $X(t)$ is manifestly gauge invariant whereas $Q_5$ is invariant 
only under ``small'' gauge transformations; the charge $Q_5$ transforms as
\begin{equation}
Q_5 \rightarrow Q_5 - 2 f \ n
\label{eqf128}
\end{equation}
where $n$ is the winding number associated with the gauge transformation $U$.
Although $Q_5$ is gauge dependent we can define a gauge invariant chirality 
$q_5$
for a given operator ${\cal O}$ through the gauge-invariant 
eigenvalues of the equal-time commutator
\begin{equation}
[ \ Q_5 \ , \ {\cal O} \ ]_{-} = - 
q_5 \ {\cal O} 
.
\label{eqf129}
\end{equation}
The gauge invariance of 
$q_5$
follows since this commutator
 appears in gauge invariant Ward Identities \cite{Crewther:1978}
 despite the gauge dependence of $Q_5$.
The time derivative of spatial components of the gluon field 
have zero chiralty 
$q_5$
\begin{equation}
[ \ Q_5 \ , \ \partial_0 A_i \ ]_- \ = \ 0
\label{eqf130}
\end{equation}
but non-zero $X$ charge
\begin{equation}
\lim_{t' \rightarrow t}
\biggl[ \ X(t') \ , \ \partial_0 A_i ({\vec x}, t) \ \biggr]_- 
\ = \ 
{i f g^2 \over 4 \pi^2} {\tilde G}_{0i} \ + \ O(g^4 \ln | t'-t| )
.
\label{eqf131}
\end{equation}
The analogous situation in QED is discussed in
\textcite{Adler:1969a,Jackiw:1969} and \textcite{Adler:1970}. 
Eq.(\ref{eqf130}) follows from the non-renormalization of the conserved
current $J_{\mu 5}^{\rm con}$. 
Eq.(\ref{eqf131}) follows from the implicit 
$A_{\mu}$ dependence of the (anomalous) gauge invariant current $J_{\mu 5}$.
The higher-order terms $g^4 \ln | t'-t|$ are caused by wavefunction
renormalization of $J_{\mu 5}$ \cite{Crewther:1978}.

This formalism generalizes readily to the definition of baryon number
in the presence of electroweak gauge fields.
The vector baryon number current is sensitive to the axial anomaly
through the parity violating electroweak interactions.
If one requires that baryon number is renormalization group invariant
and that the time derivative of the spatial components of the W boson
field have zero baryon number, then one is led to using the conserved
vector current analogy of 
$q_5$
to define the baryon number.
Sphaleron induced electroweak baryogenesis in the early Universe 
\cite{Rubakov:1996,Kuzmin:1985}
is then accompanied by the formation of a ``topological condensate''
\cite{Bass:2004}
which (probably) survives in the Universe we live in today.

Lastly, we comment on the use of the chiralities $\chi^q$ and the 
quantity $\chi^g$ to define the ``quark spin'' and ``gluon spin'' 
content of the proton.
This suggestion starts from the decomposition
\begin{equation}
g_A^{(0)} = \chi^q (0) + 3 \chi^g (0)
\label{eqf132}
\end{equation}
but is less optimal
because the separate
``quark'' and ``gluonic'' pieces is
very much infra-red sensitive and strongly dependent of the ratios of 
the light quark masses $m_u / m_d$
\cite{Cheng:1989,Veneziano:1989} -- see also \cite{Gross:1979,Ioffe:1979}.
Indeed, for the polarized real photon structure function $g_1^{\gamma}$
the quantity
$\chi^g_{\tt photon} \sim 30$ at realistic deep inelastic values of $Q^2$ 
\cite{Bass:1992} !

\section{CHIRAL SYMMETRY AND THE SPIN STRUCTURE OF THE PROTON}

Goldberger-Treiman relations relate the spin structure of 
the proton to spontaneous chiral symmetry breaking in QCD.

The isovector Goldberger-Treiman relation \cite{Adler:1968}
\begin{equation}
2 M g_A^{(3)} = f_{\pi} g_{\pi NN}
\label{eqg133}
\end{equation}
relates $g_A^{(3)}$ and therefore $(\Delta u - \Delta d)$
to the product of the pion decay constant $f_{\pi}$ and the
pion-nucleon coupling constant $g_{\pi NN}$.
This result is non-trivial.
It means that the spin structure of the nucleon measured in
high-energy, high $Q^2$ polarized deep inelastic scattering
is intimately
related
to spontaneous chiral symmetry breaking and low-energy pion physics.
The Bjorken sum rule can also be written
$\int_0^1 dx (g_1^p - g_1^n) 
 = {1 \over 6}
   \{ f_{\pi} g_{\pi NN} / 2M \}
\Bigl\{1 + \sum_{\ell\geq 1} c_{{\rm NS} \ell\,}
\alpha_s^{\ell}(Q)\Bigr\} 
$
(modulo small chiral corrections $\sim 5\%$ coming the finite light
 quark and pion masses).

The flavour-singlet generalization of the Goldberger-Treiman 
was derived independently by Shore and Veneziano \cite{Shore:1990,Shore:1992}
and Hatsuda \cite{Hatsuda:1990}.

Isoscalar extensions of the Goldberger-Treiman relation are quite 
subtle because of the axial U(1) problem
whereby gluonic degrees of freedom mix with the flavour-singlet 
Goldstone
state to increase the masses of the $\eta$ and $\eta'$ mesons.
The vacuum condensates
$\langle {\rm vac} | {\overline{q} q} | {\rm vac} \rangle$
($q=u,d,s$)
spontaneously breaks both chiral SU(3) and also axial U(1) 
symmetry.
One expects a nonet of would-be Goldstone bosons: 
the physical pions and kaons plus also octet and 
singlet states.
In the singlet channel the axial anomaly and non-perturbative
gluon topology induce a substantial gluonic mass term for the 
singlet boson.

The Witten-Veneziano mass formula \cite{Witten:1979,Veneziano:1979}
relates the gluonic mass term for the singlet boson
to the
topological susceptibility of pure Yang-Mills (glue with no quarks)
\begin{equation}
{\tilde m}_{\eta_0}^2 = - {6 \over f_{\pi}^2} \chi (0)
\end{equation}
where
$
\chi (k^2) = 
\int d^4 z \ i \ e^{ik.z} \
\langle {\rm vac} | \ T \ Q(z) Q(0) \ | {\rm vac} \rangle 
\big|_{\rm YM}
$
and
$Q(z)$
denotes
the
topological charge density.
Without this singlet gluonic mass term the $\eta$ meson would 
be approximately degenerate with the pion and the $\eta'$
meson would have a mass $\sim \sqrt{2 m_K^2 - m_{\pi}^2}$
after we take into account 
mixing between the octet and singlet bosons induced by the 
strange quark mass.

In the chiral limit the flavour-singlet Goldberger-Treiman relation 
reads
\begin{equation}
2M g_A^{(0)} = \sqrt{ \chi' (0)} \ g_{\phi_0 NN} .
\label{eqg135}
\end{equation}
Here $\chi'(0)$ is the first derivative of the topological 
susceptibility
and
$g_{\phi_0 NN}$ denotes the one particle irreducible coupling 
to the nucleon of the flavour-singlet Goldstone boson 
which would exist in a gedanken world where OZI is exact 
in the singlet axial U(1) channel. 
The $\phi_0$
is a theoretical object and not a physical state in the spectrum.
The important features of Eq.(\ref{eqg135}) 
are first that $g_A^{(0)}$ factorises
into the product of the target dependent coupling 
$g_{\phi_0 NN}$
and the target independent gluonic term $\sqrt{\chi'(0)}$.
The coupling
$g_{\phi_0 NN}$ is renormalization scale invariant and the 
scale dependence of 
$g_A^{(0)}$ 
associated 
with the renormalization group factor $E(\alpha_s)$
is carried by the gluonic term $\sqrt{ \chi' (0)}$.
Motivated by this observation,
Narison, Shore and Veneziano \cite{Narison:1995} 
conjectured that any OZI
violation in $g_A^{(0)}|_{\rm inv}$
might be carried 
by the target independent factor $\sqrt{ \chi' (0)}$
and suggested experiments to test this hypothesis by studying 
semi-inclusive polarized deep inelastic scattering in the target
fragmentation region (which allows one to vary the de facto
hadron target -- e.g. a proton or $\Delta$ resonance)
\cite{Shore:1998b}.

OZI violation associated with the gluonic topological charge density may also
be important to a host of $\eta$ and $\eta'$ interactions in hadronic physics.
We refer to \textcite{Bass:2002b} for an overview of the phenomenology.
Experiments underway at COSY-J\"ulich are measuring 
the isospin dependence of $\eta$ and $\eta'$ production close to threshold 
in proton-nucleon collisions
\cite{Moskal:2004}. 
These experiments are looking for signatures of possible OZI violation 
in the $\eta'$ nucleon interaction.
Anomalous glue may play a key role in the structure of the
light mass (about 1400-1600 MeV) exotic mesons with quantum numbers
$J^{PC} = 1^{-+}$
that have been observed in experiments at BNL and CERN.
These states might be dynamically generated resonances in 
$\eta' \pi$
rescattering \cite{Bass:2002,Szczepaniak:2003}
mediated by the OZI violating coupling of the $\eta'$.
Planned experiments at the GSI in Darmstadt will measure the $\eta$ 
mass in nuclei \cite{Hayano:1999}
and thus probe aspects of axial U(1) dynamics in the nuclear medium.

\section{CONNECTING QCD AND QCD INSPIRED MODELS OF THE PROTON SPIN PROBLEM}

We now collect and compare the various proposed explanations of 
the proton spin problem 
(the small value of $g_A^{(0)}$ extracted from polarized deep
 inelastic scattering) 
in roughly the order that they enter the derivation of the $g_1$ 
spin sum-rule:
\begin{enumerate}
\item
{\it
A ``subtraction at infinity'' in the dispersion relation for $g_1$}
perhaps generated in the 
transition from current to constituent quarks and 
involving gluon topology and the mechanism of dynamical axial U(1) 
symmetry breaking.
In the language of Regge phenomenology it is associated with a 
fixed pole in the real part of the spin dependent part of the forward
Compton amplitude.

In this scenario the strange quark polarization $\Delta s$ extracted 
from inclusive polarized deep inelastic scattering and 
neutrino proton elastic scattering would be different.
A precision measurement of $\nu p$ elastic scattering would be very useful.

Note that fixed poles play an essential role in 
the Adler and Schwinger term sum-rules - one should be on the look out !

\item
{\it SU(3) flavour breaking in the analysis of hyperon beta decays}.
Phenomenologically, SU(3) flavour symmetry seems 
to be well respected in the measured beta decays, 
including the recent KTeV measurement of the $\Xi^0$ decay \cite{KTeV:2001}.
\textcite{Leader:2003} have recently argued that even the most 
extreme SU(3) breaking 
scenarios consistent with hyperon decays will still lead to a 
negative value of the strange quark axial charge $\Delta s$ 
extracted from polarized deep inelastic data.
Possible SU(3) breaking in the large $N_c$ limit of QCD has been investigated
by \textcite{Flores:1998}.

One source of SU(3) breaking that we have so far observed is in the polarized 
sea generated through 
photon-gluon fusion where 
the strange-quark mass term is important
-- see Eq.(\ref{eqf122}) and Fig.\ref{fig:fig9}.
The effect of including SU(3) breaking in 
the parton model 
for $\Delta q_{\rm partons}$ within various
factorization schemes has been investigated in \textcite{Gluck:2001}.

\item
{\it 
Topological charge screening and target independence of 
the ``spin effect'' 
generated by a small value of $\chi'(0)$ }
in the flavour-singlet Goldberger-Treiman relation.
This scenario could be tested through semi-inclusive measurements
where a pion or D meson is detected
in the target fragmentation region, perhaps using a polarized $ep$
collider with Roman pot detectors \cite{Shore:1998b}.
These experiments could, in principle, be used to vary the target
and measure $g_1$ for e.g. $\Delta^{++}$ and $\Delta^-$ targets
along the lines of the programme that has been carried through
in unpolarized scattering
\cite{Holtmann:1994}.

\item
{\it
Non-perturbative evolution associated with the renormalization group 
factor $E(\alpha_s)$ 
between deep inelastic scales and the low-energy scale where quark 
models might, perhaps, describe the twist 2 parton distributions} 
\cite{Jaffe:1987}.
One feature of this scenario
is that 
(in the four flavour theory)
the polarized charm and strange quark 
contributions evolve at
the same rate with changing $Q^2$
since $\Delta s - \Delta c$ is flavour non-singlet
(and therefore independent of the QCD axial anomaly) 
\cite{Bass:1993a}.
Heavy-quark renormalization group 
arguments suggest that $\Delta c$ is small \cite{Bass:2002d,Kaplan:1988}
up to $1/m_c$ corrections.

\item
{\it
Large gluon polarization $\Delta g \sim 1$ at the scale 
$\mu \sim 1$GeV} could restore consistency between the
measured $g_A^{(0)}$ and quark model predictions if the
quark model predictions are associated with $\Delta q_{\rm partons}$
(the low $k_t$ contribution to $g_A^{(0)}$) in Eq.(\ref{eqf123}).
$\Delta g$ can be measured through a variety of gluon induced
partonic production processes
including 
charm production and
two-quark-jet events 
in polarized deep inelastic scattering, and
prompt photon production and jet studies in polarized proton collisions
at RHIC -- see Section IX.E below.
First attempts to extract $\Delta g$ from QCD motivated fits to the $Q^2$
dependence of $g_1$ data yield values between 0 and 2 at $Q^2 \sim 1$GeV$^2$
-- see Section IX.C.

{\it
How big should we expect $\Delta g$ to be ?
}
Working in the framework of light-cone models one finds contributions 
from ``intrinsic'' and ``extrinsic'' gluons.
Extrinsic contributions arise from gluon bremstrahlung 
$q_V \rightarrow q_V g$ of a valence quark and 
have a relatively hard virtuality.
Intrinsic gluons are associated with the physics of the nucleon 
wavefunction (for example, gluons emitted by one valence quark 
and absorbed by another quark) and have a relatively soft spectrum 
\cite{Bass:1999b}.
Light-cone models including QCD colour coherence at small Bjorken $x$ and 
perturbative QCD counting rules at large $x$
\cite{Brodsky:1990,Brodsky:1995a}
suggest values of 
$\Delta g \sim 0.6$ at low scales $\sim 1$GeV$^2$ 
-- sufficient to account for about half of the ``missing spin'' 
or measured value of $g_A^{(0)}$.

Bag model calculations give values $\Delta g \sim -0.4$ 
(note the negative sign) 
when one includes gluon exchange contributions and no
``self field'' contribution 
where the gluon is emitted and absorbed by the same quark
\cite{Jaffe:1996}
and
$\Delta g \sim 0.24$ (positive sign) when the ``self field'' contribution
is included \cite{Barone:1998}.
A QCD sum-rule calculation \cite{Saalfeld:1998} gives $\Delta g \sim 2 \pm 1$.

\item
{\it
Large negatively polarized strangeness in the quark sea (with small
$k_t$)}.
This scenario can be tested through semi-inclusive measurements of
polarized deep inelastic scattering provided that 
radiative corrections,
fragmentation functions and the experimental acceptance are under control.

\end{enumerate}

Of course, the final answer may prove to be a cocktail solution of
these possible explanations or include some new dynamics that has 
not yet been thought of.

In testing models of quark sea and gluon polarization it 
is important to understand the transverse momentum and 
Bjorken $x$ dependence of the different sea-quark dynamics.
For example, sea quark contributions to deep inelastic structure 
functions are induced by perturbative photon-gluon fusion
\cite{Efremov:1988,Altarelli:1988,Carlitz:1988}, 
pion and kaon cloud physics
\cite{Melnitchouk:1999,Koepf:1992,Cao:2003}, 
instantons \cite{Forte:1991,Schafer:2004,Nishikawa:2004,Dolgov:1999}, ...
In general, different mechanisms will produce sea with different
$x$ and $k_t$ dependence.

Lattice calculations are also making progress in unravelling
the spin structure of the proton \cite{Mathur:2000,Negele:2004}.
Interesting new results \cite{Negele:2004} suggest a value of
$g_A^{(0)}$ about 0.7 in a heavy-pion world where the pion mass
$m_{\pi} \sim 700 - 900$ MeV.
Physically, in the heavy-pion world (away from the chiral limit)
the quarks become less relativistic and it is reasonable to expect
the nucleon spin to arise from the valence quark spins. 
Sea quark effects are expected to become more important as 
the quarks become lighter and sea production mechanisms become important.
It will be interesting to investigate the behaviour of 
$g_A^{(0)}$ in future lattice calculations as these calculations approach
the chiral limit.

In an alternative approach to understanding low energy QCD 
\textcite{Witten:1983a,Witten:1983b}
noticed that in the limit that the number of colours $N_c$ 
is taken to infinity
($N_c \rightarrow \infty$ with $\alpha_s N_c$ held fixed)
QCD behaves like a system of bosons
and the baryons emerge as topological solitions called Skyrmions
in the meson fields. 
In this model the spin of the large $N_c$ ``proton'' is a 
topological quantum number.
The ``nucleon's'' axial charges turn out to be  
sensitive to which meson fields are included in the model
and the relative contribution of a quark source and pure mesons
-- 
we refer to the lectures of \textcite{Aitchison:1988} 
for a more detailed discussion of the Skyrmion approach.
\textcite{Brodsky:1988} 
found that 
$g_A^{(0)}$ 
vanishes in a particular version of the Skyrmion model with just
pseudoscalar mesons.
Non-vanishing values of $g_A^{0}$ are found using more general
Skyrmion Lagrangians 
\cite{Ryzak:1989,Cohen:1989}, 
including with additional vector mesons \cite{Johnson:1990}.

\begin{widetext}

\begin{figure}[t!]
\vspace*{8.0cm}
\includegraphics{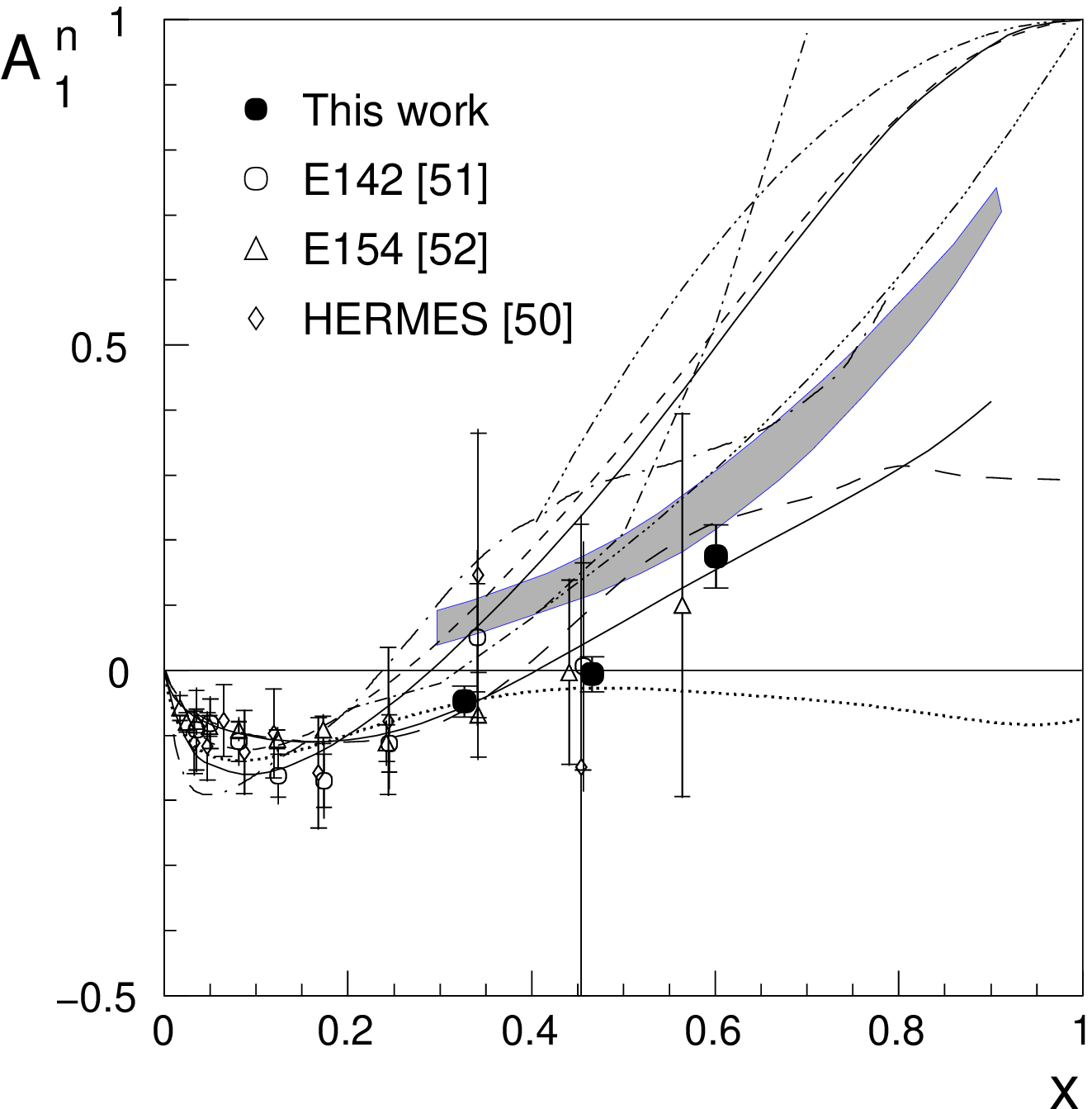}
\includegraphics{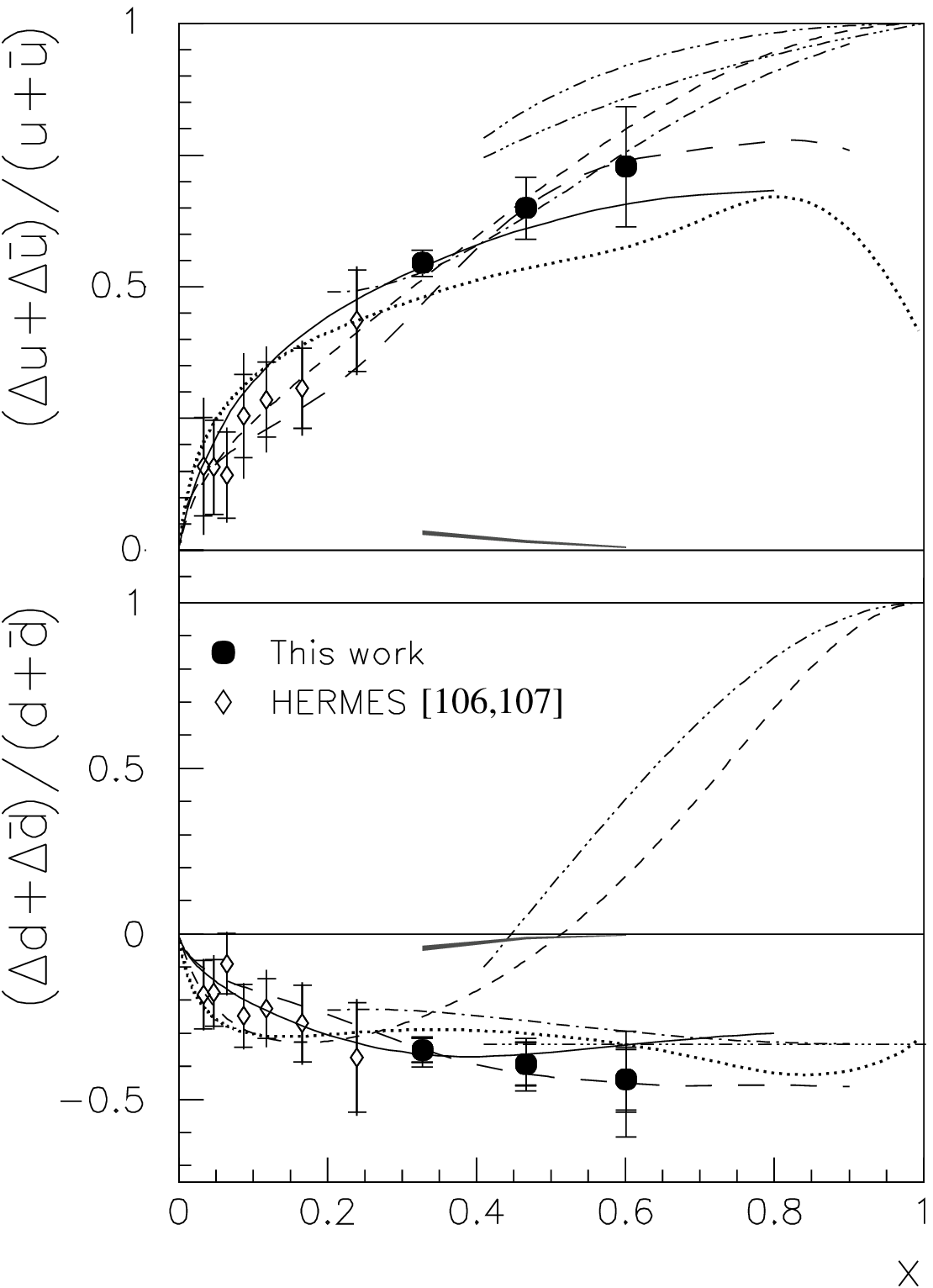}
\vspace*{4mm}
\caption[*]{Left: Recent data on ${\cal A}_1^n$ from the E99-117 experiment
\cite{Zheng:2004a,Zheng:2004b}. 
Right: extracted polarization asymmetries for $u+\bar{u}$
and $d+\bar{d}$. For more details and references on the various
model predictions, see \cite{Zheng:2004a,Zheng:2004b}. 
\label{fig:fig10}}
\end{figure}

\end{widetext}

\section{THE SPIN-FLAVOUR STRUCTURE OF THE PROTON}

\subsection{The valence region and large $x$}

The large $x$ region ($x$ close to one) is very interesting and
particularly sensitive to the valence structure of the nucleon.
Valence quarks dominate deep inelastic structure functions for
large and intermediate $x$ (greater than about 0.2).
Experiments at Jefferson Lab are making the first precision
measurements of the proton's spin structure at large $x$ --
see Fig.\ref{fig:fig10}

QCD motivated predictions for the large $x$ region exist based on 
perturbative QCD counting rules and 
quarks models of the proton's structure based on SU(6)
[flavour SU(3) $\otimes$ spin SU(2)] and scalar diquark 
dominance.
We give a brief explanation of these approaches.

\begin{enumerate}
\item
Perturbative QCD counting rules predict that the parton distributions
should
behave as a power series expansion in $(1-x)$ when $x \rightarrow 1$
\cite{Brodsky:1995a,Farrar:1975}.
The fundamental principle behind these counting rules results is that
for the leading struck quark to carry helicity polarized in the same 
direction as the proton the spectator pair should carry spin zero, 
whence they are bound through longitudinal gluon exchange.
For the struck quark to be polarized opposite to the direction of the 
proton the spectator pair should be in a spin one state, and in this
case one has also to consider the effect of transverse gluon exchange.
Calculation shows that this is supressed by a factor of $(1-x)^2$.
We use $q^{\uparrow}(x)$ and $q^{\downarrow}(x)$ to denote the parton
distributions
polarized parallel and antiparallel to the polarized proton.
One finds \cite{Brodsky:1995a}
\begin{equation}
q^{\uparrow \downarrow} (x) \rightarrow (1-x)^{2n - 1 + 2 \Delta S_z}
, \ \ \ \ (x \rightarrow 1).
\label{eqi136}
\end{equation}
Here $n$ is the number of spectators and $\Delta S_z$ is the difference
between the polarization of the struck quark and the polarization of the
target nucleon.
When $x \rightarrow 1$ the QCD counting rules predict that the structure
functions should be dominated by valence quarks polarized parallel to the
spin of the nucleon.
The ratio of polarized to unpolarized structure functions should 
go to one when $x \rightarrow 1$.
For the helicity parallel valence quark distribution
one predicts
\begin{equation}
q^{\uparrow} (x) \sim (1-x)^3, \ \ \ \ \ (x \rightarrow 1)
\end{equation}
whereas for the helicity anti-parallel distribution one obtains
\begin{equation}
q^{\downarrow} (x) \sim (1-x)^5, \ \ \ \ \ (x \rightarrow 1).
\end{equation}
Sea distributions are suppressed and the leading term starts as $(1-x)^5$.

\item
Scalar diquark dominance is based on the observation that, within 
the context of the SU(6) wavefunction of the proton 
in Eq.~(\ref{eqa9}), 
one gluon exchange
tends to make the mass of the scalar diquark pair lighter than
the vector spin-one diquark combination. 
One-gluon exchange offers an explanation of 
the nucleon-$\Delta$ mass splitting
and has the practical consequence that in model calculations of
deep inelastic structure functions
the 
scalar diquark term
${1 \over \sqrt{2}} | u \uparrow (ud)_{S=0} \rangle$
in Eq.(\ref{eqa9}) dominates the physics at large Bjorken $x$ 
\cite{Close:1988}.

\end{enumerate}

In the large $x$ region ($x$ close to one) where sea quarks and 
gluons
can be neglected the neutron and proton spin asymmetries are
given by
\begin{equation}
{\cal A}_1^n = {\Delta u + 4 \Delta d \over u + 4d}, \ \ \ \ \ \ \ \ \ \
{\cal A}_1^p = {4 \Delta u + \Delta d \over 4u + d}
.
\label{eqi137}
\end{equation}
Rearranging these expressions one obtains formulae 
for the separate up and down quark distributions in the proton:
\begin{eqnarray}
{\Delta u \over u} &=&
{4 \over 15} {\cal A}_1^p (4 + {d \over u}) 
- {1 \over 15} {\cal A}_1^n (1 + 4 {d \over u})
\nonumber \\
{\Delta d \over d} &=&
{4 \over 15} {\cal A}_1^n (4 + {u \over d}) 
- {1 \over 15} {\cal A}_1^p (1 + 4 {u \over d})
.
\label{eqi138}
\end{eqnarray}
The predictions of perturbative QCD counting rules and scalar diquark
dominance models for the large $x$ limit of these asymmetries are given
in Table \ref{tab:table1}.
On the basis of both perturbative QCD {\it and} SU(6),
one expects the ratio of polarized to unpolarized structure functions,
${\cal A}_{1n}$, should approach 1 as $x \rightarrow 1$ 
\cite{Melnitchouk:1996,Isgur:1999}.
It is vital to test this prediction. If it fails we understand nothing 
about the valence spin structure of the nucleon.

\begin{table}[h]
\caption{\label{tab:table1}
QCD motivated model predictions for the large $x$ limit of 
deep inelastic spin asymmetries and parton distributions.
}
\begin{ruledtabular}
\begin{tabular}{lccccr}
Model & $\Delta u / u$ & $\Delta d / d$ & ${\cal A}_1^p$ 
& ${\cal A}_1^n$ & $d/u$ \\
\hline
SU(6) & ${2 \over 3}$ & -${1 \over 3}$ & ${5 \over 9}$ & 0 & ${1 \over 2}$ \\
Broken SU(6), scalar diquark  & 1 &  -${1 \over 3}$ & 1 & 1 & 0 \\
QCD Counting Rules            & 1 & 1 & 1 & 1 & ${1 \over 5}$ \\
\end{tabular}
\end{ruledtabular}
\end{table}

Interesting new data from the Jefferson Laboratory Hall A Collaboration 
the neutron asymmetry ${\cal A}_1^n$ 
\cite{Zheng:2004a} are shown in Fig.~\ref{fig:fig10}. 
This data shows a clear trend for ${\cal A}_1^n$ 
to become positive at large $x$. 
The crossover point where ${\cal A}_1^n$ changes sign is 
particularly interesting because the value of $x$ where
this occurs in the neutron asymmetry is the result of a
competition between 
the SU(6) valence structure
\cite{Close:1988} and 
chiral corrections \cite{Schreiber:1988,Steffens:1995}.
Fig.~\ref{fig:fig10} also
shows the extracted valence polarization asymmetries. 
The data are consistent with constituent quark models 
with scalar diquark dominance
which predict $\Delta d/d\to -1/3$ at large $x$, 
while
perturbative QCD counting rules predictions
(which neglect quark orbital angular momentum)
give $\Delta d/d\to 1$ and tend to deviate from
the data, unless the convergence to 1 sets in very late.

A precision measurement of ${\cal A}_{1n}$ up to $x \sim 0.8$
will be possible following the 12 GeV upgrade of Jefferson Laboratory
\cite{Meziani:2002}.

\subsection{The isovector part of $g_1$}

\begin{figure}[t] 
\includegraphics{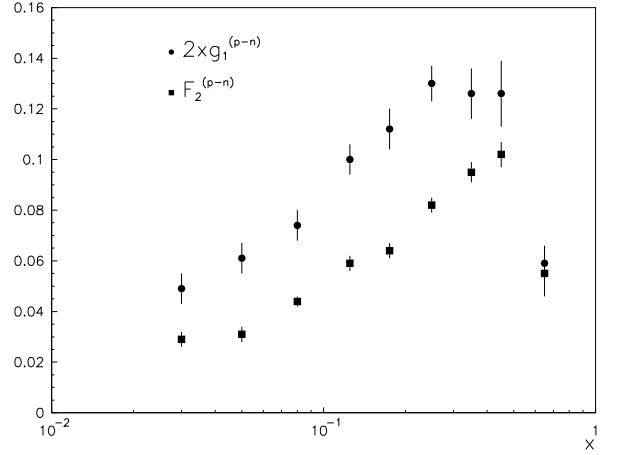} 
\begin{center} 
\vspace{7.0cm} 
\parbox{8.0cm} 
{\caption[Delta]
{The isovector structure functions $2x g_1^{(p-n)}$ (SLAC data) and 
$F_2^{(p-n)}$ (NMC) from \textcite{Bass:1999}.}
\label{fig:fig11}} 
\end{center} 
\end{figure}

\begin{figure}[h] 
\includegraphics{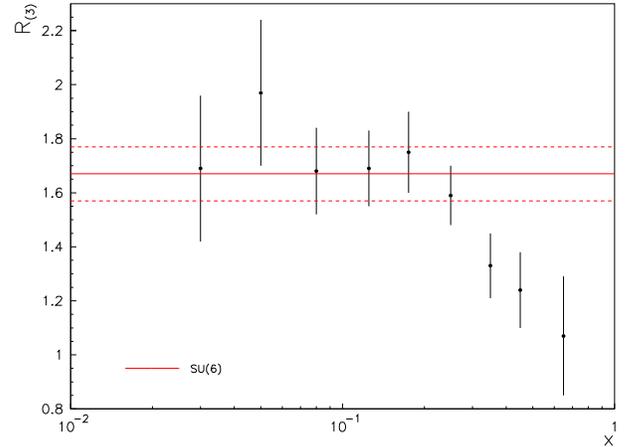}
\begin{center} 
\vspace{7.0cm} 
\parbox{8.0cm} 
{\caption[Delta]
{The ratio $R_{(3)} = 2x g_1^{(p-n)}/F_2^{(p-n)}$ from \textcite{Bass:1999}. }
\label{fig:fig12}} 
\end{center} 
\end{figure}

Constituent quark model predictions for $g_1$ are observed to work 
very well in the 
isovector channel. 
First, as highlighted in Section III.B above, the Bjorken sum rule 
which relates the first moment of the isovector part of $g_1$, 
$(g_1^p - g_1^n)$, 
to the isovector axial charge $g_A^{(3)}$ 
has been confirmed in polarized deep inelastic scattering experiments
at the level of 10\% \cite{Windmolders:1999}. 
Second, looking beyond the first moment, one finds the following 
intriguing observation about the shape of $(g_1^p - g_1^n)$. 
Figure \ref{fig:fig11} 
shows $2 x (g_1^p - g_1^n)$ 
(SLAC data) 
together with the isovector structure function $(F_2^p - F_2^n)$ 
(NMC data).
The ratio 
$R_{(3)} = 2 x (g_1^p - g_1^n) / (F_2^p - F_2^n)$ is plotted in 
Fig. \ref{fig:fig12}.
It measures the ratio of polarized to unpolarized isovector quark 
distributions.
In the QCD parton model
\footnote{
In a full description one should also include the perturbative QCD
Wilson coefficients for the non-singlet spin difference and spin
averaged cross-sections.
However, the affect of these coefficients
makes a 
non-negligible contribution to the deep inelastic structure
functions only at $x < 0.05$ and is small in the kinematics
where there is high $Q^2$ spin data.
There is no gluonic or singlet pomeron contributions
to the isovector
structure functions $(g_1^p-g_1^n)$ and $(F_2^p-F_2^n)$.
}
\begin{equation}
2x (g_1^p - g_1^n) =
{1 \over 3} x
\Biggl[ (u + {\overline u})^{\uparrow} - (u + {\overline u})^{\downarrow}
      - (d + {\overline d})^{\uparrow} + (d + {\overline
         d})^{\downarrow}
 \Biggr] 
\label{eqi139}
\end{equation}
and
\begin{equation}
(F_2^p - F_2^n)
=
{1 \over 3} x
\Biggl[ (u + {\overline u})^{\uparrow} + (u + {\overline u})^{\downarrow}
      - (d + {\overline d})^{\uparrow} - (d + {\overline
         d})^{\downarrow}
 \Biggr]. 
\label{eqi140}
\end{equation}

The data reveal a large isovector contribution in $g_1$
and the ratio
$R_{(3)}$ is observed to be approximately constant 
(at the value $\sim 5/3$ predicted by SU(6) constituent quark models) 
for $x$ between 0.03 and 0.2, and goes towards one when $x \rightarrow 1$
(consistent with the prediction of both QCD counting rules and scalar
 diquark dominance models).
The small $x$ part of this data is very interesting.
The area under $(F_2^p-F_2^n)/2x$ is determined by the Gottfried integral 
\cite{Gottfried:1967,Arneodo:1994} and 
is about 25\% suppressed relative to the simple SU(6) 
prediction (by the pion cloud, Pauli blocking, ... ).
The area under $(g_1^p-g_1^n)$ is fixed by the Bjorken sum rule 
(and is also about 25\% suppressed relative to the SU(6) prediction
 -- the suppression here being driven
 by relativistic effects in the nucleon and by perturbative QCD 
 corrections to the Bjorken sum-rule).
Given that perturbative QCD counting rules or scalar diquark models work 
and assuming that the ratio $R_{(3)}$ takes the constituent quark 
prediction at the
canonical value of $x \sim {1 \over 3}$, one finds \cite{Bass:1999} 
that the observed shape of $g_1^p - g_1^ n$ is almost required
to reproduce the area under the Bjorken sum rule (which is determined 
by the physical value of $g_A^{(3)}$ --- a non-perturbative constraint)!
The constant ratio in the low to medium $x$ range contrasts  
with 
the naive Regge prediction 
using $a_1$ exchange
(and no hard-pomeron $a_1$ cut)
that the ratio $R_{(3)}$ should fall and be roughly proportional 
to $x$ as $x \rightarrow 0$.
It would be very interesting to have precision measurements of $g_1$
at high energy and low $Q^2$ from a future polarized $ep$ collider 
to test the various scenarios how small $x$
dynamics might evolve through the transition region and the application
of spin dependent Regge theory.

\subsection{QCD fits to $g_1$ data}

In deep inelastic scattering experiments the different $x$ data
points on $g_1$ are each measured at different values of $Q^2$,
viz. $x_{\rm expt.}(Q^2)$.
One has to evolve these experimental data points
to the same value of $Q^2$
in order to test the Bjorken \cite{Bjorken:1966,Bjorken:1970} and 
Ellis-Jaffe \cite{Ellis:1974} sum-rules.
DGLAP evolution is frequently used in analyses of polarized deep 
inelastic data to achieve this.

The $\lambda^2$ dependence of the parton distributions is given
by the DGLAP equations \cite{Altarelli:1977}
\begin{eqnarray}
{d \over dt} \Delta \Sigma(x,t) &=& {\alpha_s(t) \over 2 \pi}
\biggl[
\int_x^1 {dy \over y} \Delta P_{qq}({x \over y}) \Delta \Sigma (y,t)
\nonumber \\
& & \ \ \ \ \ \ \ \ \ \ + \ 2 f
\int_x^1 {dy \over y} \Delta P_{qg}({x \over y}) \Delta g (y,t)
\biggr]
\nonumber \\
{d \over dt} \Delta g (x,t) &=& {\alpha_s(t) \over 2 \pi}
\biggl[
\int_x^1 {dy \over y}
\Delta P_{gq}({x \over y}) \Delta \Sigma (y,t)
\nonumber \\
& & 
\ \ \ \ \ \ \ \ \ \ \ \ \ 
+ 
\int_x^1 {dy \over y} \Delta P_{gg}({x \over y}) \Delta g (y,t) \ \biggr]
\nonumber \\
\label{eqi141}
\end{eqnarray}
where $\Sigma(x,t) = \sum_q \Delta q(x,t)$ and $t = \ln \lambda^2$.
The splitting functions $P_{ij}$ in Eq.(63) have been calculated at
leading-order by Altarelli and Parisi \cite{Altarelli:1977}
and at
next-to-leading order 
by Mertig, Zijlstra and van Neerven \cite{Mertig:1996,Zijlstra:1994} 
and Vogelsang \cite{Vogelsang:1996}.

\begin{figure}[!b]
\vspace*{5.5cm}
\includegraphics{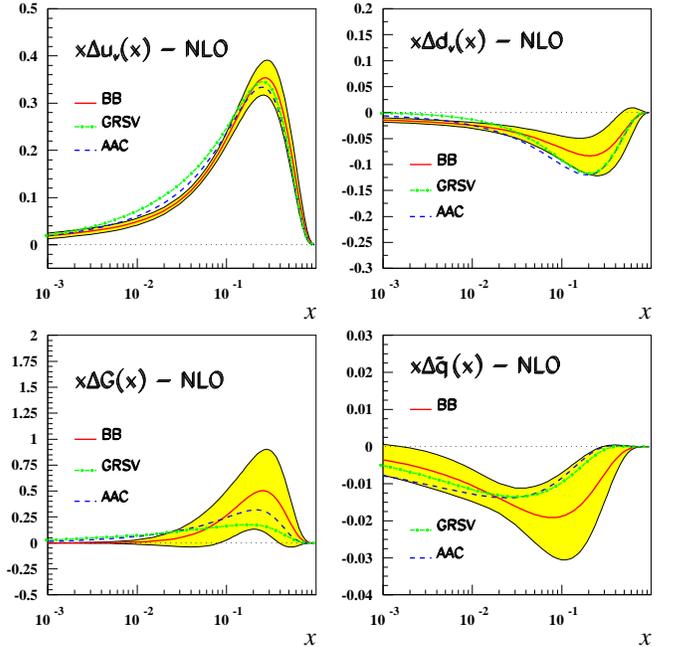}
\vspace*{4.2cm}
\caption[*]{Polarized parton distribution functions from NLO pQCD
(${\overline {\tt MS}}$) fits at $Q^2 =4$GeV$^2$ 
using SU(3) flavour assumptions
\cite{Stoesslein:2002}.
\label{fig:fig13}}
\end{figure}

Similar to the analysis that is carried out on unpolarized data, global 
NLO perturbative QCD analyses
have been performed on the polarized structure function data sets.
The aim is to extract the polarized quark and gluon parton distributions. 
These QCD fits are performed within a given factorization scheme,
e.g.
the ``AB'', chiral invariant (CI) or JET and $\overline{\rm MS}$ schemes.

Let us briefly review these different factorization schemes.

Different factorization schemes correspond to different procedures 
for separating the phase space for photon-gluon fusion into ``hard''
and ``soft'' contributions in the convolution formula Eq. (\ref{eqd80}).
In the QCD parton model analysis of photon gluon fusion that we discussed 
in Section XI.D
using the cut-off on the transverse momentum squared, 
the polarized gluon contribution to the first moment of $g_1$ 
is associated with two-quark jet events carrying $k_t^2 \sim Q^2$.
The gluon coefficient function
is given by
$C^{(g)}_{{\rm PM}} = g_1^{(\gamma^{*} g)}|_{\rm hard}$ 
where
$g_1^{(\gamma^{*} g)}|_{\rm hard}$ is taken 
from Eq.(\ref{eqf121}) 
with $Q^2 \geq \lambda^2$ and $\lambda^2 \gg P^2, m^2$.
This transverse-momentum cut-off scheme is sometimes called 
the ``chiral invariant'' (CI) \cite{Cheng:1996} or JET \cite{Leader:1998} 
scheme.

Different schemes can be defined relative to this $k_t$ cut-off scheme 
by the transformation
\begin{eqnarray}
& & 
C^{(g)} \biggl( x, {Q^2 \over \lambda^2}, \alpha_s(\lambda^2) \biggr) 
\nonumber \\
& & 
\rightarrow \
C^{(g)} \biggl( x, {Q^2 \over \lambda^2}, \alpha_s(\lambda^2) \biggr) 
-
{\tilde C}^{(g)}_{\rm scheme} \biggl( x, \alpha_s(\lambda^2) \biggr) .
\nonumber \\
\end{eqnarray}
Here ${\tilde C}^{(g)}_{\rm scheme}$ 
shall be ${\alpha_s \over \pi}$ times a polynomial in $x$.
The parton distributions transform as
\begin{eqnarray}
& & \Delta \Sigma (x,\lambda^2)_{\rm scheme} 
    = \Delta \Sigma (x,\lambda^2)_{\rm PM} 
\nonumber \\
& & 
\ \ \ \ \ \ \ \ \ \ + 
  f 
  \int_x^1 {dz \over z} \Delta g({x \over z},\lambda^2)_{\rm PM} 
  {\tilde C}^{(g)}_{\rm scheme} (z, \alpha_s(\lambda^2)) 
\nonumber \\
& & 
\Delta g (x,\lambda^2)_{\rm scheme} = \Delta g(x,\lambda^2)_{\rm PM}
\end{eqnarray}
so that the physical structure function $g_1$ is left invariant under 
the change of scheme.
The virtuality and invariant-mass cut-off versions of the parton model 
that we discussed in Section XI.D correspond to different choices of scheme.

The ${\overline {\rm MS}}$ and AB schemes are defined as follows.
In the ${\overline {\rm MS}}$ scheme the gluonic hard scattering 
coefficient is calculated using 
the operator product expansion with ${\overline {\rm MS}}$ 
renormalisation \cite{tHooft:1972}. 
One finds \cite{Cheng:1996,Bass:1992a}:
\begin{equation}
C^{(g)}_{{\overline{\rm MS}}} 
=
C^{(g)}_{\rm PM}
\ + \ 
{\alpha_s \over \pi} (1-x) .
\end{equation}
In this scheme
$\int_0^1 dx \ C^{(g)}_{\overline {\rm MS}} =0$ 
so that
$\int_0^1 dx \ \Delta g(x, \lambda^2)$ decouples from $\int_0^1 dx g_1$.
This result corresponds to the fact that there is no gauge-invariant 
twist-two, spin-one, gluonic operator with $J^P = 1^+$ 
to appear in 
the operator product expansion for the first moment of $g_1$.
In the ${\overline {\rm MS}}$ scheme the 
contribution of $\int_0^1 dx \ \Delta g$
to the first moment of $g_1$ is included 
into 
$\int_0^1 dx \ \Sigma_{{\overline{\rm MS}}} (x,\lambda^2)$. 
The AB scheme \cite{Ball:1996} is defined by the formal operation of
adding the $x$-independent term $-{\alpha_s \over 2 \pi}$ to the 
${\overline {\rm MS}}$ gluonic coefficient,
{\it viz.}
\begin{equation}
C^{(g)}_{\rm AB}(x) = 
C^{(g)}_{{\overline{\rm MS}}}
\ - \ {\alpha_s \over 2 \pi}.
\end{equation}

In the $\overline{\rm MS}$ scheme the polarized gluon distribution
does not contribute explictly to the first moment of $g_1$.
In the AB and JET schemes on the other hand the polarized gluon
(axial anomaly contribution)
$\alpha_s \Delta g$ does contribute explicitly to the first moment
since
$\int_0^1 dx \ C^{(g)} = - {\alpha_s \over 2 \pi}$.

For the SMC data
one finds for the  $\overline{\rm MS}$ (AB) scheme at a $Q^2$ of
1 GeV$^2$ \cite{Adeva:1998b}:
$\Delta \Sigma = 0.19\pm 0.05 (0.38\pm0.03)$ and
$\Delta g = 0.25^{+0.29}_{-0.22} (1.0^{+1.2}_{-0.3})$ 
where
$\Delta \Sigma = (\Delta u + \Delta d + \Delta s)$.
The main source of error in the QCD fits
comes from lack of knowledge about $g_1$ in the small $x$
region and (theoretical)
the functional form chosen for the quark and gluon distributions in
the fits.
Note that these QCD fits in both the AB and
${\overline {\rm MS}}$ schemes give values of
$\Delta \Sigma$ which are smaller than the Ellis-Jaffe value 0.6.

New fits are now being produced taking into account all 
the available
data including new data from polarized semi-inclusive
deep inelastic scattering.
Typical polarized distributions extracted from the fits
are shown in Fig.~\ref{fig:fig13}.
Given the uncertainties in the fits, values of $\Delta g$
are extracted ranging between about zero and +2.
In these pQCD analyses one ends up with a consistent picture of
the proton spin:
the low value of $\Delta \Sigma$ may be compensated by a
large polarized gluon.
The precision on $\Delta g$ is however still rather modest.
Moreover, it is vital to validate this model with {\it direct}
measurements of $\Delta g$, as we discuss in Section IX.E below.
Also, the first moments depend on integrations from $x = 0$ to 1.
Perhaps there is an additional component at very small $x$?

\subsection{Polarized quark distributions 
            and semi-inclusive polarized deep inelastic scattering}

As noted above, there are several possible mechanisms for producing
sea quarks in the nucleon:
photon-gluon fusion, the meson cloud of the nucleon, instantons, ...
In general the different dynamics will produce polarized sea with
different $x$ and transverse momentum dependence.

Semi-inclusive measurements of fast pions and kaons in the current
fragmentation region with final state particle identification can
be used to reconstruct the individual up, down and strange quark
contributions to the proton's spin \cite{Close:1978,Close:1991,Frankfurt:1989}.
In contrast to inclusive polarized deep inelastic scattering
where the $g_1$ structure function is deduced by detecting only
the scattered lepton, the detected particles in the semi-inclusive
experiments are high-energy (greater than 20\% of the energy of the
incident photon) charged pions and kaons in coincidence with 
the scattered lepton.
For large energy fraction $z=E_h/E_{\gamma} \rightarrow 1$
the most probable occurence is that the detected $\pi^{\pm}$ and
$K^{\pm}$
contain the struck quark or antiquark in their valence Fock state.
They therefore act as a tag of the flavour of the struck quark 
\cite{Close:1978}.

\begin{figure}[!t]
\vspace*{6.4cm}
\includegraphics{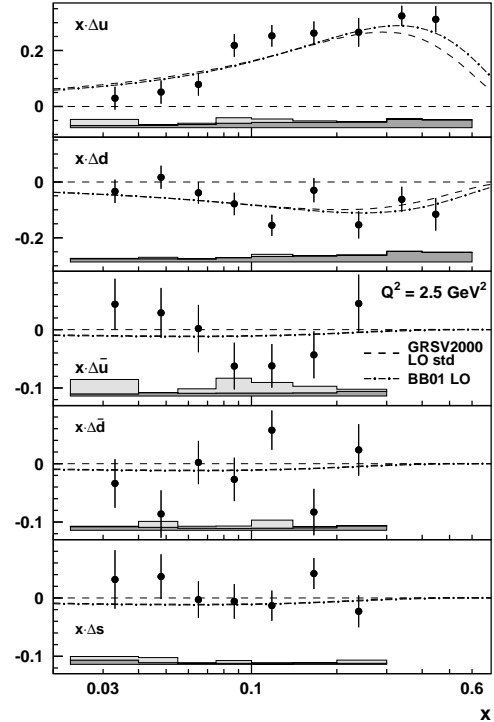}
\vspace*{3.7cm}
\caption[*]{Recent {\sc Hermes} results \cite{Airapetian:2004a}
for the quark and antiquark
polarizations extracted from semi-inclusive DIS.  
\label{fig:fig14}}
\end{figure}

In leading order (LO) QCD 
the double-spin asymmetry for the production of hadrons $h$ in 
semi-inclusive polarized $\gamma^*$ polarized proton collisons 
is:
\begin{equation}
A_{1p}^{h} (x,Q^2) \simeq
{
\sum_{q,h} e_q^2 
\Delta q(x,Q^2) \int_{z_{\rm min}}^1 D_q^h (z, Q^2) 
\over
\sum_{q,h} e_q^2 
q(x,Q^2) \int_{z_{\rm min}}^1 D_q^h (z, Q^2)
}
\label{eqi142}
\end{equation}
where $z_{\rm min} \sim 0.2$.
Here
\begin{equation}
D_q^h (z, Q^2) = \int d k_t^2 D_q^h (z, k_t^2, Q^2)
\label{eqi143}
\end{equation}
is the fragmentation function for the struck quark or antiquark 
to produce a hadron $h$ ($=\pi^{\pm}, K^{\pm}$) carrying energy 
fraction $z=E_h/E_{\gamma}$ in the target rest frame;
$\Delta q(x,Q^2)$ 
is the quark (or antiquark) 
polarized parton 
distribution and $e_q$ is the quark charge.
Note the integration over the transverse momentum $k_t$ of the 
final-state hadrons \cite{Close:1991}. 
(In practice this integration over $k_t$ is determined 
 by the acceptance of the experiment.)
Since pions and kaons have spin zero, the fragmentation functions 
are the same for both polarized and unpolarized leptoproduction.
NLO corrections to Eq.~(\ref{eqi142}) 
are discussed in 
\textcite{deFlorian:1998,deFlorian:2000}.

This programme for polarized deep inelastic scattering was pioneered 
by the SMC \cite{Adeva:1998c} and HERMES 
\cite{Ackerstaff:1999}
Collaborations
with new recent measurements
from HERMES reported in \textcite{Airapetian:2004a,Airapetian:2004b}.
Fig.~\ref{fig:fig14} shows the latest results on the flavor separation 
from HERMES \cite{Airapetian:2004a}, which were obtained using a 
leading-order 
(naive parton model) Monte-Carlo code based ``purity'' analysis. 
The polarization of the up and down quarks are positive and 
negative respectively, 
while the sea polarization data are consistent with zero and
not inconsistent with the negative sea polarization suggested 
by inclusive deep inelastic data within the measured $x$ range 
\cite{Gluck:2001,Blumlein:2002}.
However,
there is also no evidence from this semi-inclusive analysis 
for a large negative strange quark polarization.
For the region $0.023 < x < 0.3$
the extracted $\Delta s$
integrates to the value $+0.03 \pm 0.03 \pm 0.01$
which contrasts with the negative value for the polarized
strangeness
(\ref{eqa2}) extracted from inclusive measurements of $g_1$.
It will be interesting to see whether this effect persists 
in forthcoming semi-inclusive data from COMPASS.
The HERMES data also favour an isospin symmetric sea 
$\Delta {\bar u} - \Delta {\bar d}$, but with large uncertainties.

An important issue for semi-inclusive measurements is the angular
coverage of the detector \cite{Bass:2003a}.
The non-valence spin-flavour structure of the proton extracted
from semi-inclusive measurements of polarized deep inelastic scattering
may depend strongly on the transverse momentum (and angular)
acceptance of the detected final-state hadrons which are used
to determine the individual polarized sea distributions.
The present semi-inclusive experiments detect final-state hadrons
produced only at small angles from the incident lepton beam
(about 150 mrad angular coverage)
The perturbative QCD ``polarized gluon interpretation'' 
\cite{Efremov:1988,Altarelli:1988} of 
the inclusive measurement (\ref{eqa2}) involves physics at the maximum 
transverse momentum \cite{Carlitz:1988,Bass:2003a} and large angles -- 
see Fig.\ref{fig:fig9}.
Observe the small value for the light-quark sea polarization at 
low transverse momentum and
the positive value for the integrated
strange sea polarization at low $k_t^2$:
$k_t < 1.5$GeV at the HERMES $Q^2=2.5$GeV$^2$.
When we relax the transverse momentum cut-off, 
increasing the acceptance of the experiment,
the measured strange sea polarization changes sign and becomes
negative (the result implied by fully inclusive deep inelastic measurements).
For HERMES the average transverse momentum of the detected 
final-state fast hadrons is less than about 0.5 GeV whereas 
for SMC the $k_t$ of the detected fast pions was less than about 1 GeV.
Hence, there is a question whether the leading-order 
sea quark 
polarizations 
extracted from semi-inclusive experiments with 
limited angular resolution fully include the effect of the axial anomaly or 
not.

Recent theoretical studies motivated by this data include also possible 
effects associated 
spin dependent fragmentation functions \cite{Christova:2001}, 
possible higher twist effects in semi-inclusive deep inelastic
scattering, 
and possible improvements in the Monte Carlo \cite{Kotzinian:2003}.

\begin{figure}[!t]
\vspace*{6.4cm}
\includegraphics{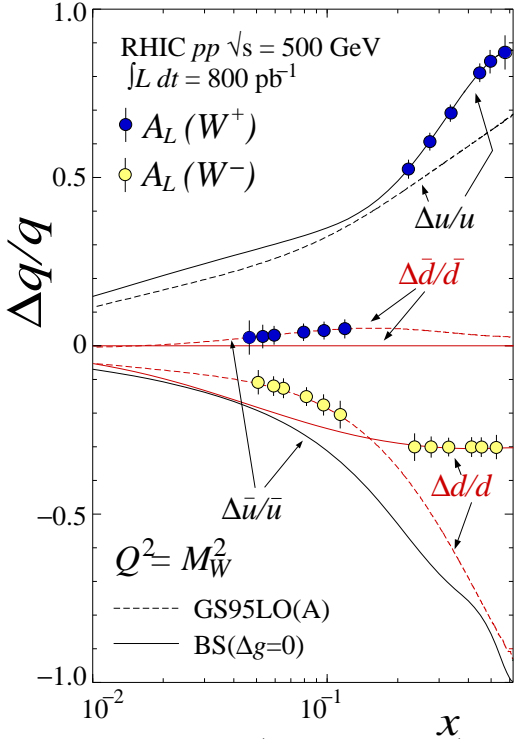}
\vspace*{3.7cm}
\caption[*]
{Expected sensitivity \cite{Bunce:2000}
for the flavor decomposition
of quark and anti-quark polarizations at RHIC. 
Reprinted, with permission, from the Annual Reviews of
Nuclear and Particle Science, Volume 50 (c)2000 by Annual
Reviews {\tt www.annualreviews.org.}
\label{fig:fig15}}
\end{figure}

The dependence on the details of the fragmentation process 
limits the accuracy of the method above. 
At RHIC~\cite{Bunce:2000}
the polarization of the 
$u, \overline{u}, d$ and $\overline{d}$ 
quarks in the proton will be measured directly and 
precisely using $W$ boson production in 
$u\overline{d} \rightarrow W^+$ and $d\overline{u} \rightarrow W^-$.
The charged weak boson is produced through a pure V-A coupling
and the chirality of the quark and anti-quark in the reaction is fixed.
A parity violating asymmetry for $W^+$ production in $pp$
collisions can be expressed as
\begin{equation}
 A(W^+) =
\frac{\Delta u(x_1)\overline{d}(x_2) -\Delta\overline{d}(x_1) u(x_2)}
{ u(x_1)\overline{d}(x_2) + \overline{d}(x_1) u(x_2)} 
.
\label{eqi144}
\end{equation}
For $W^-$ production $u$ and $d$ quarks should be exchanged.
The expression converges to
$\Delta u(x)/u(x)$ and $- \Delta\overline{d}(x)/\overline{d}(x)$
in the limits
$x_1 \gg x_2$ and $x_2 \gg x_1$ respectively.
The momentum fractions are calculated as
$x_1= \frac{M_W}{\sqrt{s}} e^{y_W}$ and
$x_2= \frac{M_W}{\sqrt{s}} e^{-y_W}$, with $y_W$ the rapidity of
the $W$.
The experimental difficulty is that the $W$ is observed through its
leptonic decay $W \rightarrow l \nu$ and only the charged lepton is
observed.
With the assumed integrated luminosity of 800 pb$^{-1}$ at $\sqrt{s}
= 500 $ GeV, one can expect about 5000 events each for $W^+$ and $W^-$.
The resulting measurement precision is shown in Fig.~\ref{fig:fig15}.

It has also been pointed out that neutrino factories would be an ideal 
tool for polarized quark flavour decomposition studies. 
These would allow one to collect large data samples of charged current 
events, in the kinematic region $(x,Q^2)$ of present fixed target 
data \cite{Forte:2001}. A complete separation of all four 
flavours and anti-flavours would become possible, including $\Delta s(x,Q^2)$.

\subsection{The polarized gluon distribution $\Delta g(x,Q^2)$}
\vspace{1mm}
\noindent

Motivated by the discovery of \textcite{Altarelli:1988} and 
\textcite{Efremov:1988} that polarized glue makes a scaling 
contribution to the first 
moment of $g_1$,
$\alpha_s \Delta g \sim {\tt constant}$, 
there has been a 
vigorous and ambitious programme to measure $\Delta g$.
The NLO QCD motivated fits to the inclusive $g_1$ data 
are suggestive that, 
perhaps, 
the net polarized glue might be positive but more direct 
measurements involving glue sensitive observables are needed 
to really extract
the magnitude of $\Delta g$ and 
the shape of $\Delta g (x, Q^2)$
including any possible nodes in the distribution function.
Possible channels include gluon mediated processes 
in
semi-inclusive polarized deep inelastic scattering
and
hard QCD processes in high energy polarized proton-proton
collisions at RHIC.

\begin{figure}[h]
\includegraphics{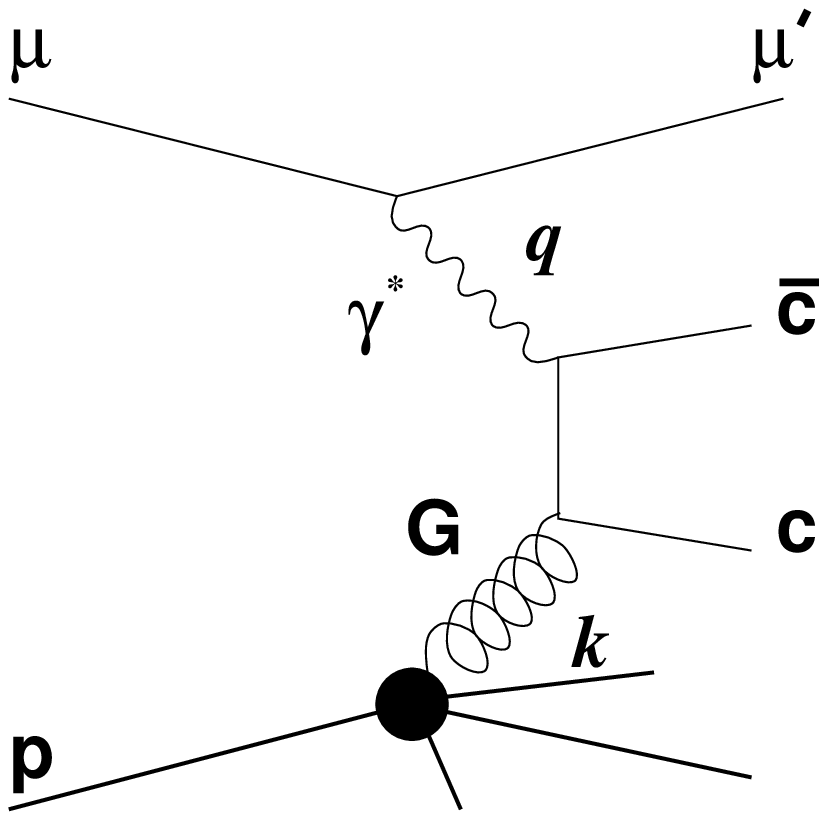}
\vspace{8.5cm}
\parbox{8.0cm}
{\caption[Delta] 
{c $\overline{\rm c}$ production in Photon Gluon Fusion}
\label{fig:fig16}}
\end{figure}

COMPASS has been conceived to measure $\Delta g$ via the study of 
the photon-gluon fusion process, as shown in Fig.~\ref{fig:fig16}.
The cross section of this process is directly related to the
gluon density at the Born level. 
The experimental technique consists of the reconstruction of charmed 
mesons in the final state.
COMPASS will also use the same process 
with high $p_t$ particles 
instead of charm to access $\Delta g$.
This may lead to samples with larger statistics, but these
have larger background contributions, 
namely from QCD Compton processes and fragmentation.
The expected sensitivity on the measurement of $\Delta g/g$
from these experiments
is estimated to be about $\delta(\Delta g/g) =0.1$ at $x_g \sim 0.1$.

HERMES was the first to attempt to measure $\Delta g$ using high $p_t$ 
charged particles, as proposed for COMPASS above, and 
nearly real photons $\langle Q^2 \rangle = 0.06$GeV$^2$.
The measurement is at the limit of where a
perturbative treatment of the data can be expected to be valid, 
but the result is interesting:
$\Delta g/g = 0.41 \pm 0.18 \pm 0.03$ at an average $<x_g>= 0.17$
\cite{Airapetian:2000}.
The SMC Collaboration have performed a similar analysis for their own 
data keeping $Q^2 > 1$GeV$^2$.
An average gluon polarization was extracted 
$\Delta g / g = -0.20 \pm 0.28 \pm 0.10$ at an average gluon momentum 
$x_g = 0.07$ \cite{Adeva:2004}.

The hunt for $\Delta g$ is also one of the main physics drives for
polarized RHIC.  The key processes used here are high-$p_t$ prompt
photon production
$pp \rightarrow \gamma X$,
jet production
$pp \rightarrow $ jets $+ X$,
and heavy flavour production
$pp \rightarrow c\overline{c}X, b\overline{b}X, J/\psi X$.
Due to the first stage detector capabilities most emphasis
has so far been put on the prompt photon channel.
Measurements of $\Delta g/g$ are expected in the gluon $x$ 
range $0.03 < x_g < 0.3$.

These anticipated RHIC measurements of $\Delta g$ have inspired
new theoretical developments 
aimed
at implementing higher-order calculations of
partonic cross-sections into global
analyses of polarized parton distribution functions, which will
benefit the analyses of future polarized $pp$ data to measure $\Delta g$.
Hard polarized reactions at RHIC and the polarized parton distributions
that 
they probe are summarized in Table II.

\begin{table}
\caption{\label{tab:table2}
Polarized partons from RHIC }
\begin{ruledtabular}
\begin{tabular}{lccr}
reaction & LO subprocesses & partons probed & $x$ range \\
\hline
$pp \rightarrow {\rm jets} X$  &  
$q {\overline q}, qq, qg, gg \rightarrow {\rm jet} X$
& $\Delta q, \Delta g$  &  $x > 0.03$ \\
$pp \rightarrow \pi X$  &  $q {\overline q}, qq, gg \rightarrow \pi X$
& $\Delta q, \Delta g$  &  $x > 0.03$ \\
$pp \rightarrow \gamma X$  &  
$q g \rightarrow q \gamma$, $q {\overline q} \rightarrow g \gamma $ 
& $ \Delta g$  &  $x > 0.03$ \\
$pp \rightarrow Q {\overline Q} X$  &  
$ gg \rightarrow Q {\overline Q}$, 
$q {\overline q} \rightarrow Q {\overline Q} $ &
$\Delta g $ &  $x > 0.01$ \\
$pp \rightarrow W^{\pm} X$  &  $q {\overline q'} \rightarrow W^{\pm}$
& $\Delta u, \Delta {\overline u}, \Delta d, \Delta {\overline d}$  &  
$x > 0.06$ \\
\end{tabular}
\end{ruledtabular}
\end{table}

\begin{figure}[h]
\includegraphics{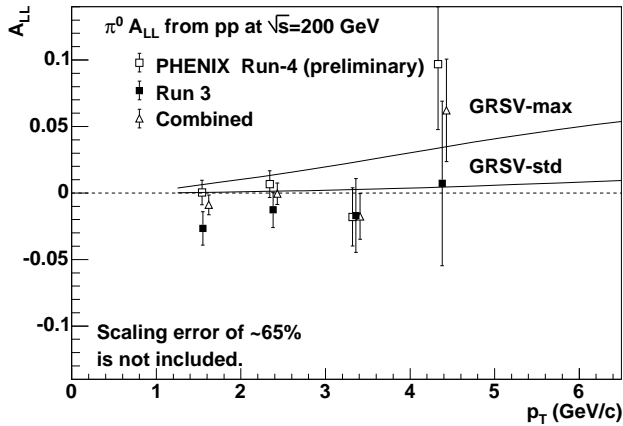}
\vspace{6.5cm}
\parbox{8.0cm}
{\caption[Delta] 
{The PHENIX (preliminary Run-4) data \cite{Fukao:2005}
 for the spin asymmetry ${\cal A}_{LL}^{\pi}$
 along with NLO perturbative QCD predictions for various $\Delta g$
 \cite{Jager:2004} }
\label{fig:fig17}}
\end{figure}

In the first runs at RHIC the longitudinal double spin asymmetry for 
production of a leading pion $\pi^0$
with large transverse momentum has been used as a 
surrogate jet 
to investigate possible gluon polarization in the proton.
NLO perturbative QCD corrections 
to this process have been calculated in 
\textcite{deFlorian:2003}
and
\textcite{Jager:2003}.
The data from PHENIX \cite{Adler:2004,Fukao:2005}
are shown in Fig.~17 and
are consistent with a significant 
(up to a few percent) 
negative asymmetry ${\cal A}^{\pi}_{LL}$ 
for pion transverse momentum $1< p_t <4$GeV in contrast 
with the predictions of leading twist 
perturbative QCD calculations
which do provide a good description of the unpolarized cross-section 
in the same kinematics.
It will be interesting to see whether this effect survives 
more precise data.
The NLO perturbative QCD analysis of \textcite{Jager:2004} 
suggests that 
the leading power in $p_t$ contribution to
${\cal A}^{\pi}_{LL}$ 
cannot be large and negative in the measured 
range of 
$p_t$ within the framework of perturbative QCD
independent of the sign of $\Delta g$. 
One would need to invoke power suppressed contributions 
(though the leading power term seems to describe the 
 corresponding unpolarized data) and/or non-perturbative effects.
Increase in precision and data at 
higher pion $p_t$
(up to about 12 GeV) are expected from future runs.

Future polarized $ep$ colliders could add information in two ways: 
by extending the kinematic range for measurements of $g_1$ 
or by direct measurements of $\Delta g$.
A precise measurement of $\Delta g$ is 
crucial for a full understanding of the proton spin problem.
HERA has shown
that large centre of mass energy allows 
several processes to be used to extract the unpolarized gluon
distribution.
These include jet and high $p_t$ hadron production, 
charm production both in DIS and photoproduction, and 
correlations between multiplicities of the current and target 
hemisphere of the events in the Breit frame.
The most promising process for a direct extraction of $\Delta g$ 
is di-jet production ~\cite{DeRoeck:1999,Radel:2002}.
The underlying idea is to isolate boson-gluon 
fusion events
where the gluon distribution enters at the Born level.

\section{TRANSVERSITY}

There are three species of twist-two quark distributions in QCD.
These are the spin independent distributions $q(x)$ measured in 
the unpolarized structure functions $F_1$ and $F_2$, 
the spin dependent distributions $\Delta q(x)$ measured in $g_1$
and the transversity distributions $\delta q(x)$.

The transversity distributions describe the density of 
transversely polarized quarks inside a transversely polarized proton 
\cite{Barone:2002}.
Measuring transversity is an important experimental challenge in QCD
spin physics.
We briefly describe the physics of transversity and the programme to
measure it.

\begin{figure}[!h]
\vspace*{6.8cm}
\includegraphics{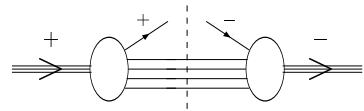}
\vspace*{-4.8cm}
\caption[*]{Transversity in helicity basis. 
\label{fig:fig18}}
\end{figure}

The twist-two transversity distributions 
\cite{Ralston:1979,Artru:1990,Jaffe:1992}
can be interpreted in parton language as follows.
Consider a nucleon moving with (infinite) momentum in the $\hat
e_{3}$-direction, but polarized along one of the directions transverse
to $\hat e_{3}$.  
Then $\delta q (x,Q^{2})$ counts the quarks with flavour $q$,
momentum fraction $x$ and
their spin parallel to the spin of a nucleon minus the number antiparallel.  
That is, in analogy with Eq.~(\ref{eqb28}),
$\delta q (x)$
measures the distribution 
of partons with transverse polarization in a transversely polarized nucleon,
{\it viz.}
\begin{equation}
\delta q (x,Q^2)=q^{\uparrow}(x) - q^{\downarrow}(x) 
.
\label{eqj145}
\end{equation}
In a helicity basis, transversity corresponds 
to the helicity-flip structure shown in Fig.~\ref{fig:fig18}
making
transversity a probe of chiral symmetry breaking \cite{Collins:1993b}.
The first moment of the transversity distribution is proportional 
to the nucleon's C-odd tensor charge, 
{\it viz.}
$\delta q = \int_0^1 dx \delta q(x)$ with
\begin{equation}
\langle p, s | \ {\bar q} i \sigma_{\mu \nu} \gamma_5 q \ | p,s \rangle
= 
(1 / M) ( s_{\mu} p_{\nu} - s_{\nu} p_{\mu} ) \delta q.
\end{equation}
Transversity is C-odd and chiral-odd.

If quarks moved non-relativistically in the nucleon $\delta q$ and 
$\Delta q$ would be identical 
since rotations and Euclidean boosts commute and a series of boosts 
and rotations can convert a longitudinally polarized
nucleon into a transversely polarized nucleon at infinite momentum.
The difference between the transversity and helicity distributions
reflects the relativistic character of quark motion in the nucleon.
In the MIT Bag Model this effect is manifest as follows.
The lower component of the Dirac spinor enters the relativistic 
spin depolarization factor with the opposite sign to $\Delta q$
because of the extra factor of $\gamma_{\mu}$ in the tensor charge
\cite{Jaffe:1992}.
That is, the relativistic depolarization
factor 
$N^2 \int_0^R dr r^2 (f^2 - {1 \over 3} g^2)$
for $\Delta q$ mentioned in Section I.A
is replaced 
by 
$N^2 \int_0^R dr r^2 (f^2 + {1 \over 3} g^2)$ for $\delta q$ 
where
$\psi = 
{N \over \sqrt{4 \pi}}
\biggl({ f \atop i \sigma .{\hat{r}} g }\biggr)$ 
is the Dirac spinor.

Little is presently known about the shape of the transversity distributions.
However some general properties can be deduced from QCD arguments.
The spin distributions $\Delta q(x)$ and $\delta q(x)$ 
have opposite charge conjugation properties: 
$\Delta q(x)$ is C-even whereas $\delta q(x)$ is C-odd.
The spin dependent quark and gluon helicity distributions
($\Delta q$ and $\Delta g$) mix under $Q^{2}$-evolution.
In contrast,
there is no analog of gluon transversity in the nucleon so $\delta q$ 
evolves without mixing, like a non-singlet parton distribution function. 
Not coupling to glue or perturbative $\bar q q$ pairs, 
$\delta q (x)$ and the tensor charge promise
to be more quark-model--like than the singlet 
axial-charge 
(though they are both scale dependent)
and should be an interesting contrast \cite{Jaffe:2001}.
Under QCD evolution the moments
$\int_0^1 dx x^n \delta q(x,Q^2)$ decrease with increasing $Q^2$.
In leading order QCD the transversity distributions
are bounded above by Soffer's inequality \cite{Soffer:1995}
\begin{equation}
 | \delta q (x, Q^2) | 
  \leq 
  {1 \over 2}
  \biggl[  q (x,Q^2) + \Delta q (x,Q^2) \biggr] .
\label{eqj146}
\end{equation}

Experimental study of transversity distributions at leading-twist requires 
observables which are the product of two objects with odd 
chirality 
-- the transversity distribution and 
either a second transversity distribution or a chiral odd fragmentation 
function.
In proton-proton collisions the transverse double spin 
asymmetry, ${\cal A}_{TT}$, is proportional to $\delta q \delta\bar{q}$ 
with even chirality.
However the asymmetry is small requiring very large luminosity samples 
because of the large background from gluon induced processes in unpolarized 
scattering.
The most promising process to measure this double spin asymmetry is perhaps 
Drell-Yan production.

Transverse single spin asymmetries ${\cal A}_N$ are also being studied with 
a view to extracting information about transversity distributions.
Here the focus is on single hadron production with a transversely polarized
proton beam or target in $pp$ and $ep$ collisions.
The key process is
\begin{equation}
A(p,{\vec s}_t) + B(p') \rightarrow C(l) + X
\label{eqj147}
\end{equation}
where $C$ is typically a pion produced at large transverse momentum $l_t$.

Several mechanisms for producing these transverse single spin asymmetries 
have been discussed in the literature.
The asymmetries ${\cal A}_N$ are powered suppressed in QCD.
Leading $l_t$ behaviour of the produced pion can occur from the 
Collins \cite{Collins:1993b}
and Sivers \cite{Sivers:1991} effects plus twist-3 mechanisms \cite{Qiu:1999}.
The Collins effect involves 
the chiral-odd 
twist-2 transversity distribution 
in combination with a chiral-odd fragmentation function for the high $l_t$
pion in the final state.
It gives a possible route to measuring transversity.
The Sivers effect is associated with intrinsic quark transverse momentum in 
the initial state.
The challenge is to disentangle these effects from experimental data.

Factorization for transverse single spin processes 
in proton-proton collisions 
has been derived by \textcite{Qiu:1999} 
in terms of the convolution of a twist-two parton distribution from the 
unpolarized hadron, a twist-three quark-gluon correlation function from
the polarized hadron, and a short distance partonic hard part calculable
in perturbative QCD.
We refer to \textcite{Anselmino:2004b} 
for a discussion of factorization for processes such as the Collins and
Sivers effects involving unintegrated transverse momentum dependent parton 
and fragmentation functions.

We next outline the Collins and Sivers effects.

The Collins effect \cite{Collins:1993b} uses properties of fragmentation
to probe transversity.
The idea is that a pion produced in fragmentation will have some transverse 
momentum with respect to the momentum $k$ of the transversely polarized 
fragmenting parent quark. 
One finds a correlation of the form 
$\;i\vec{s}_t \cdot  (\vec{l}^{(\pi)} \times \vec{k}_t)$.
The Collins fragmentation function associated 
with this correlation is chiral-odd and T-even.
It combines with the chiral-odd transversity distribution to contribute to
the transverse single spin asymmetry.

For the Sivers effect \cite{Sivers:1991} the $k_t$ distribution of a quark 
in a transversely polarized hadron can generate an azimuthal 
asymmetry through the correlation 
$\,\vec{s}_t \cdot  (\vec{p} \times \vec{k}_t)$.
In this process final state interaction (FSI) of the active quark
produces the asymmetry before it fragments into hadrons
\cite{Brodsky:2002,Burkardt:2004a,Yuan:2003,Bachetta:2004}.
This process involves a $k_t$ unintegrated quark distribution function in
the transversely polarized proton.
The dependence on intrinsic quark transverse momentum 
means that this Sivers process is
related to quark orbital angular momentum in the proton
\cite{Burkardt:2002}.
The Sivers distribution function is chiral-even and T-odd.
The possible role of quark orbital angular momentum in understanding 
transverse single-spin
asymmetries is also discussed in \textcite{Boros:1993}.

The Sivers process is associated with the gauge link in operator definitions 
of the parton distributions.
The gauge link factor is trivial and equal to one for the usual $k_t$ 
integrated parton distributions measured in inclusive polarized deep inelastic
scattering. 
However, for $k_t$ unintegrated distributions the gauge link survives 
in a transverse direction at light-cone component $\xi^- = \infty$.
The gauge-link plays a vital role in the Sivers process \cite{Burkardt:2004}.
Without it (e.g. in the pre-QCD ``naive'' parton model)
time reversal invariance implies vanishing Sivers effect 
\cite{Ji:2002,Belitsky:2003a}.
The Sivers distribution function has the interesting 
property that it has the opposite sign in 
deep inelastic scattering and Drell-Yan reactions \cite{Collins:2002}.
It thus violates the universality of parton distribution functions.

\begin{figure}[!h]
\vspace*{5.5cm}
\includegraphics{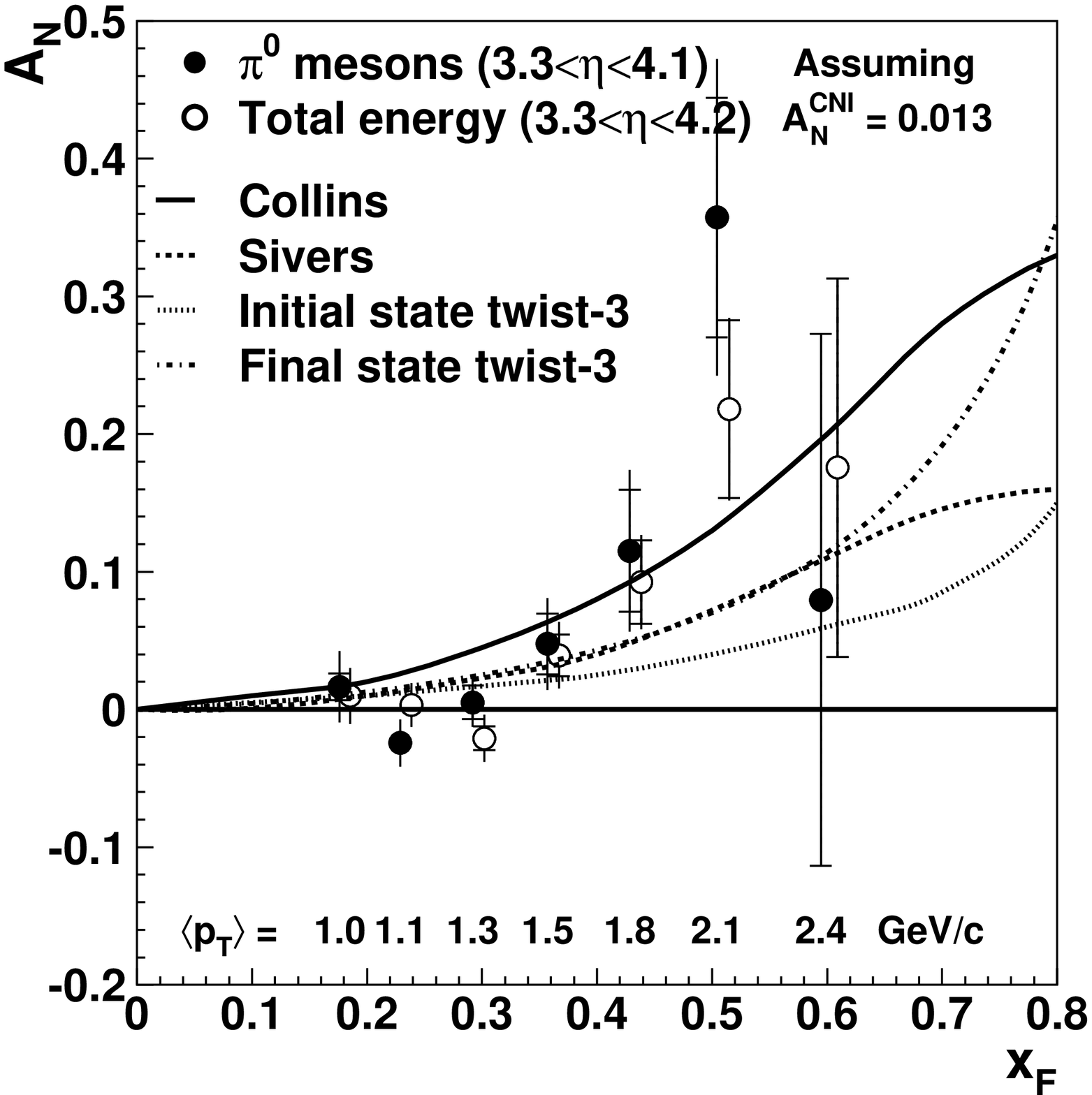}
\vspace*{26mm}
\caption[*]{Recent {\sc Star} results for the asymmetry $A_{\mathrm{N}}$
in $pp\to \pi^0X$ in the forward Feynman-$x_F$
region \cite{Adams:2004}. 
\label{fig:fig19}}
\end{figure}

The FermiLab experiment E704 found large transverse single-spin asymmetries 
${\cal A}_N$ for $\pi$ and $\Lambda$ production 
in proton-antiproton collisions at centre of mass energy $\sqrt{s} = 20$GeV
\cite{Adams:1991a,Adams:1991b,Bravar:1996}.
Large transverse single-spin asymmetries have also been observed in recent
data from the STAR collaboration at RHIC in proton-proton collisions at 
centre of mass energy $\sqrt{s} = 200$GeV \cite{Adams:2004}
-- see Figure \ref{fig:fig19} which also shows various theoretical predictions.
In a recent paper \textcite{Anselmino:2004b}
take into account intrinsic parton motion in the distribution and 
fragmentation functions as well as in the elementary dynamics and
argue that the Collins mechanism may be strongly suppressed at large 
Feynman $x_F$ in this process.
The Sivers effect is not suppressed and remains a candidate to explain 
the data. 
Higher-twist contributions \cite{Qiu:1999} from quark-gluon correlations 
may also be important.

The HERMES experiment has taken measurements of charged pion production 
in $ep$ scattering with transverse target polarization 
\cite{Airapetian:2004c}.
This data has been analysed for possible contributions from the Collins 
and Sivers effects.
The azimuthal distribution of the final state pions with respect to the 
virtual photon axis is expected to carry information about transversity 
through the Collins effect and 
about intrinsic transverse momentum in the proton through the Sivers effect.
In this analsis
one first writes the transverse single-spin asymmetry ${\cal A}_{N}$ as the
sum
\begin{equation}
{\cal A}_{N}(x,z)
=
{\cal A}_{N}^{\rm Collins} +  {\cal A}_{N}^{\rm Sivers} + ...
\label{eqj148}
\end{equation}
where
\begin{equation}
{\cal A}_{N}^{\rm Collins} \propto
| {\vec s}_t | 
\sin (\phi + \phi_S) 
\frac{
\sum_q e_q^2 \delta q(x) H_1^{\perp,q}(z)
}
{\sum_q e^2_q  q(x)D_q^{\pi}(z)}
\label{eqj149}
\end{equation}
and
\begin{equation}
{\cal A}_{N}^{\rm Sivers} \propto
| {\vec s}_t | 
\sin (\phi - \phi_S) 
\frac{
\sum_q e_q^2 f_{1 T}^{\perp,q} D_q^{\pi} (z)
}
{\sum_q e^2_q  q(x) D_q^{\pi}(z)}
\label{eqj150}
\end{equation}
denote the contributions from the Collins and Sivers effects.
Here $\phi$ is the angle between the lepton direction and 
the 
$(\gamma^* \pi)$ plane and $\phi_S$ is the angle between
the lepton direction and the transverse target spin;
$H_1^{\perp q}$ is the Collins function for a quark of 
flavour $q$, 
$f_{1 T}^{\perp,q}$ is the Sivers distribution function, and 
$D_q^{\pi}$ is the regular spin independent fragmentation function.
When one projects out the two terms with different azimuthal angular 
dependence the HERMES analysis suggests that both the Collins and 
Sivers effects 
are present in the data.
Futhermore, the analysis suggests the puzzling result that 
the ``favoured'' (for $u \rightarrow \pi^+$)
and ``unfavoured'' (for $d \rightarrow \pi^+$) 
Collins fragmentation functions may contribute with equal weight
(and opposite sign)
\cite{Airapetian:2004c}.

Other processes and experiments will help to clarrify the importance of 
the Collins and Sivers processes.
Additional studies of the Collins effect have been proposed in $e^+ e^-$ 
collisions using 
the high statistics data samples of BABAR and BELLE.
The aim is to measure two relevant fragmentation functions:
the Collins function $H_1^{\perp}$ and
the interference fragmentation functions $\delta \hat{q}^{h_1,h_2}$.
For the first, one measures
the fragmentation of a transversely polarized quark into a charged pion
and the azimuthal distribution of the final state pion with respect to
the initial quark momentum (jet-axis).
For the second, one measures the fragmentation of transversely polarized
quarks into
pairs of hadrons in a state which is the superposition of two different
partial wave amplitudes; e.g. $\pi^+,\pi^-$ pairs in the $\rho$ and $\sigma$
invariant mass region \cite{Collins:1994,Jaffe:1998}.
The high luminosity and particle identification capabilities of detectors
at B-factories makes these measurements possible.

The Sivers distribution function might be measurable through the 
transverse single spin asymmetry ${\cal A}_N$ for D meson
production generated 
in
$p^{\uparrow} p$ scattering
\cite{Anselmino:2004a}.
Here the underlying elementary processes guarantee the absence of
any polarization in the final partonic state so that there is no
contamination from Collins like terms.
Large dominance of the process $gg \rightarrow c {\bar c}$ process
at low and intermediate $x_F$ offers a unique opportunity to measure
the gluonic Sivers distribution function.
The gluonic Sivers function could also be extracted from back-to-back 
correlations in the azimuthal angle of
jets in collisions of unpolarized and transversely polarized proton
beams at RHIC \cite{Boer:2004}.

Measurements with transversely polarized targets have a bright future and 
are already yielding surprises.
The results promise to be interesting and 
to teach us about transversity and 
about the role of quark transverse momentum in the structure of the proton 
and fragmentation processes.

\section{DEEPLY VIRTUAL COMPTON SCATTERING 
AND EXCLUSIVE PROCESSES}

\subsection{Orbital angular momentum}

\begin{figure}[t!]
\includegraphics{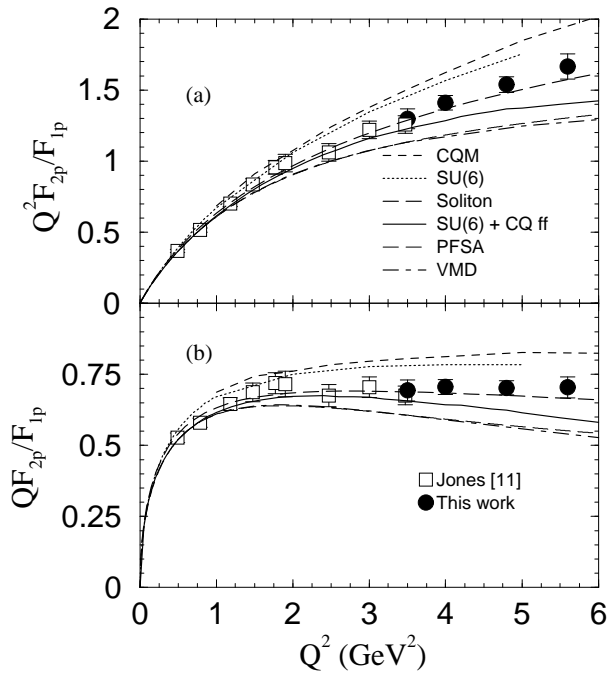}
\vspace{8.5cm}
\parbox{8.5cm}
{\caption[Delta]
{
Jefferson Lab data on the the ratio of the proton's Pauli to Dirac 
form-factors \cite{Gayou:2002}.}
}
\label{fig:fig20}
\end{figure}

So far in this review we have concentrated on intrinsic spin in the proton.
The orbital angular momentum structure of the proton is also of
considerable interest and much effort has gone into devising
ways to measure it.
The strategy involves the use of hard exclusive reactions and
the formalism of generalized parton distributions (GPDs) 
which 
describes
deeply virtual Compton scattering (DVCS) and meson production (DVMP).
Possible hints of quark orbital angular momentum are also suggested 
by recent form-factor measurements at Jefferson Laboratory
\cite{Jones:2000,Gayou:2002} -- see Fig.20.
The ratio of the spin-flip Pauli form-factor to the Dirac form-factor
is observed to have a $1/\sqrt{Q^2}$
behaviour in the measured region 
in contrast with the $1/Q^2$
behaviour predicted by QCD Counting Rules 
(helicity conservation neglecting angular momentum),
{\it viz.}
$F_1 \sim 1/Q^4$ and $F_2 \sim 1/Q^6$
\cite{Brodsky:1981}.
However, this data can also be fit with the formula
\begin{equation}
\frac{F_2 (Q^2)}{F_1 (Q^2)}
=
\frac{\mu_A}{1 + (Q^2/c)\ln^b(1+Q^2/a)}
\label{eqk151}
\end{equation}
(with $\mu_A=1.79$, $a=4 m_{\pi}^2 =0.073$ GeV$^2$, 
 $b=-0.5922$, $c=0.9599$ GeV$^2$)
prompting the question at which $Q^2$ the Counting Rules 
prediction is supposed to work 
and at which $Q^2$ higher twist effects can be neglected
\cite{Brodsky:2002a}.
At this point it is interesting to recall that the simple 
Counting Rules prediction 
 fails to describe 
 the large $x$ behaviour of $\Delta d / d$ 
 in the presently measured JLab kinematics -- Section IX.A.

A $1/Q$ behaviour for $\frac{F_2 (Q^2)}{F_1 (Q^2)}$
is found in a light-front Cloudy Bag 
calculation
\cite{Miller:2002a,Miller:2002b} and in
quark models with orbital angular momentum \cite{Ralston:2004,Ralston:2002}.
A new perturbative QCD calculation which takes into account orbital
angular momentum
\cite{Belitsky:2003b}
gives
$F_2/F_1 \sim (\log^2  Q^2 / \Lambda^2)/Q^2$
and also fits the Jefferson Lab data well.
The planned 12 GeV upgrade at Jefferson Laboratory will enable 
us to measure these nucleon form-factors at higher $Q^2$ and the 
inclusive spin asymmetries at values of 
Bjorken $x$ closer to one, 
and thus probe deeper into the kinematic regions where QCD Counting Rules 
should apply.
This data promises to be very interesting!

Deeply virtual Compton scattering (DVCS) provides a possible experimental
tool to access the quark total angular momentum, $J_q$, in the proton
through the physics of generalized parton 
distributions (GPDs) ~\cite{Ji:1997a,Ji:1997b}.
The form-factors which appear in the forward limit $(t \rightarrow 0)$
of the second moment of the spin-independent generalized quark parton
distribution in the (leading-twist) spin-independent part of the DVCS
amplitude project out the quark total angular momentum defined through
the proton matrix element of the QCD angular-momentum tensor.
We explain this physics below.

DVCS studies have to be careful to chose the kinematics not to be
saturated
by a large Bethe-Heitler (BH) background where the emitted real photon
is radiated from the electron rather than the proton.
The HERMES and Jefferson Laboratory experiments
measure in the kinematics where they expect to be dominated by
the DVCS-BH
interference term and observe the $\sin \phi$
azimuthal angle and
helicity dependence expected for this contribution -- see Fig.21.
First measurements of the single spin asymmetry
have been
reported in ~\textcite{Airapetian:2001,Stepanyan:2001},
which have the characteristics expected from the DVCS-BH interference.

\begin{widetext}

\begin{figure}[t!]
\includegraphics{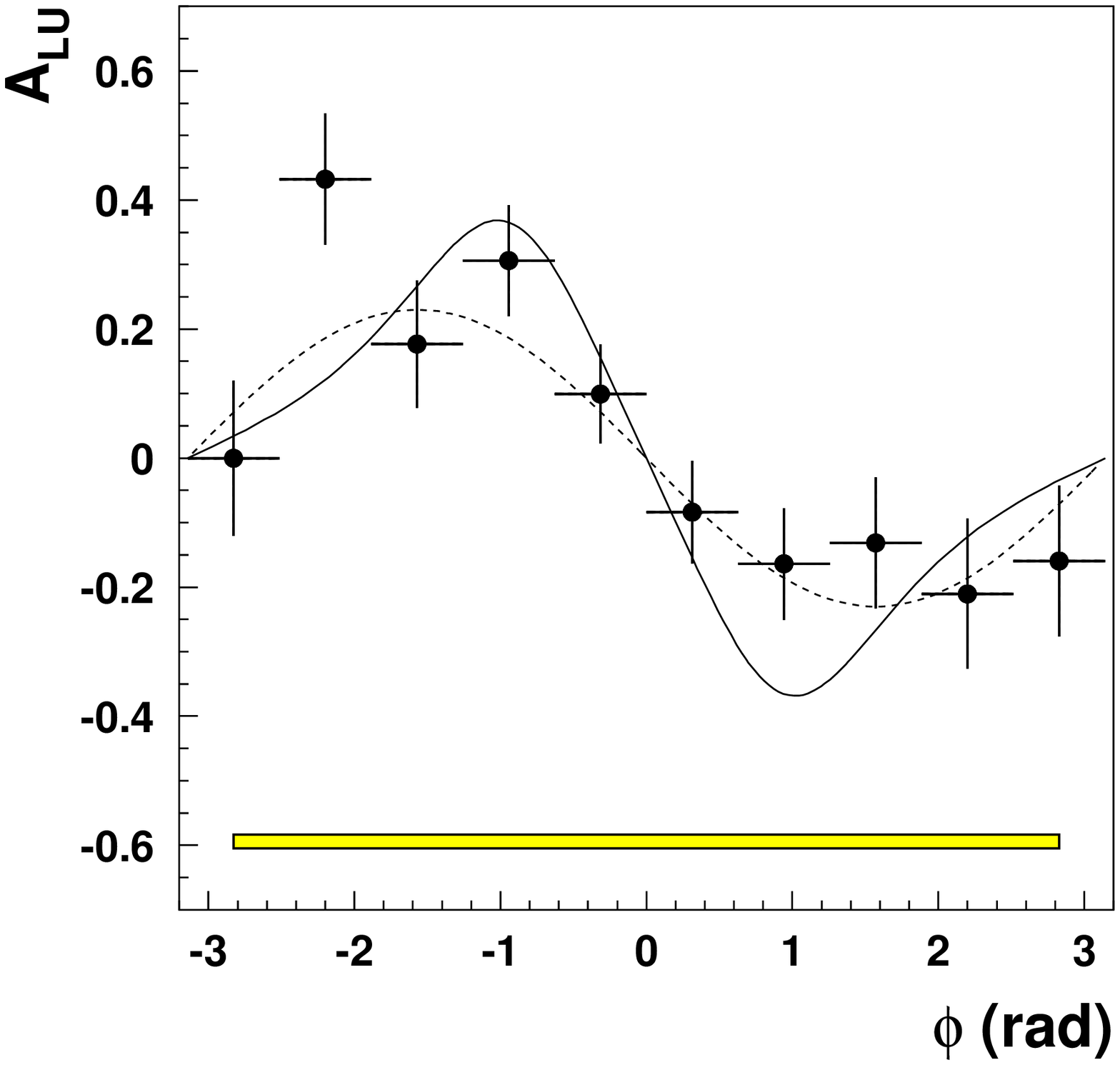}
\includegraphics{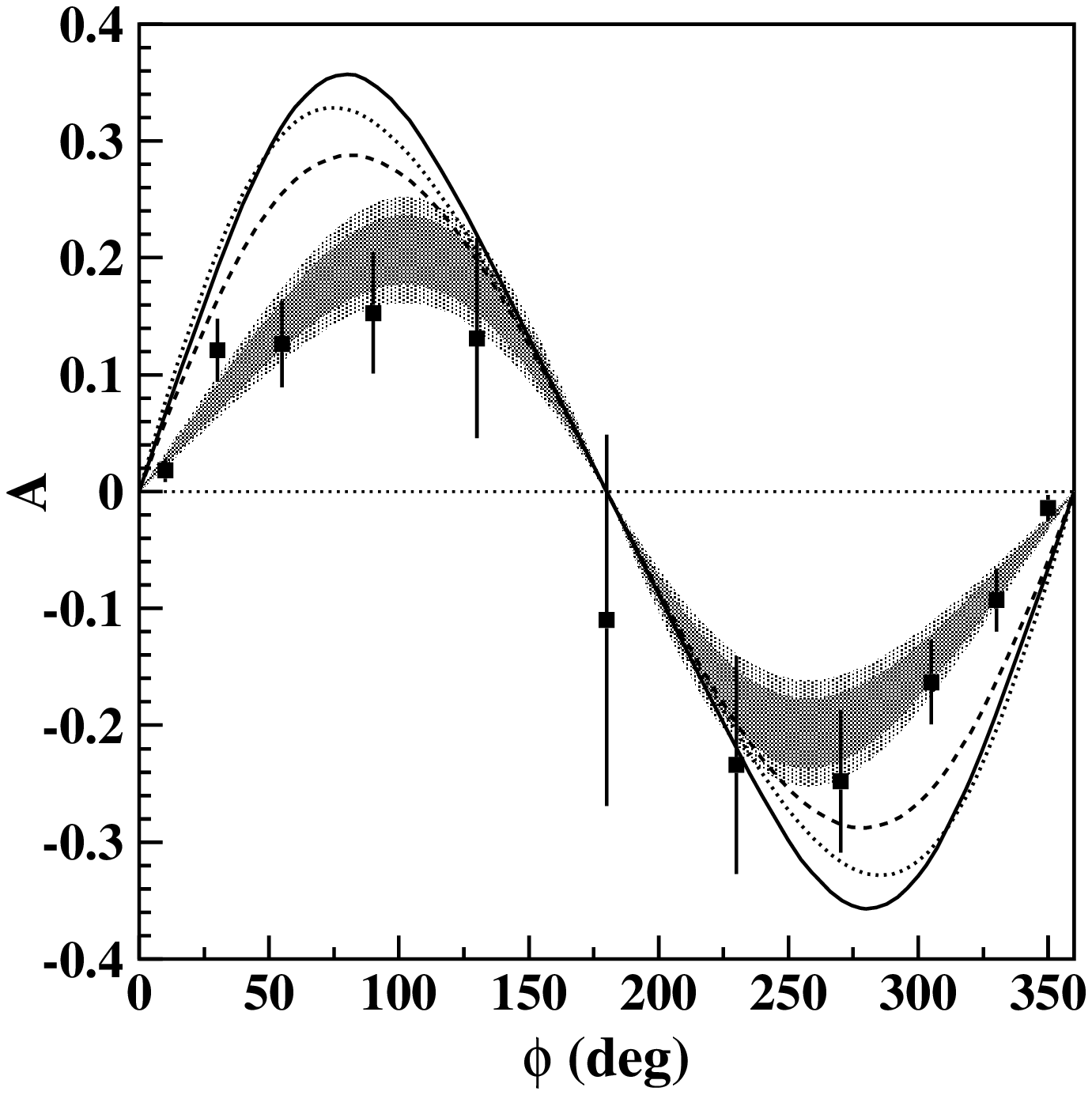}
\vspace{8.5cm}
\parbox{17.5cm}
{\caption[Delta]
{
Recent data from HERMES (left) and the CLAS experiment at 
Jefferson Laboratory (right)
in the realm of DVCS Bethe-Heitler interference.
The $\sin \phi$ azimuthal dependence of the single spin asymmetry is
clearly visible in the data \cite{Airapetian:2001,Stepanyan:2001}.
}
}
\label{fig:fig21}
\end{figure}

\end{widetext}

\subsection{Generalized parton distributions}

For exclusive processes such as DVCS or hard meson production 
the generalized parton distributions
involve non-forward proton matrix elements
\cite{Goeke:1998,Ji:1998,Diehl:2003,Radyushkin:1997,Vanderhaeghen:1998}.
The important kinematic variables are the virtuality of the hard
photon $Q^2$,
the momenta $p - \Delta/2$ of the incident proton and 
$p + \Delta /2$ of the outgoing proton,
the invariant four-momentum transferred to the target $t = \Delta^2$,
the average nucleon momentum $P$, 
the generalized Bjorken variable $k^+ = x P^+$
and 
 the light-cone momentum transferred to the target proton
$\xi = - \Delta^+/2p^+$.
The generalized parton distributions are defined as 
the light-cone Fourier transform of the point-split 
matrix element
\begin{eqnarray}
& & 
{P_+ \over 2 \pi} 
\int d y^- e^{-ixP^+y^-}
\langle p' | {\bar \psi}_{\alpha}(y) \psi_{\beta}(0) |p \rangle 
_{y^+ = y_{\perp} = 0}
\nonumber \\
& & 
=
{1 \over 4} \gamma^-_{\alpha \beta}
\biggl[ H(x,\xi,\Delta^2) {\bar u}(p') \gamma^+ u(p) 
\nonumber \\
& & 
\ \ \ \ \ \ \ \ \ \ \ \ \ \ \ \ \ \ \ \ \ \
+
        E(x,\xi,\Delta^2) {\bar u}(p') \sigma^{+ \mu} {\Delta_{\mu} \over 2M}
        u(p) \biggr]
\nonumber \\
& & 
\ \ \ 
+ {1 \over 4} ( \gamma_5 \gamma^-)_{\alpha \beta}
\biggl[ {\tilde H} (x,\xi,\Delta^2) {\bar u}(p') \gamma^+ \gamma_5 u(p) 
\nonumber \\
& &
\ \ \ \ \ \ \ \ \ \ \ \ \ \ \ \ \ \ \ \ \ \ \ \ \ 
+
        {\tilde E} (x,\xi,\Delta^2) 
        {\bar u}(p') \gamma_5 {\Delta^+ \over 2M} u(p) \biggr]
.
\nonumber \\
\label{eqk152}
\end{eqnarray}
(Here we work in the light-cone gauge $A_+=0$
 so that the path-ordered gauge-link becomes trivial and equal 
 to one to maintain gauge invariance through-out.)

The physical interpretation of the generalized parton distributions
(before worrying about possible renormalization effects and higher
 order corrections) is the following.
Expanding out the quark field operators in (\ref{eqk152}) 
in terms of light-cone quantized creation and annihilation operators
one finds that
for $x > \xi$ ($x < \xi$) the GPD is 
the amplitude to take a quark 
(anti-quark) of momentum $k - \Delta / 2$ out of the proton and reinsert
a quark (anti-quark) of momentum $k+\Delta/2$ into the proton some distance
along the light-cone to reform the recoiling proton.
In this region the GPD is a simple generalization of the usual 
parton distributions studied in inclusive and semi-inclusive scattering.
In the remaining region $-\xi < x < \xi$ the GPD involves taking out (or
inserting) a $q {\bar q}$ pair with momentum $k - \Delta / 2$ and 
$-k - \Delta / 2$ (or $k + \Delta /2$ and $-k + \Delta / 2$) respectively.
Note that the GPDs are interpreted as probability amplitudes rather than
densities.

In the forward limit the GPDs $H$ and ${\tilde H}$ are related to the forward 
parton distributions studied in (polarized) deep inelastic scattering:
\begin{eqnarray}
H (x,\xi,\Delta^2)|_{\xi=\Delta^2=0} &=& q (x)
\nonumber \\
{\tilde H} (x,\xi,\Delta^2)|_{\xi=\Delta^2=0} &=& \Delta q(x)
\label{eqk153}
\end{eqnarray}
whereas the GPDs $E$ and ${\tilde E}$ have no such analogue.
In the fully renormalized theory the spin dependent distributions 
${\tilde H}$ and ${\tilde E}$ will be sensitive to the physics of 
the axial anomaly and, in this case, 
it is not easy to separate off an ``anomalous component'' because 
the non-forward matrix elements of the gluonic Chern-Simons current
are non-gauge-invariant even in the light-cone gauge $A_+=0$.
Integrating over $x$ the first moments of the GPDs are related
to the nucleon form-factors:
\begin{eqnarray}
\int_{-1}^{+1} dx H (x,\zeta, \Delta^2) &=& F_1 (\Delta^2)
\nonumber \\
\int_{-1}^{+1} dx E (x,\zeta, \Delta^2) &=& F_2 (\Delta^2)
\nonumber \\
\int_{-1}^{+1} dx {\tilde H} (x,\zeta, \Delta^2) &=& G_A (\Delta^2)
\nonumber \\
\int_{-1}^{+1} dx {\tilde E} (x,\zeta, \Delta^2) &=& G_P (\Delta^2)
.
\label{eqk154}
\end{eqnarray}
Here
$F_1$ and $F_2$ are the Dirac and Pauli form-factors of the nucleon,
and 
$G_A$ and $G_P$ are the axial and induced-pseudoscalar form-factors
respectively.
(The dependence on $\xi$ drops out after integration over $x$.)

The GPD formalism allows one, in principle, to extract information
about quark angular momentum from hard exclusive reactions \cite{Ji:1997a}.
The current associated with Lorentz transformations is
\begin{equation}
M_{\mu \nu \lambda} 
= z_{\nu} T_{\mu \lambda} - z_{\lambda} T_{\mu \nu}
\label{eqk155}
\end{equation}
where $T_{\mu \nu}$ is the QCD energy-momentum tensor.
Thus, the total angular momentum operator is related 
to the energy-momentum tensor through the equation
\begin{equation}
J_{q,g}^z = 
\langle p', {1 \over 2} | \
\int d^3z \ ( \vec{z} \ {\rm x} \ {\vec T}_{q,g} )^{z} \ 
| p, {1 \over 2} \rangle
.
\label{eqk156}
\end{equation}
The form-factors corresponding to the energy-momentum tensor can 
be projected out by taking the second moment with respect to $x$)
of the GPD.
One finds Ji's sum-rule for the total quark angular momentum
\begin{equation}
J_q = {1 \over 2}
\int_{-1}^{+1} dx x 
\biggl[ H(x,\zeta,\Delta^2=0) + E(x,\zeta,\Delta^2=0) \biggr]
.
\label{eqk157}
\end{equation}
The gluon ``total angular momentum'' could then be obtained
through the equation
\begin{equation}
\sum_q J_q + J_g = {1 \over 2}
.
\label{eqk158}
\end{equation}
In principle, it could also be extracted from precision measurements of 
the $Q^2$ dependence of hard exclusive processes like DVCS and meson 
production at next-to-leading-order accuracy where the quark GPD's mix 
with glue under QCD evolution.

To obtain information about the ``orbital angular momentum'' $L_q$
we need to subtract the value of the ``intrinsic spin'' measured 
in polarized deep inelastic scattering 
(or a future precision measurement of $\nu p$ elastic scattering) 
from the total quark angular momentum 
$J_q$.
This means that $L_q$ is scheme dependent with different schemes 
corresponding to different physics content depending on how the scheme 
handles information about the axial anomaly, large $k_t$ physics and 
any possible ``subtraction at infinity'' in the dispersion relation
for $g_1$.
The quark total angular momentum $J_q$ is anomaly free in QCD so that
QCD axial anomaly effects occur with equal magnitude and opposite sign
in $L_q$ and $S_q$
\cite{Bass:2002a,Shore:2000}.
The ``quark orbital angular momentum'' $L_q$ is measured by the proton 
matrix element of $[{\bar q} ({\vec z} \ {\rm x} \ {\vec D})_3 q](0)$.
The gauge covariant derivative means that $L_q$ 
becomes sensitive to gluonic degrees of freedom 
in addition
to the axial anomaly,
--- for a recent discussion see \textcite{Jaffe:2001}.
A first attempt to extract the valence contributions 
to the energy-momentum form-factors entering 
Ji's sum rule is reported in \textcite{Diehl:2004}.

The study of GPDs is being pioneered in experiments at HERMES, 
Jefferson Laboratory and COMPASS.
Proposals and ideas exist for dedicated studies using a 12 GeV
CEBAF machine, a possible future polarized $ep$ collider (EIC)
in connection with RHIC or JLab,
and 
a high luminosity polarized proton-antiproton collider at GSI.
To extract information about quark total angular momentum one needs 
high luminosity, plus measurements over a range of kinematics
$Q^2$, $x$ and $\Delta$ 
(bearing in mind the need to make reliable extrapolations
 into unmeasured kinematics).
There is a challenging programme to disentangle the GPDs 
from the formalism and to undo the convolution integrals
which relate the GPDs to measured cross-sections, and to 
check (experimentally) the kinematics where twist 2 dominates.
Varying the photon or meson in the final state will give access to
different spin-flavour combinations of GPDs even with unpolarized
beams and targets.
Besides yielding possible information about the spin structure of
the proton,
measurements of hard exclusive processes will, in general, 
help to constrain our understanding of the structure of the proton.

\section{POLARIZED PHOTON STRUCTURE FUNCTIONS}

Deep inelastic scattering from photon targets reveals many novel effects 
in QCD. The unpolarized photon structure function has been well studied 
both theoretically and experimentally.
The polarized photon spin structure function is an ideal
(theoretical) laboratory to study the QCD dynamics associated with the
axial anomaly.

The photon structure functions are observed experimentally in
$e^{+}$ $e^{-} \rightarrow$ hadrons where for example 
a hard photon (large $Q^{2}$)
probes the quark structure of a soft photon ($P^{2} \sim 0$).
For any virtuality $P^{2}$ of the target photon the measured structure 
functions receive a contribution both from contact photon-photon 
fusion and also a hadronic piece, 
which is commonly associated
 with vector meson dominance (VMD) of the soft target photon. 
The hadronic term scales with $Q^{2}$ whilst
the contact term behaves as $\ln Q^{2}$ as we 
let $Q^{2}$ tend to $\infty$.
This result was discovered by \textcite{Witten:1977}
for the unpolarized 
structure function $F_2^{\gamma}$
and extended to the polarized case in Refs. \cite{Manohar:1989,Sasaki:1980}.
The $\ln Q^{2}$ scaling behaviour mimics the leading-order 
box diagram prediction but the coefficient of 
the logarithm receives a finite renormalization in QCD.
From the viewpoint of the renormalization group the essential
detail
discovered by Witten is that the coefficient functions of the 
photonic and singlet hadronic operators will mix under QCD evolution.
The hadronic matrix elements are of leading order in
$\alpha$ 
whilst the photon operator matrix elements are $O(1)$. 
Since the hadronic coefficient functions are $O(1)$ and 
the photon coefficient functions start at $O(\alpha)$ 
the photon structure functions
receive leading order contributions in $\alpha$ 
from both the hadronic and photonic channels.

In polarized scattering the first moment of $g_1^{\gamma}$ is especially 
interesting.
First, consider a real photon target
(and assume no fixed pole correction).
The first moment of $g_1^{\gamma}$ vanishes
\begin{equation}
\int_0^1 \ dx  \ g_1^{\gamma} (x, Q^2) = 0
\label{eql159}
\end{equation}
for a real photon target
independent of the virtuality $Q^2$ of the photon that it is probed with
\cite{Bass:1998b,Bass:1992}.
This result is non-perturbative.
To understand it,
consider the real photon as the beam and the virtual photon as the target.
Next
apply the Gerasimov-Drell-Hearn sum rule.
The anomalous magnetic moment of a photon vanishes to all orders because of
Furry's theorem whence one obtains the sum-rule. 
The sum rule (\ref{eql159}) holds to all orders in perturbation theory and
at every twist \cite{Bass:1998b}.

The interplay of QCD and QED dynamics here can be seen through the
axial anomaly equation
\begin{equation}
\partial^{\mu} J_{\mu 5} 
= 
2 m_q {\overline q} i \gamma_5 q
+ 
{\alpha_s \over 4 \pi} G_{\mu \nu} {\tilde G}^{\mu \nu}
+ 
{\alpha \over 2 \pi}   F_{\mu \nu} {\tilde F}^{\mu \nu}
\label{eql160}
\end{equation}
(including the QED anomaly).
The gauge-invariantly renormalized axial-vector current
can then be written as the sum of the partially conserved
current plus QCD and abelian QED Chern-Simons currents
\begin{equation}
J_{\mu 5} = J_{\mu 5}^{\rm con} + K_{\mu} + k_{\mu}
\label{eql161}
\end{equation}
where $k_{\mu}$ is the anomalous Chern Simons current in QED.

The vanishing first moment of $\int_0^1 dx g_1^{\gamma}$ is the
sum of a contact term 
$-{\alpha \over \pi} \sum_q e_q^2$
measured by the QED Chern Simons current and a 
hadronic term associated with the two QCD currents in Eq.(\ref{eql161}).
The contact term is associated with high $k_t$ leptons and
two quark jet events (and no beam jet) in the final state.
For the gluonic contribution associated with polarized glue
in the hadronic component of the polarized photon, the two
quark jet cross section is associated with an extra soft ``beam jet''.

For a virtual photon target one expects the first moment 
to exhibit 
similar
behaviour to that suggested by the tree box graph amplitude 
--
that is, for $P^2 \gg m^2$ the first moment tends to equal
just the QED anomalous contribution and the hadron term vanishes.
However, here the mass scale $m^2$ 
is expected to be set by the $\rho$ meson mass corresponding
to a typical hadronic scale and vector meson dominance 
of the soft photon instead of the light-quark mass or the pion mass
\cite{Shore:1993a,Shore:1993b}.
Measurements of $g_1^{\gamma}$ 
might be possible with a polarized $e \gamma$ collider 
\cite{DeRoeck:2001}.
The virtual photon target could be investigated through the study of 
resolved photon contributions to polarized deep inelastic scattering
from a nucleon target 
\cite{Stratmann:1998}.
Target mass effects in the polarized virtual photon structure function
are discussed in
\textcite{Baba:2003}.

\section{CONCLUSIONS AND OPEN QUESTIONS}

The exciting challenge to understand 
the
{\it Spin Structure of the Proton}
has produced many unexpected surprises in experimental data
and inspired much theoretical activity and new insight into
QCD dynamics and
the interplay between spin and chiral/axial U(1) symmetry breaking in QCD.

There is a vigorous global programme in experimental spin physics 
spanning
(semi-)inclusive polarized deep inelastic scattering, 
photoproduction experiments,
exclusive measurements over a broad kinematical region,
polarized proton-proton collisions,
fragmentation studies in $e^+ e^-$ collisions and $\nu p$ elastic
scattering.

In this review we surveyed the present (and near future) experimental 
situation and
the new theoretical understanding that spin experiments have inspired.
New experiments (planned and underway) will surely produce more surprises
and exciting new challenges for theorists as we continue our quest to
understand the internal structure of the proton and QCD confinement related
dynamics.

We conclude with a summary of key issues and open problems in QCD 
spin physics where the next generation of 
present and future experiments should yield vital information:
\begin{itemize}
\item
What happens to ``spin''
in the transition from current to constituent
quarks through dynamical axial U(1) symmetry breaking ?
\item
How large is the gluon spin polarization in the proton ?
If $\Delta g$ is indeed large, what would this mean for models of
the structure of the nucleon ?
What dynamics could produce a large $\Delta g$ ?
\item
Are there fixed pole corrections to spin sum rules for polarized
photon nucleon scattering ?
If yes, which ones ?
\item
Is gluon topology important in the spin structure of the proton ?
\item
What is the $x$ and $k_t$ dependence of the (negative) polarized
strangeness 
extracted from inclusive and 
semi-inclusive polarized deep inelastic scattering ?
\item
How (if at all) do the effective intercepts for small $x$ physics 
change in the transition region between polarized photoproduction 
and polarized deep inelastic scattering ?
\item
In which kinematics, if at all, does the magnitude of the isosinglet 
component of 
$g_1$ at small $x$ exceed the magnitude of the isovector component ?
\item
How does $\Delta d / d$ behave at $x$ very close to one ?
\item
Does perturbative QCD factorization work for spin dependent processes ?
--
{\it viz.} 
will the polarized quark and gluon distributions extracted
 from the next generation of experiments prove to be process independent ?
\item
Is the small value of $g_A^{(0)}$ extracted from polarized deep inelastic
scattering  ``target independent'', e.g. through topological charge screening ?
\item
Can we find and oberserve processes in the $\eta'$ nucleon interaction 
which are also sensitive to dynamics which underlies the singlet axial
charge ?
\item
How large is quark (and gluon) ``orbital angular momentum'' in the proton ?
\item
Transversity measurements are sensitive to $k_t$ dependent effects in 
the proton and fragmentation processes. 
The difference between the C-odd transversity distribution $\delta q(x)$
and 
the C-even spin distribution $\Delta q(x)$, 
{\it viz.}
$(\delta q - \Delta q)(x)$,
probes relativistic dynamics in the proton.
Precision measurements at large Bjorken $x$ where just the valence quarks 
contribute would allow a direct comparison and teach us about relativistic 
effects in the confinement region.
\end{itemize}

\section*{Acknowledgments}

My understanding of the issues discussed here has benefited 
from collaboration, conversations and correspondence with 
many colleagues.
It is a pleasure to thank 
M. Brisudova, S.J. Brodsky, R.J. Crewther, A. De Roeck, A. Deshpande, 
B.L. Ioffe, P.V. Landshoff, N.N. Nikolaev, I. Schmidt, F.M. Steffens 
and A.W. Thomas
for collaboration and sharing their insight on the spin structure of
the proton.
In addition I have benefited particularly from discussions with 
M. Anselmino, B. Badelek, N. Bianchi, V.N. Gribov, R.L. Jaffe, 
P. Kienle, W. Melnitchouk, P. Moskal, G. R\"adel, G.M. Shore, 
J. Soffer, P. van Baal, W. Vogelsang and R. Windmolders.
I thank C. Jarlskog for her enthusiasm and support and A. Bialas 
for his invitation to lecture at the 2003 Cracow School of Theoretical 
Physics in Zakopane. 
Those lectures laid the foundation for the present review.
I thank
C. Aidala, M. Amarian, E. Beise, P. Bosted, K. Helbing, G. Mallot, 
Z.E. Meziani, 
U. Stoesslein, R. Tayloe and G. van der Steenhoven
for helpful 
communications about experimental data during the writing of this article.
This work was supported in part by the Austrian Science Fund 
(FWF grants M770-N08 and P17778-N08).
I thank R. Rosenfelder for helpful comments on the manuscript.

\newpage

\bibliographystyle{apsrmp}


\end{document}